\documentclass{IEEE-Con-Sys-mag}
\usepackage[utf8]{inputenc}
\usepackage[T1]{fontenc}
\usepackage{graphicx}
\graphicspath{{./Figs/}}
\usepackage{grffile}
\usepackage{longtable}
\usepackage{makecell}
\usepackage{wrapfig}
\usepackage{rotating}
\usepackage[normalem]{ulem}
\usepackage{textcomp}
\usepackage{amsmath,mathtools}
\usepackage{amsfonts,amssymb}
\usepackage{empheq}
\usepackage{enumerate}
\usepackage{capt-of}
\usepackage{hyperref}
\usepackage{tikz}\usetikzlibrary{matrix,arrows,positioning,fit}
\usepackage{graphicx}
\usepackage{color}
\usepackage{epstopdf}
\usepackage{savesym}
\savesymbol{iint}
\savesymbol{iiint}
\usepackage{times}
\usepackage{mathptmx}
\usepackage{multicol}
\usepackage{array}
\usepackage{mathdots} % for \iddots
\usepackage{url} % [hyphens]
\usepackage{algorithm}
\usepackage{algorithmic}

\renewcommand{\algorithmicrequire}{\textbf{Input:}}
\renewcommand{\algorithmicensure}{\textbf{Output:}}

\renewcommand{\hat}{\widehat}
\renewcommand{\tilde}{\widetilde}

\newcommand{\la}[1]{\label{#1}}

\newcommand{\bmx}{\begin{bmatrix}}
\newcommand{\emx}{\end{bmatrix}}
\newcommand{\bsm}{\left[\begin{smallmatrix}}
\newcommand{\esm}{\end{smallmatrix}\right]}
\newcommand{\norm}[1]{\|#1\|}

\newcommand{\longpage}{\enlargethispage{\baselineskip}}

	       {\begin{list}{}{\renewcommand{\makelabel}[1]{\textsf{##1}\hfil}%
		     \settowidth{\labelwidth}{\textsf{#1}}%
		     \setlength{\leftmargin}{\labelwidth+\labelsep}}}%
	       {\end{list}}

\definecolor{jgreen}{rgb}{0.0, 0.5, 0.0}

\newcommand{\hermconj}  {^{\mathsf{H}}}

\usepackage[customcolors]{hf-tikz}
\usepackage[compress]{cite}
\usepackage[font=footnotesize]{caption}
\usepackage[font=footnotesize]{subcaption}
\usepackage{comment}

\usepackage{xcolor}
\definecolor{backgroundred}{rgb}{1, 0.6, 0.6}
\definecolor{backgroundblue}{rgb}{0.6, 0.8, 1}

\jvol{XX}
\jnum{XX}
\paper{X}
\jmonth{October}
\jname{IEEE CONTROL SYSTEMS}
\pubyear{2026}

\begin{document}

% \title{\stitle{Decentralized Stability Certificates: Towards Modular and Scalable Analysis of Renewable Power Systems}}
\title{\stitle{Decentralized Stability Certificates for Interoperability in Power-Electronics-Dominated Power Systems}}
\author{{Linbin Huang}, Verena Häberle, Ruohan Leng, Huanhai Xin, Hua Geng, Florian Dörfler, and Xiuqiang He \vspace{5mm}}

\affil{{L. Huang, R. Leng, and H. Xin} are with the College of Electrical Engineering, Zhejiang University, Hangzhou 310058, China (e-mails: hlinbin@zju.edu.cn, lengruohan@zju.edu.cn, xinhh@zju.edu.cn), \linebreak V. Häberle and F. Dörfler are with the Automatic Control Laboratory, ETH Z\"urich, 8092 Zürich, Switzerland (e-mails: verenhae@control.ee.ethz.ch, dorfler@ethz.ch), \linebreak H. Geng and X. He are with the Department of Automation, Tsinghua University, Beijing 100084, China (e-mails: genghua@tsinghua.edu.cn, xhe@tsinghua.edu.cn). 
\vspace{-7mm}}

\maketitle

% \dois{}{}

\begin{summary}

% \summaryinitial{T}he stability analysis of large-scale multi-agent network systems is a challenging task, especially when the agents exhibit heterogeneous dynamics, such as in power systems. Decentralized stability conditions are promising solutions to this task, as they allow for assessing stability in a modular and scalable manner. The passivity condition and the small-gain condition are representative examples, and recent advances include the developments of small-phase theorem, scaled-relative graphs (SRGs), Davis-Wielandt (DW) shells, and so on. In this self-contained tutorial reviews, we revisit different decentralized stability conditions and investigate their connections. We show that the decentralized stability condition derived from the three-dimensional DW shells is the least conservative one among existing conditions in the literature, and the projection of DW shells onto different planes and axes recovers many other conditions such as passivity. Moreover, we present the application of decentralized stability conditions in power electronics-dominated power systems and demonstrate how they enable parametric conditions for guiding control design. 

\summaryinitial{T}he rapid integration of power electronics converters has fundamentally altered the power grid’s dynamic landscape,
making modularity of operation and interoperability of heterogeneous devices two critical hurdles for power system reliability.
In control language, these challenges can be paraphrased as decentralized certificates on the input/output behavior of every
device, assured by local control, so that stability and guaranteed performance of the interconnected system emerge. For the
power system application, such certificates should eventually be part of the rule book, the so-called grid code, and thus be
generic and minimally restrictive for the control architecture of each device. Classical system-theoretic angles of
attack to these questions go back to dissipativity concepts, which have historically been co-developed in automatic
control, circuit theory, and power systems. Since standard dissipativity (e.g., passivity or small gain) criteria fall short in terms of conservatism, a range of more flexible and advanced tools have been developed to assure less conservative
decentralized certificates guaranteeing stability under interconnection as well as dynamic performance following a disturbance. In this
article, we review a range of methods that have been put forward to tackle these questions, compare and validate them, and point towards practically relevant research gaps which are amenable to control-theoretic approaches. Moreover, we provide a tutorial exposition of practically relevant power systems/electronics issues accessible to control engineers.
\end{summary}

\section{}

% Power electronics (PE) converters are becoming one of the most important, albeit complex, components in modern power systems, as they serve as the grid interface for renewable energy sources, data centers, energy storage systems, electric vehicles, and high-voltage DC transmission systems. Many countries have witnessed a dramatic growth in wind and solar generation in the past decade. For instance, in 2025, wind and solar combined generated 30\% of all EU electricity, officially surpassing fossil fuels (29\%) for the first time in history; in 2026, solar generation surpassed coal-fired power to become China's largest installed power source. With the rapid integration of renewable energy sources, modern power systems are evolving towards PE-dominated power systems. 
\vspace{-7mm}
Power electronics (PE) converters have emerged as a critical, albeit profoundly complex, component of modern power systems. Serving as the indispensable grid interface for renewable energy sources, energy storage systems, high-voltage direct current transmission systems, and electronic loads such as electric vehicles and data centers, their presence is expanding at an unprecedented rate~\cite{milano2018foundations,kroposki2017achieving}. For instance, wind and solar combined generated 30\% of all EU electricity in 2025, officially surpassing fossil fuels (29\%) for the first time in history; in 2026, solar generation surpassed coal-fired power to become China's largest installed power source. 
Driven by this rapid integration of renewables, traditional power grids are fundamentally evolving into PE-dominated power systems.

\begin{pullquote}
Modern grid codes already impose passivity requirements on converters over specified frequency ranges to ensure stable grid integration.
\end{pullquote}

The dynamic behavior of PE converters markedly differs from that of conventional fossil-fuel-based power plants. They have low inertia and weak capability to support the grid voltage and frequency, rely on multiple nested control loops to realize synchronization and power transfer, subject to tight state constraints, operating-point dependent behavior, and are distributed and often non-dispatchable~\cite{wang2018harmonic,ellabban2014renewable}. 
Moreover, the design principles and physical constraints of PE converters are different across different application scenarios, resulting in heterogeneous dynamics and performance under grid disturbances. For instance, the PE converters in photovoltaics aim to regulate the dc-link voltage, while the PE converters in energy storage systems focus on regulating active power. The heterogeneity, on the other hand, also comes from the variety of control strategies of converters such as the traditional grid-following (GFL) control and the emerging grid-forming (GFM) control~\cite{rocabert2012control,wen2015impedance,lasseter2019grid,huang2019grid,xin2024many}.
The large-scale integration of heterogeneous PE converters has profoundly altered power system dynamics and posed major challenges for stability analysis and reliable operation. Indeed, many real-world grid accidents involve (though are not caused by) high PE penetrations, such as the 2019 power outage in the UK, the 2016 South Australian blackout, and the 2025 Iberian Peninsula blackout. Moreover, oscillation and instability events have been frequently observed in many wind farms worldwide and even in data centers, which are caused by the complex interaction between converters and the power grid~\cite{liu2017subsynchronous,Cheng2023,force2025characteristics}. 

% It is worth mentioning that after an oscillation occurs, the system operators can make extensive efforts to reproduce the event in simulations by dedicated tuning and recalibration and perform ex post analysis, but they generally cannot predict or prevent the events beforehand. The reason behind this is that conventional approaches such as eigenvalue analysis and frequency-domain analysis based on the (generalized) Nyquist criterion can hardly analyze high-dimensional PE-dominated power systems effectively due to the lack of scalability. To be specific, the stability of PE-dominated power systems is determined by the closed-loop interaction between the power network and the large number of heterogeneous converters (each converter usually has more than 10 state variables), and conventional approaches like eigenvalue analysis require building, e.g., high-dimensional state-space matrices, of the full interconnected systems, which cannot clearly explain how the heterogeneous converters interact with the power network and why such interaction may result in instabilities.

Following such instability events, system operators often invest considerable effort in reproducing the observed phenomena retrospectively by meticulously tuning, identifying, and recalibrating electromagnetic transient (EMT) simulation models. Yet, they remain largely incapable of predicting or preventing these events \textit{a priori}. This critical vulnerability stems from the fact that conventional stability analysis techniques, such as eigenvalue computation or the generalized Nyquist criterion, lack the scalability required for high-dimensional, PE-dominated power systems. 
Because the stability of such a grid is dictated by the closed-loop interactions between the power network and a large number of heterogeneous converters (each introducing a multitude of state variables), centralized full models become analytically and computationally intractable, and they are not even available due to data silos amongst the system operators and the vendors. More importantly, these monolithic models obscure the network interconnection, failing to explain how individual converters interact with the power network and why specific interactions trigger instability. It is therefore preferable to have model-agnostic interface specifications for enforcing interconnection stability, without centralized models and analysis.

On the control design side, PE converters have much higher controllability and flexibility than conventional synchronous machines, which, via appropriate design, can support system stability and prevent oscillations~\cite{yang2020placing,wang2015,Poolla2019,ma2024assessing}. However, the sheer scale, complexity of dynamic grid-device interactions, and model opacity render  holistic stability guarantees elusive. Furthermore, centralized analysis tools provide virtually no actionable guidance on how individual converters should be locally controlled to ensure the stability of the global interconnected system.

To simultaneously overcome these analytical bottlenecks, data silos, and design hurdles, the concept of decentralized stability certificates has gained significant traction~\cite{huang2020h,huang2024gain,feng2025unified,Baron2025,wang2026d,Niehues2026, pates2019robust,gorbunov2021identification,vorobev2019decentralized}. This line of research traces back to the seminal works by Moylan and Hill in the 1970s analyzing the stability of multi-machine power systems using dissipativity~\cite{moylan1977stability,moylan1978stability}.
By specifying rigorous bounds on the local input--output behavior of each device, decentralized stability certificates mathematically guarantee the stability and dynamic performance of the interconnected system. Under this paradigm, global stability analysis is reduced to verifying that each individual device satisfies a local criterion, a profoundly scalable approach that is indifferent to the network's sheer size and can cope with only locally available device models. Concurrently, it offers explicit design guidelines: local controllers should simply be designed to adhere to these local certificates. Ultimately, such modular, ideally little conservative, and minimally restrictive certificates are envisioned to form the technical foundation of future grid codes in PE-dominated power systems, ensuring the modularity and interoperability of heterogeneous devices. In fact, these ideas (particularly passivity) have drawn great attention from the power industry and have been included in some grid codes. For instance, NERC requires that GFM converters should present a non-negative resistance and present a passive characteristic to the grid within a wide frequency range (0-300~Hz) to prevent adverse interactions~\cite{hoke2023white}; UNIFI requires that GFM converters should be passive from approximately 4~Hz to 40~Hz and have low gain from 10~Hz to 40~Hz~\cite{kroposki2026unifi}; Fingrid requires that GFM converters shall present a positive resistance to the grid within frequency ranges 0–47~Hz and 53-250~Hz to prevent adverse interactions~\cite{antti2023specific}.

Classic dissipativity conditions, such as passivity and small gain, are examples of decentralized stability certificates, which are popular in analyzing converter dynamics~\cite{Harnefors2015Passivity,Harnefors2015Controller,chen2015passivity}. However, these two conditions could be conservative and difficult for converters to directly satisfy in the whole frequency range~\cite{huang2024gain,chen2024limitations}. To this end, a range of more flexible and advanced tools have been developed to reduce the conservatism.
The so-called small-phase theorem, as a counterpart to the small-gain theorem, was established in~\cite{chen2024phase, wang2020phases}, which extends the passivity condition and reduces its conservatism. 
The small-phase theorem builds upon the concept of matrix phases, which is defined using the numerical range and requires that the numerical range does not include the origin, i.e., the matrix is sectorial. The passivity condition is a special case of the small-phase theorem, where the matrix phases are constrained within $\pm 90^\circ$. 
Moreover, the small-phase theorem can be combined with gain information, leading to a mixed small gain-phase theorem, which provides more flexibility in analyzing system stability~\cite{zhao2022small, woolcock2023mixed}. 
Advancing this trajectory, recent literature~\cite{feng2025unified,huang2025geometric,leng2026geometric} has demonstrated that the Davis-Wielandt (DW) shell, an elegant mathematical tool dating back to the 1950s~\cite{davis1968shell,wielandt1955eigenvalues,lestas2012large}, enables a unified and even more generalized set of decentralized certificates. Operating as a 3-dimensional (3D) geometric extension of the 2D numerical range, the DW shell explicitly captures both the phase and gain characteristics of a complex matrix and enables stability analysis from a geometric perspective. We will further demonstrate in this article how the DW shell is compatible with scaled relative graphs (SRGs) and integral quadratic constraints (IQCs) which have been leveraged for power system analysis~\cite{Baron2025,hallinan2026decentralised,liu2026scalable}.

In the remainder of this article, we elucidate how the DW shell enables decentralized stability certificates and bridge this advanced concept with established results like passivity, small gain, small phase, SRGs, and IQCs. The interpretability, computational tractability, and conservatism of different certificates will be compared and discussed. We give a tutorial presentation of complex PE-dominated power system models from the perspective of structured closed-loop interactions so that they are accessible to control theorists and engineers. Furthermore, we demonstrate how these decentralized certificates facilitate scalable stability and dynamic performance analysis and guarantee interoperability in PE-dominated power systems from both parametric and non-parametric (model-free) perspectives. We begin by revisiting the classical concept of dissipativity.

\section{Dissipativity for Decentralized Stability Certificates}
\label{sec:passivity}
% Reviewing passivity and passivity in the frequency domain, and then also in the time domain for nonlinear system analysis

%Passivity-based analysis has been popular in the stability analysis of power electronics converters~\cite{harnefors2015passivity}.

Dissipativity provides a general framework for characterizing how a dynamical system stores, dissipates, and exchanges energy with other systems. The particular relevance of dissipativity to PE-dominated power systems stems from its \emph{compositionality}: a stability certificate can be established from local subsystem conditions without explicitly analyzing the full interconnected system. This makes dissipativity particularly attractive for large-scale systems consisting of heterogeneous power converters and networks. In ``\nameref{sb-history-passivity}'', we review the development and application of dissipativity and passivity theory in circuit theory and power system stability analysis.

\begin{sidebar}{}
\setcounter{sequation}{0}
\renewcommand{\thesequation}{S\arabic{sequation}}
\setcounter{stable}{0}
\renewcommand{\thestable}{S\arabic{stable}}
\setcounter{sfigure}{0}
\renewcommand{\thesfigure}{S\arabic{sfigure}}
\section{Dissipativity and Passivity in Circuits and Power Systems: A Historical Perspective}
\label{sb-history-passivity}

Passivity theory originated from circuit theory and was developed into a general systems theory, laying the foundation for its subsequent application to power system stability and control. In the 19th century, circuit theory established the physical distinction between energy storage in inductors and capacitors versus energy dissipation in resistors. During the first half of the 20th century, passive network theory and positive-real functions provided a mathematical framework for characterizing passive electrical networks \cite{Desoer1969}. In the 1970s, Willems formalized dissipative dynamical systems using storage functions and supply rates \cite{Willems1972-1,Willems1972-2}, followed by Hill and Moylan’s work on their input-output and state-space properties, which clarified the relationship between dissipativity and stability in interconnected feedback systems \cite{Hill1980}. In parallel, passivity-based control (PBC) developed as a systematic control methodology, with early contributions by Ortega and Spong in the late 1980s \cite{Ortega1989} and subsequent applications to electrical systems through energy shaping and damping injection.

At the circuit level, passivity has a direct physical interpretation in terms of port power and energy dissipation. With the passive sign convention, the instantaneous power entering a circuit is $p(t)=v(t)^\top i(t)$, where $v(t)$ and $i(t)$ denote the port voltage and current, respectively. For an idealized circuit, the energy balance can be expressed as $\dot V(x)=p(t)-p_{\mathrm{diss}}(t)\leq p(t)$, where $V(x)$ is the stored energy and $p_{\mathrm{diss}}(t)\geq0$ is the dissipated power. More generally, passivity requires that the rate of change of stored energy does not exceed the supplied power, as expressed by the input-output condition in \eqref{eq:passivity}. For circuit impedance models in the frequency domain, passivity requires the real part of the impedance (or its Hermitian part in multi-input multi-output systems) to be positive semidefinite at all frequencies as in \eqref{eq:positive_real}. This property is known as \textit{positive-realness} in control theory and commonly referred to as \textit{positive resistance} in power electronics.

The physical interpretation of passivity in circuits and its suitability for analyzing interconnected systems have motivated its application to power system stability. For example, Harnefors \textit{et al.} investigated input-admittance shaping to improve the passivity of grid-connected converters \cite{Harnefors2007}, while Pates and Mallada \cite{pates2019robust}, Spanias and Lestas \cite{Spanias2019}, and Monshizadeh and Lestas \cite{Monshizadeh2019} developed decentralized stability criteria based on positive-realness and related input-output properties. To quantify the degree of passivity, Yang \textit{et al.} adopted output-differential passivity indices \cite{Yang2020} to derive stability conditions for power systems with heterogeneous nonlinear bus dynamics, while He \textit{et al.} established an output-feedback passivity (OFP) index for grid-connected converters and used the network's input-feedforward passivity (IFP) property to derive decentralized stability conditions \cite{he2024passivity}. Furthermore, H\"aberla \textit{et al.} established decentralized parametric stability certificates by leveraging dynamic loop-shifting techniques to compensate for the lack of passivity in the network dynamics \cite{Haeberle2026}. More general energy inequalities are captured by dissipativity theory: Moylan and Hill applied this framework to multimachine power-system stability \cite{moylan1977stability}, while \textit{QSR}-dissipativity has been used in decentralized control of nonlinear DC microgrids under plug-and-play operations \cite{Wang2024}. In the frequency domain, integral quadratic constraints (IQCs) provide a flexible framework for stability assessment, enabling to unify existing stability criteria and reduce conservatism in the analysis of power-electronic-dominated power systems \cite{Chen2025,Laib2023,hallinan2026decentralised,liu2026scalable}. This progress reflects the continued development of decentralized stability analysis toward less restrictive conditions that accommodate heterogeneous component dynamics and complex network interconnections.

\end{sidebar}

\begin{sidebar}{}
\setcounter{sequation}{0}
\renewcommand{\thesequation}{S\arabic{sequation}}
\setcounter{stable}{0}
\renewcommand{\thestable}{S\arabic{stable}}
\setcounter{sfigure}{0}
\renewcommand{\thesfigure}{S\arabic{sfigure}}
\section{Connection Between Dissipativity and Small-Gain Property}
\label{sb-passivity-small-gain}

In the dissipativity framework, a finite $L_2$-gain property can be established via the quadratic supply rate. Let us define the supply rate as
\begin{equation}
    w(u,y)=\gamma^2 u^\top u - y^\top y,
\end{equation}
with $\gamma > 0$. Suppose the system is dissipative with respect to this supply rate. That is, there exists a nonnegative storage function $V(x)$ (let $V(x(0)) = 0$) such that for all $T \ge 0$,
\begin{equation}
    V(x(T)) \le \int_0^T \left(\gamma^2 u^\top u - y^\top y\right)\mathrm{d}t.
\end{equation}
Since $V(x(T)) \ge 0$, rearranging the dissipativity inequality yields
\begin{equation}
    \int_0^T y^\top y \, \mathrm{d}t \le \gamma^2 \int_0^T u^\top u \, \mathrm{d}t - V(x(T)) \le \gamma^2 \int_0^T u^\top u \, \mathrm{d}t.
\end{equation}
Taking the limit as $T \to \infty$ and applying the monotone convergence theorem to the left-hand side, we obtain
\begin{equation}
    \|y\|_2 \le \gamma \|u\|_2.
\end{equation}
Therefore, dissipativity with respect to the quadratic supply rate $w(u,y)=\gamma^2 u^\top u - y^\top y$ serves as a sufficient condition for the system to have a finite $L_2$ gain $\gamma$.

For two subsystems with gains $\gamma_1$ and $\gamma_2$,
\begin{equation}
\dot V_i\leq \gamma_i^2u_i^\top u_i-y_i^\top y_i,
\qquad i=1,2,
\end{equation}
we consider the feedback interconnection as in Figure~\ref{fdbk}: $u_1=-y_2$, $u_2=y_1$.
Introducing positive multipliers $\lambda_1,\lambda_2>0$ and defining
\begin{equation}
V=\lambda_1V_1+\lambda_2V_2,
\end{equation}
gives
\begin{equation}
\begin{aligned}
\dot V
&\leq
(\lambda_2\gamma_2^2-\lambda_1)y_1^\top y_1
+(\lambda_1\gamma_1^2-\lambda_2)y_2^\top y_2.
\end{aligned}
\end{equation}
Hence, the interconnected system is dissipative with a nonpositive supply rate if there exist $\lambda_1,\lambda_2>0$ such that
\begin{equation}
\lambda_2\gamma_2^2<\lambda_1,\qquad
\lambda_1\gamma_1^2<\lambda_2.
\end{equation}
Such positive multipliers exist if and only if
\begin{equation}
\gamma_1\gamma_2<1,
\end{equation}
which is precisely the classical small-gain condition.

Under a common dissipativity framework, passivity and small gain differ primarily in their choice of supply rate: passivity uses the port-power supply $u^\top y$, whereas small gain uses the quadratic supply $\gamma^2u^\top u-y^\top y$.

Another connection between passivity and the small-gain condition can be established through the {\em scattering transformation}, beyond the choice of different supply rates in dissipativity theory. Specifically, for a passive transfer function $G(s)$, the positive-real condition, $G(j\omega)+G(j\omega) \hermconj \succeq0,\ \forall\omega \in \mathbb{R}$, is equivalent to the small-gain condition $\|(I-G(j\omega))(I+G(j\omega))^{-1}\|\leq 1,\ \forall\omega \in \mathbb{R}$. Graphically speaking, the scattering transformation maps the positive-realness condition into a unit-disc constraint, thereby establishing a direct link between passivity-based and small-gain-based stability analysis.

\end{sidebar}

\subsection{Dissipativity and Passivity for Nonlinear Systems}

Consider a nonlinear system
\begin{equation*}
\dot{x}=f(x,u),\quad y=h(x,u),
\end{equation*}
where $u$ and $y$ denote the inputs and outputs, respectively. Let $w(u,y)$ be a supply rate that characterizes the ``instantaneous power'' supplied to the system. The system is said to be \emph{dissipative} with respect to $w$ if there exists a nonnegative storage function $V(x)$ such that, along all system trajectories,
\begin{equation}
\dot{V}(x)\leq w(u,y).
\label{eq:dissipation_inequality}
\end{equation}
Thus, the storage function $V(x)$ measures the energy accumulated in the subsystem, while the supply rate accounts for the energy exchanged through its input-output ports. 

\emph{Passivity} is the special case of dissipativity, where
\begin{equation}
w(u,y)=u^\top y,
\label{eq:passivity}
\end{equation}
leading to $\dot{V}(x)\leq u^\top y$.

Dissipativity further allows the supply rate to encode margins of passivity. A commonly used form is
\begin{equation}
w(u,y)=u^\top y-\nu\,u^\top u-\rho\,y^\top y,
\label{eq:passivity_index_supply}
\end{equation}
where $\nu$ and $\rho$ are the input-feedforward passivity (IFP) and output-feedback passivity (OFP) indices, respectively \cite{khalil2002nonlinear}. Positive indices quantify a strict passivity margin, whereas negative indices characterize a shortage of passivity. This provides a useful quantitative description of how much non-passive behavior a subsystem can exhibit before interconnection stability is compromised.

More generally, a quadratic supply rate can be written as
\begin{equation}
w(u,y)=\begin{bmatrix}y\\u\end{bmatrix}^{\!\top}\begin{bmatrix}Q&S\\S^\top&R\end{bmatrix}\begin{bmatrix}y\\u\end{bmatrix},
\label{eq:general_supply_rate}
\end{equation}
where the matrices $Q$, $S$, and $R$ define the desired dissipativity property (also termed \textit{QSR}-dissipativity), including passivity and passivity indices as special cases. The \textit{QSR}-dissipativity can also be linked to the well-known small-gain property. Specifically, choosing $Q = I,\ S = 0,\ R = \gamma^2I$ yields the supply rate $w(u,y)=\gamma^2 u^\top u - y^\top y$. If the system is dissipative with respect to this supply rate, its $L_2$ gain is bounded above by $\gamma$; see ``\nameref{sb-passivity-small-gain}'' for further details.

The dissipativity framework also relates to the integral quadratic constraint (IQC) framework, which is described in ``\nameref{sb-IQC}''. In particular, the quadratic supply rate in \eqref{eq:general_supply_rate} corresponds to an IQC when written as the integral of $w(u,y)$. The main additional flexibility of IQCs is that the quadratic form can be frequency-dependent, i.e.,
\begin{equation}
\int_{-\infty}^{\infty}
\begin{bmatrix} y(j\omega) \\  u(j\omega)\end{bmatrix}\hermconj
\Pi(j\omega)
\begin{bmatrix} y(j\omega) \\  u(j\omega)\end{bmatrix}d\omega
\geq 0,
\end{equation}
where $\Pi(j\omega)$ is an IQC multiplier, and ${\hermconj}$ denotes the conjugate transpose of a complex matrix. Thus, IQCs naturally extend quadratic dissipativity when frequency-dependent input-output characteristics or structured uncertainties need to be captured.

\subsection{Passivity for Linear Systems}

For a linear time-invariant (LTI) system
\begin{equation*}
\dot{x}=Ax+Bu,\quad y=Cx+Du,
\label{eq:lti_system}
\end{equation*}
its passivity can first be formulated directly in the time domain through a quadratic storage function
\begin{equation*}
V(x)=\frac{1}{2}x^\top Px,\quad P=P^\top\succeq0.
\end{equation*}
For the supply rate in \eqref{eq:passivity}, the passivity inequality \eqref{eq:dissipation_inequality} can be converted into a linear matrix inequality (LMI) in $P$. Specifically,
\begin{equation}
\begin{bmatrix}
A^\top P + PA & PB - C^\top \\
B^\top P - C & -(D+D^\top)
\end{bmatrix} \preceq0.
\label{eq:dissipativity_lmi}
\end{equation}
Hence, passivity can be certified computationally through convex feasibility conditions.

For LTI systems, passivity also admits a frequency-domain description. In particular, the \emph{Kalman--Yakubovich--Popov (KYP) lemma} \cite[Lemma 6.2]{khalil2002nonlinear} establishes, under standard assumptions, the equivalence between \eqref{eq:dissipativity_lmi} and the frequency-domain \emph{positive-realness} condition of the transfer function $G(s)$ from $u$ to $y$, which requires
\begin{equation}
G(j\omega)+G(j\omega) \hermconj \succeq0,\quad\forall\omega \in \mathbb{R}.
\label{eq:positive_real}
\end{equation}
Thus, passivity can also be interpreted as a \textit{positive-realness} property in the frequency domain, also colloquially called \textit{positive resistance} in power systems and electronics lingo.

This dual interpretation is particularly useful for PE converters. Time-domain passivity naturally connects with Lyapunov analysis and nonlinear models, whereas frequency-domain positive-realness provides an intuitive interpretation in terms of impedance or admittance characteristics. In this sense, passivity forms a bridge between energy-based nonlinear stability analysis and impedance-based frequency-domain analysis.

\begin{figure}[!]%[b]
\centering
\includegraphics{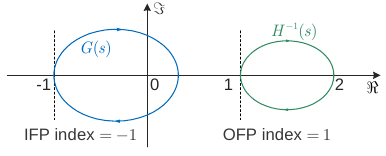}
\caption{The leftmost points of the Nyquist plots of $G(s)$ and $H^{-1}(s)$ determine the input-feedforward passivity (IFP) and output-feedback passivity (OFP) indices, respectively, for $G(s) = \frac{-s+1}{s+3},\ H(s) = \frac{s+2}{2s+2}$.} \label{fig:passivity-index}
\vspace{4mm}
\end{figure}

For a linear system $G(s)$, the passivity indices can likewise be defined and computed in the frequency domain~\cite[p.~28]{bao2007process}, in addition to the time-domain definition in terms of the supply rate $w(u,y)$ in \eqref{eq:passivity_index_supply}. For a stable linear system $G(s)$, its input-feedforward passivity (IFP) index is given by
\begin{equation}\label{eq:IFPindex}
\nu = \frac{1}{2} \min_{\omega \in \mathbb{R}} \,\,\lambda_{\min}\bigl(G(j\omega) + G(j\omega) \hermconj\bigr),
\end{equation}
where $\lambda_{\min}(\cdot)$ denotes the minimum eigenvalue of a Hermitian matrix. For the single-input-single-output (SISO) case, this expression reduces to $\nu = \min_{\omega} \Re[G(j\omega)]$ where $\Re(\cdot)$ denotes the real part. Thus, we can test the IFP index directly using the leftmost point of the Nyquist plot, as exemplified in Figure~\ref{fig:passivity-index}. For a stable and minimum-phase $H(s)$, its output-feedback passivity (OFP) index can be derived as
\begin{equation}\label{eq:OFPindex}
\rho = \frac{1}{2} \min_{\omega \in \mathbb{R}} \,\,\lambda_{\min}\bigl(H(j\omega)^{-1} + (H(j\omega)^{-1})^{\mathsf{H}}\bigr),
\end{equation}
For SISO systems this simplifies to $\rho = \min_{\omega} \Re\bigl[H^{-1}(j\omega)\bigr]$; hence the leftmost point on the Nyquist curve for $H^{-1}(s)$ gives the OFP index value, as shown in Figure~\ref{fig:passivity-index}. The passivity indices of $G(s)$ and $H(s)$ can be used to derive sufficient conditions for the stability of their feedback interconnection.

\setlength{\unitlength}{1mm}
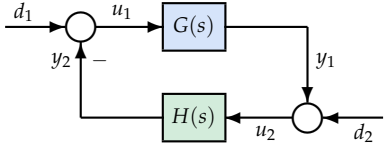
\begin{figure}[t!]
\begin{center}
\small
\begin{picture}(50,20)
\thicklines
\put(0,17){\vector(1,0){8}}
\put(10,17){\circle{4}}
\put(12,17){\vector(1,0){9}}

\put(21,14){\textcolor[RGB]{214,228,250}{\rule{8mm}{6mm}}}
\put(21,14){\framebox(8,6){$G(s)$}}

\put(29,17){\line(1,0){11}}
\put(40,17){\vector(0,-1){10}}
\put(38,5){\vector(-1,0){9}}
\put(40,5){\circle{4}}
\put(50,5){\vector(-1,0){8}}

\put(21,2){\textcolor[RGB]{211,236,219}{\rule{8mm}{6mm}}}
\put(21,2){\framebox(8,6){$H(s)$}}

\put(21,5){\line(-1,0){11}}
\put(10,5){\vector(0,1){10}}
\put(5,10){\makebox(5,5){$y_2$}}
\put(40,10){\makebox(5,5){$y_1$}}
\put(0,16.5){\makebox(5,5){$d_1$}}
\put(45,0.5){\makebox(5,5){$d_2$}}
\put(13,16.5){\makebox(5,5){$u_1$}}
\put(32,0.5){\makebox(5,5){$u_2$}}
\put(10,10){\makebox(5,5){$-$}}
\end{picture}

\caption{A standard closed-loop feedback-interconnected system $G(s)\#H(s)$.}
\label{fdbk}
\end{center}
\end{figure}

\subsection{Dissipativity and Stability of Interconnected Systems}

The key advantage of dissipativity for large-scale interconnected systems is its compositionality and application to their stability analysis. Hill and Liu recently reviewed the development of dissipativity theory with application to stability analysis \cite{hill2022dissipativity}. 

Consider two subsystems interconnected through power-preserving ports, as shown in Figure~\ref{fdbk} with $d_1 = d_2 = 0$ for now. Suppose subsystem $i$, $i \in \{1,\,2\}$, satisfies
\begin{equation*}
    \dot{V}_i\leq w_i(u_i,y_i).
\end{equation*}
Define the aggregate storage function $V(x)=V_1(x_1) + V_2(x_2)$. It follows that $\dot{V}\leq w_1(u_1,y_1) + w_2(u_2,y_2)$. If the interconnection satisfies 
\begin{equation}\label{eq:dissip}
    w_1(u_1,y_1) + w_2(u_2,y_2)\leq0 \,,
\end{equation}
the interconnected system inherits the dissipation inequality $\dot{V}\leq0$. Consequently, stability can be established from \emph{local certificates and interconnection constraints}, without constructing a monolithic Lyapunov function for the entire network.

Particularly, for dissipative subsystems characterized by passivity indices, suppose subsystem 1 is input-feedforward passive with $w_1(u_1,y_1)=u_1^\top y_1-\nu\,u_1^\top u_1$, while subsystem 2 is output-feedback passive with $ w_2(u_2,y_2) = u_2^\top y_2-\rho\,y_2^\top y_2$. Under the power-preserving interconnection $y_1 = u_2$ and $y_2 = -u_1$, as illustrated in Figure~\ref{fdbk}, the port-power terms cancel, yielding
\begin{align*}
w_1(u_1,y_1) + w_2(u_2,y_2) &= u_1^\top y_1 + u_2^\top y_2 -\nu\,u_1^\top u_1 - \rho\,y_2^\top y_2\\
&= u_1 ^\top u_2 + u_2 ^\top (-u_1) -(\rho + \nu) u_1^\top u_1 \\
&= -(\rho + \nu) u_1^\top u_1.
\end{align*}
Therefore, $\rho + \nu \geq 0$ ensures that the composite storage function satisfies $\dot{V}\leq0$. If the overall system further satisfies an appropriate strict storage function decay or detectability condition,  then asymptotic stability follows.

The same condition can be established in the frequency domain: if the IFP index $\nu$ of $G(s)$ and the OFP index $\rho$ of $H(s)$ satisfy $\rho + \nu \geq 0$, the feedback interconnection is stable. We can verify this condition and the stability using the example in Figure~\ref{fig:passivity-index}. In particular, for two passive systems, $\rho = \nu = 0$, and the condition is automatically satisfied.

The compositional property of dissipativity is especially attractive for PE-dominated power systems, where individual converters, network segments, and auxiliary devices may be modeled and certified independently. Adding or replacing a subsystem then requires only a local certificate and an updated interconnection check, rather than repeating a centralized stability analysis of the entire system. However, requiring each converter to satisfy the passivity condition individually and across the entire frequency range may be restrictive. For example, Zhao \textit{et al.} demonstrated that a generic grid-forming converter is non-passive over a frequency range around the fundamental frequency \cite{zhao2022low}. Next, we present more flexible decentralized stability certificates.

%(Xiuqiang's part)

%Should include (feel free to modify):

%- passivity for linear systems (positive real in the frequency domain)

%- dissipativity/passivity for nonlinear systems (relation to dissipativity, energy-function like analysis)

%- dissipativity/passivity index (extending the traditional passivity condition)

%- application to grid-forming stability analysis

%others if of interest...

% The Gershgorin disc theorem is also a good method for small-signal analysis.

% Dissipativity (xiuqiang); small-gain, SRG (linbin), Gershgorin (to be added somewhere), IQC (to be mentioned somewhere), Vector Lyapunov (needed?)

\section{Decentralized Stability Certificates Based on DW Shell and its Projections}
\label{sec:DWshell}
The concept of passivity for LTI systems is pertinent to the phase of complex numbers and matrices. For instance, it requires the phases of the Nyquist plot to be within $[-90^\circ,90^\circ]$ so that a SISO component is passive. However, it is nontrivial to define the phase of complex matrices and multiple-input multiple-output (MIMO) systems. Recently, Chen~\textit{et~al.} proposed to define the phase of complex matrices based on the concept of numerical ranges, which enables the development of the small-phase theorem for stability analysis~\cite{chen2024phase, wang2023phases}. The numerical range of a complex matrix can be thought of as the projection of the Davis-Wielandt (DW) shell onto the $x$-$y$ plane, and it turns out that the DW shell, a classic concept in operator theory to study algebraic and spectral properties~\cite{davis1968shell,wielandt1955eigenvalues,lestas2012large,huang2025geometric}, can be useful in the stability analysis of dynamical systems, as demonstrated in what follows.

\begin{figure}[!]%[b]
\centering
\vskip-.01cm
\includegraphics[width=.4\textwidth]{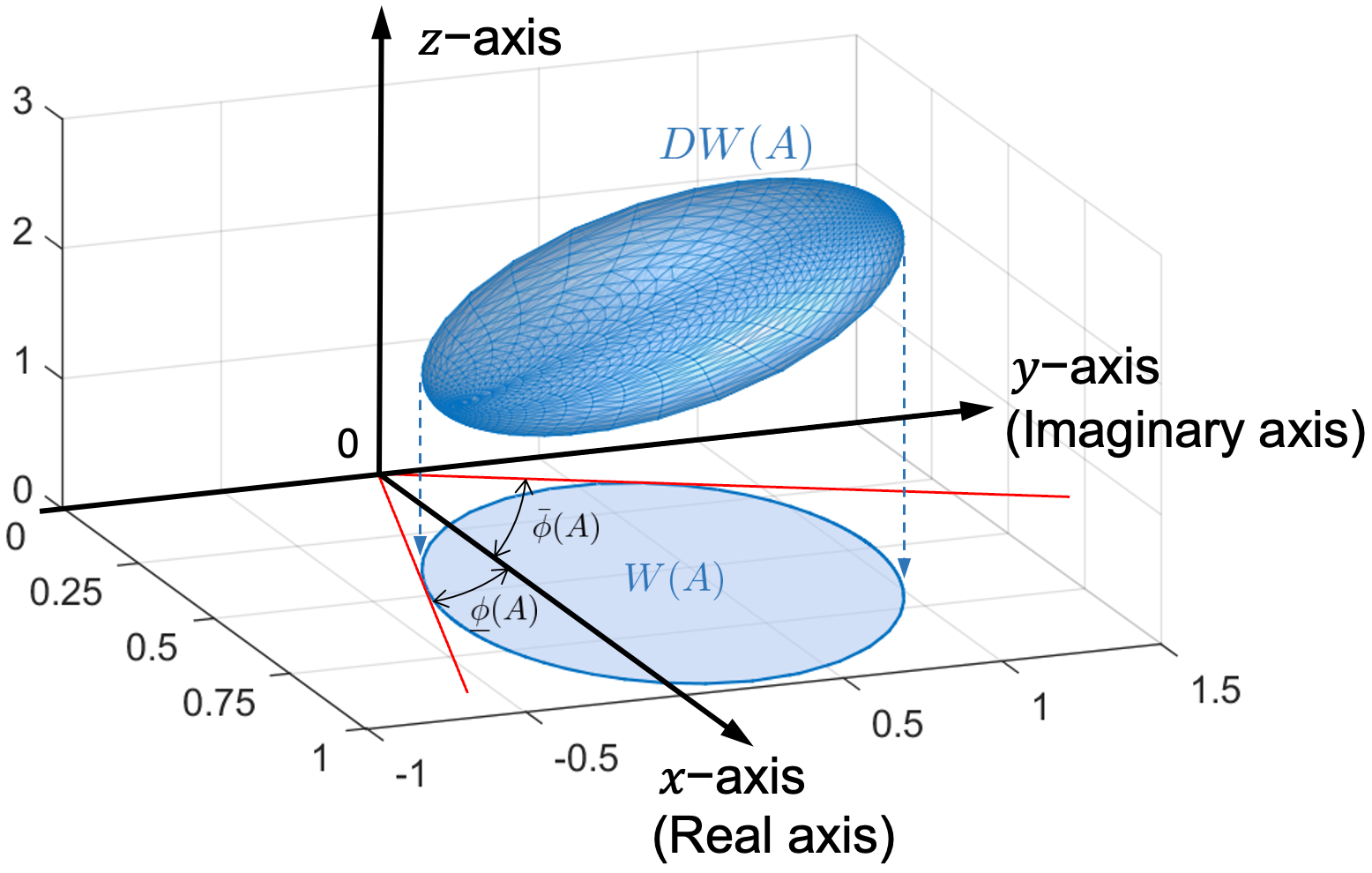}
\caption{Illustration of a DW shell in 3D space.} \la{Fig_DW_shell}
\vspace{3mm}
\end{figure}

\begin{sidebar}{}
\setcounter{sequation}{0}
\renewcommand{\thesequation}{S\arabic{sequation}}
\setcounter{stable}{0}
\renewcommand{\thestable}{S\arabic{stable}}
\setcounter{sfigure}{0}
\renewcommand{\thesfigure}{S\arabic{sfigure}}
\section[Connection Between DW Shell and Scaled Relative Graph (SRG)]{Connection Between DW Shell and Scaled Relative Graph (SRG)}
\label{sb-DW-SRG}
\sdbarinitial{T}he concept of scaled relative graphs (SRGs) has recently gained attention in analyzing dynamical systems~\cite{SR_1} and power systems in particular~\cite{SR_2} and~\cite{SR_3}, which combines the gain and phase information of an operator simultaneously. The {SRG} of a complex matrix $A \in \mathbb{C}^{n \times n}$ (i.e., a linear operator) is
\begin{equation}\label{eq:SRG1}
    SRG(A) = \left\{ \frac{\|y\|}{\|x\|}e^{\pm j \angle (x,y)}: y = Ax, x\in \mathbb{C}^n, x \ne 0 \right\} \,,
\end{equation}
where $\angle (x,y) = {\rm arccos}\left( \frac{\Re(y{\hermconj} x)}{\|x\| \|y\|}\right)$. We note that without loss of generality, one can consider $\|x\| = 1$ so that~\eqref{eq:SRG1} becomes
\begin{equation}\label{eq:SRG2}
\begin{split}
    SRG(A) =& \left\{ \|Ax\|e^{\pm j{\rm arccos}\left( \frac{\Re(x{\hermconj}Ax)}{\|Ax\|}\right)}: x\in \mathbb{C}^n, \|x\|=1 \right\}\\ 
    =& \left\{  
    \begin{array}{ll}
       c \in \mathbb{C}: & \Re(c) = \Re(x{\hermconj}Ax), |c| = \|Ax\|,  \\
       &x\in \mathbb{C}^n, \|x\|=1
    \end{array}
    \right\} \,.
\end{split}
\end{equation} 
The SRG is defined on the complex plane and is symmetric about the real axis. We denote the positive part of the SRG of matrix~$A$ as $SRG_+(A)$. As investigated in~\cite{zhang2025phantom}, there exists certain connections between the SRG and the DW shell.

Figure~\ref{Fig_connection_graphs} shows how we can obtain the numerical range, $x$-$z$ graph, and SRG by projecting the DW shell of a complex matrix $A$. It can be seen that the numerical range $W(A)$ can be simply obtained by projecting $DW(A)$ onto the $x$-$y$ plane, and the so-called \textit{$x$-$z$ graph} $P_{xz}(A)$ can be obtained by projecting $DW(A)$ onto the $x$-$z$ plane. As a matter of fact, any DW shell is contained in the solid paraboloid $z \ge x^2+y^2$ (since $\| Ax\|^2 \geq \|x{\hermconj}Ax \|^2$ for $\| x \| = 1$), and thus any $x$-$z$ graph is contained in the solid parabola $z \ge x^2$. We now consider the following two steps of projections: 1) project the $x$-$z$ graph $P_{xz}(A)$ onto the surface of the paraboloid $z=x^2+y^2$ along the positive $y$-axis and obtain a curved surface as
\begin{equation}
\begin{split}
    \Pi_\otimes (P_{xz}(A)) &= \{(x,y,z):(x,z) \in P_{xz}(A),z = x^2+y^2\}\\
    &=\{(x,\sqrt{z-x^2},z):(x,z) \in P_{xz}(A)\} \,,
\end{split}
\end{equation}
and 2) further project $\Pi_\otimes (P_{xz}(A))$ onto the $x$-$y$ plane and obtain the positive part of SRG as
\begin{equation}
\begin{split}
    SRG_{+}(A) =& \Pi_{xy}(\Pi_\otimes (P_{xz}(A)))\\
    =& \{x+j\sqrt{z-x^2} \in \mathbb{C}:(x,z) \in P_{xz}(A)\}\\
    =& \{x+j\sqrt{z-x^2} \in \mathbb{C}: x = \Re(x^*Ax), z = \|Ax\|^2,\\
    & \,\,\, x\in \mathbb{C}^n, \|x\|=1\}\,.
    % &= \left\{\left(\Re(x^*Ax),\sqrt{\|Ax\|^2-\Re(x^*Ax)}\right):x\in \mathbb{C}^n, \|x\|=1\right\} 
\end{split}\label{eq:SRG_PI}
\end{equation}
Indeed, for $c\in SRG_{+}(A)$, \eqref{eq:SRG2} gives $\Re(c)=x$ and $|c|^2=z$. Since $\Im(c)\geq0$, it follows that $\Im(c)=\sqrt{|c|^2-\Re(c)^2}=\sqrt{z-x^2}$, which verifies~\eqref{eq:SRG_PI}. Figure~\ref{Fig_connection_graphs} shows the above projection process, namely, how to obtain SRG from the $x$-$z$ graph. We also notice that $SRG_{+}(A)$ can be mapped to the $x$-$z$ graph of $A$ through a bijective function $f_{s \rightarrow xz}:\mathbb{C} \rightarrow \mathbb{R}\times \mathbb{R}$ as
\begin{equation}\label{eq:bijective}
\begin{split}
f_{s\rightarrow xz}(c)
&=\left(\Re(c),|c|^2\right)\\
&=\left(x,x^2+\left(\sqrt{z-x^2}\right)^2\right)\\
&=(x,z),
\quad
c=x+\mathrm{j}\sqrt{z-x^2}\in SRG_{+}(A).
\end{split}
\end{equation}
Hence, there exists a one-to-one mapping between the SRG and the $x$-$z$ graph. However, the two graphs have different geometric shapes and different properties, e.g., the $x$-$z$ graph is always convex while the SRG is not. Indeed, in practice  separation of complex SRGs is often established by overapproximating each SRG by a convex containing set. We found that sometimes it is easier to derive SRG-based stability results using the properties of the DW shell and the convex $x$-$z$ graph.

For instance, for $A,B \in \mathbb{C}^{m \times m}$, we can conclude from the above connection between SRG and $x$-$z$ graph that, $SRG(A)$ is separated from $SRG(B)$ if and only if the $x$-$z$ graph $P_{xz}(A)$ is separated from $P_{xz}(B)$, and we have the following result by referring to the stability condition in~\eqref{eq:P_sep}.

The closed-loop system \( G(s)\#H(s) \) in Figure~\ref{fdbk} is stable if, for each \( \omega \in [0,\infty) \), the SRG of \( G(j\omega) \) is separated from the SRG of \( -\frac{1}{\tau} H^{-1}(j\omega) \), that is,
\begin{equation}\label{eq:SRG_sep}
    SRG(G(j\omega)) \cap SRG\left(-\tfrac{1}{\tau} H^{-1}(j\omega)\right) = \varnothing ,
\end{equation}
which should hold for all homotopy parameters $\tau \in (0,1]$. Moreover, Eq.~\eqref{eq:SRG_sep} implies the DW shell separation in~\eqref{eq:DW_sep}. We will later also introduce how SRG enables a decentralized stability certificate, analogous to $x$-$z$ graph.

% We will next show how the connection between SRG and the $x$-$z$ graph enables a more general decentralized stability condition.

\sdbarfig{%
\centering
\includegraphics[width=.48\textwidth]{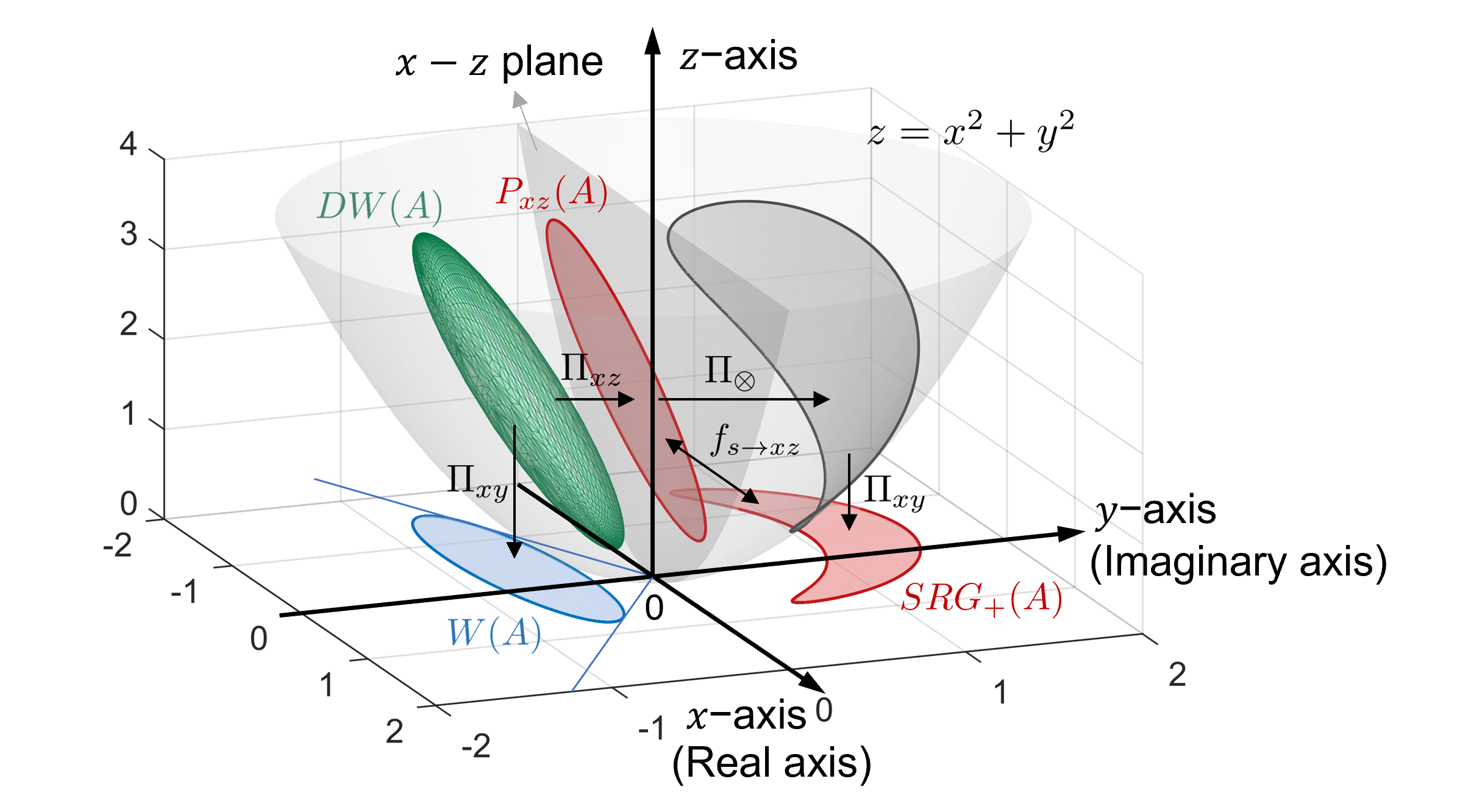}
}{Connections among DW shell $DW(A)$, numerical range $W(A)$, $x$-$z$ graph $P_{xz}(A)$, and SRG $SRG_+(A)$. \label{Fig_connection_graphs}}

\end{sidebar}

\begin{pullquote}
The DW shell provides a unified framework to connect different results like passivity, small-phase theorem, small-gain theorem, and SRGs.
\end{pullquote}

\subsection{DW Shell and Numerical Range}

The {DW shell} of a complex matrix $A \in \mathbb{C}^{n \times n}$ is
\begin{equation}\label{eq:DWA}
DW(A)=\{(x{\hermconj}Ax,\|Ax\|^2): x\in \mathbb{C}^n, \|x\|=1\},
\end{equation}
as illustrated in Figure~\ref{Fig_DW_shell}. Here, $\|x\|$ denotes the two-norm of $x$.
Any DW shell is embedded in $\mathbb{C} \times \mathbb{R}$, which is isomorphic to the 3D real space $\mathbb{R}^3$ with coordinates $(x,y,z)$. Note that we abuse the symbol $x$ to denote both the vector in \eqref{eq:DWA} and the $x$-coordinate in $\mathbb{R}^3$, with the meaning clear from the context.

The {numerical range} of a complex matrix $A \in \mathbb{C}^{n \times n}$ is 
\begin{equation}\label{eq:WA}
W(A)=\{x{\hermconj}Ax: x\in \mathbb{C}^n, \|x\|=1\},
\end{equation}
which is embedded in $\mathbb{C}$ and contains the eigenvalues of $A$. By comparing~\eqref{eq:WA} with~\eqref{eq:DWA}, we notice that $W(A)$ and $DW(A)$ have the very same shape on the complex plane (i.e., the $x$-$y$ plane), and the numerical range can be considered as the projection of the 3D DW shell onto the 2D $x$-$y$ plane, as shown in Figure~\ref{Fig_DW_shell}. The $z$-axis values $\|Ax\|^2$ of the DW shell are related to the gain of matrix~$A$, with the smallest and largest $z$-axis values corresponding to the squares of the smallest and largest singular values, respectively.

We revisit some important {properties of DW shell}~\cite{zhang2025phantom}, which are essential for the following stability analysis. Consider $A \in \mathbb{C}^{n \times n}$, we have the following 5 Properties:
\begin{enumerate}
    \item $DW(A) = DW(U{\hermconj}AU)$ holds for any unitary matrix $U$;
    \item for a block-diagonal matrix $A={\rm blkdiag}(A_1,A_2,...A_n)$ where $A_i$ is the $i$-th block, we have 
\[
DW(A) = {\rm conv}\{DW(A_1),DW(A_2),...,DW(A_n)\}\,,
\]
where ${\rm conv}\{\cdot\}$ denotes the convex hull;
    \item $DW(A) \subset \{(c = x+jy \in \mathbb{C},z \in \mathbb{R}): z \ge x^2+y^2 \}$, where $x$ and $y$ denote the real part and imaginary part of $c$;
    \item if $DW(A) \cap DW(-B) = \varnothing$, then ${\rm det}(A+B) \ne 0$;
    \item $DW(A)$ is convex for $n=1$ or $n \ge 3$, and if $n=2$, $DW(A)$ is a hollow ellipsoid;
    \item if $(z,\nu) \in DW(A)$, then $(\frac{z\hermconj}{\nu},\frac{1}{\nu}) \in DW(A^{-1})$.
\end{enumerate}

\subsection{Stability Certificates Based on DW Shell and Its Projections}

Let \( G\), \( H \in \mathcal{RH}_{\infty}^{m\times m}\), i.e., be real, rational, proper, square, and stable \( m\times m \) transfer function matrices, and let $H$ be invertible. Consider their feedback interconnection as shown in Fig.~\ref{fdbk}. The closed-loop system is said to be stable if the \textit{Gang of Four matrix} describing all input-output maps
    \begin{align*}
    G(s)\#H(s)=\begin{bmatrix}
      (I + HG)^{-1} & (I + HG)^{-1}H\\
      G(I + HG)^{-1} & G(I + HG)^{-1}H
    \end{bmatrix}
    \end{align*}
    is stable, which can be checked by using the generalized Nyquist criterion (GNC). To be specific, \( G(s)\#H(s) \) is stable if and only if \( \det(I + G(j\omega)H(j\omega)) \) does not encircle or cross the origin. A well-known sufficient version of the GNC is as follows: \( G(s)\#H(s) \) is stable if, for each \( \omega \in [0,\infty) \), the following condition holds~\cite{huang2025geometric}:
    \begin{equation}\label{eq:suff_GNC}
    \det\left(I + \tau G(j\omega) H(j\omega)\right) \neq 0, \quad \forall \tau \in (0, 1],
    \end{equation}
   which ensures that the Nyquist plot of $\det\!\left(I+G(\mathrm{j}\omega)H(\mathrm{j}\omega)\right)$ does not intersect the origin or the negative real axis, as graphically illustrated in Fig.~\ref{Fig_Nyquist-tau}.

\begin{figure}[!]%[b]
\centering
\vskip-.01cm
\includegraphics[width=.4\textwidth]{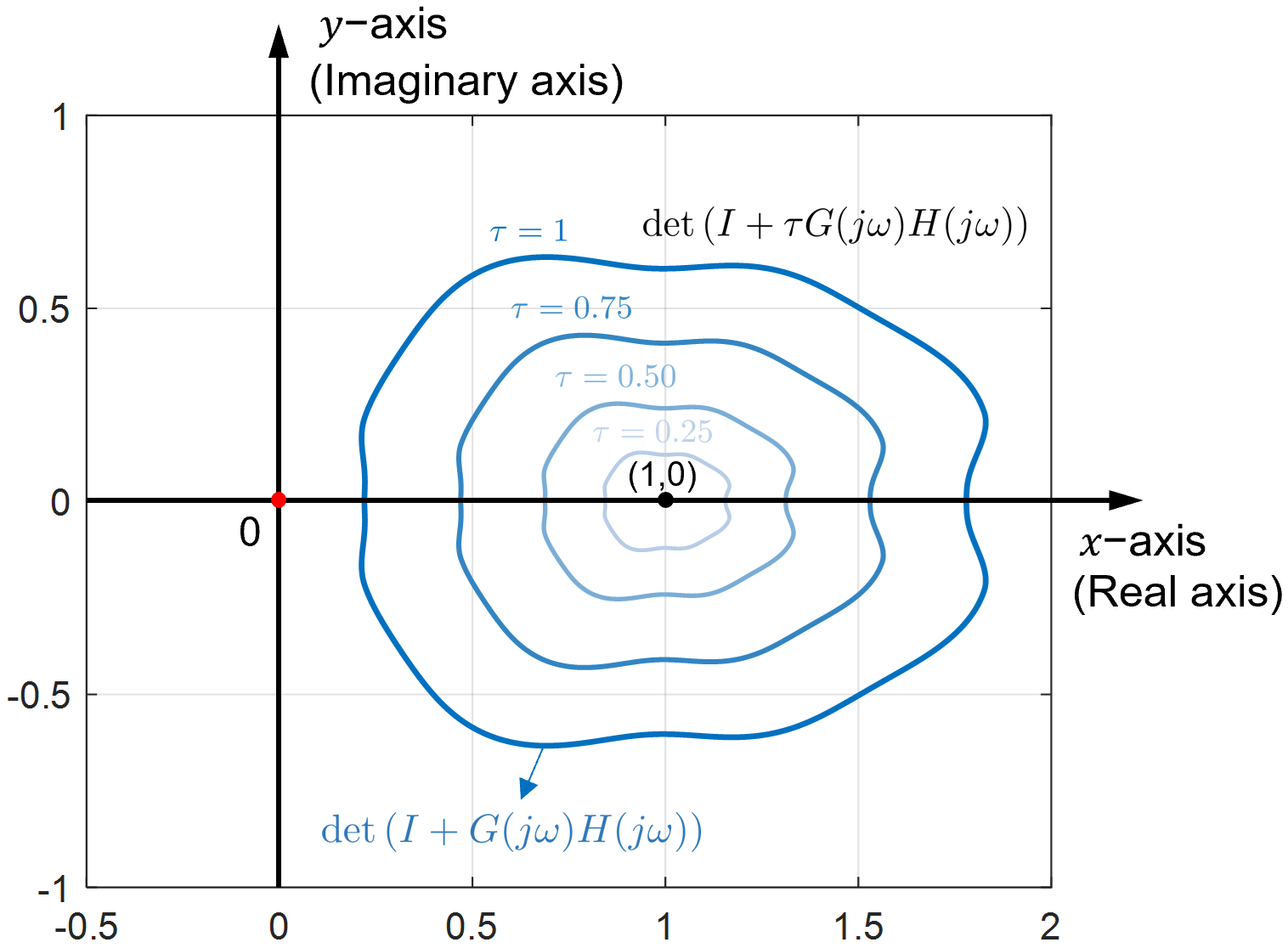}
\caption{Nyquist plots of
$\det\!\left(I+\tau G(\mathrm{j}\omega)H(\mathrm{j}\omega)\right)$
for different values of $\tau\in(0,1]$.} \la{Fig_Nyquist-tau}
\vspace{3mm}
\end{figure}

However, neither the GNC nor the sufficient condition~\eqref{eq:suff_GNC} is scalable or tractable when handling high-dimensional systems, due to the multiplication of $G(j\omega)$ and $H(j\omega)$ and the need to compute the determinant, even if $G(s)$ is block-diagonal in the context of multi-agent systems. To separate~\eqref{eq:suff_GNC} into decentralized conditions, as a first step, we consider
\begin{equation*}
\begin{split}
    &\det\left(I + \tau G(j\omega) H(j\omega)\right)\\
    =&\det\left(\tfrac{1}{\tau}H^{-1}(j\omega) + G(j\omega) \right) \det(\tau H(j\omega))
\end{split}
\end{equation*}
since $H$ is invertible. Then, we can rewrite~\eqref{eq:suff_GNC} as
\begin{equation}\label{eq:suff_GNC1}
    \det\left(\tfrac{1}{\tau}H^{-1}(j\omega) + G(j\omega) \right) \neq 0, \quad \forall \tau \in (0, 1],
\end{equation}
which avoids the multiplication of $G$ and $H$. Moreover, according to Property~4 of DW shell, the condition in~\eqref{eq:suff_GNC1} holds if the DW shell of $-\tfrac{1}{\tau}H^{-1}(j\omega)$ is separated from the DW shell of $G(j\omega)$, and thus we have the following result.

\begin{highlightbox}
\textbf{Separation of DW shells}

Consider \( G\), \( H \in \mathcal{RH}_{\infty}^{m\times m}\).
The closed-loop system \( G(s)\#H(s) \) in Figure~\ref{fdbk} is stable if, for each \( \omega \in [0,\infty) \), the DW shell of \( G(j\omega) \) is separated from the DW shell of \( -\frac{1}{\tau} H^{-1}(j\omega) \), that is,
\begin{equation}\label{eq:DW_sep}
    DW(G(j\omega)) \cap DW(-\tfrac{1}{\tau} H^{-1}(j\omega)) = \varnothing ,
\end{equation}
which should hold for all $\tau \in (0,1]$.
\end{highlightbox}

Instead of computing \( \det(I + G(j\omega)H(j\omega)) \) and using the GNC, the stability condition in~\eqref{eq:DW_sep} directly compares the DW shells of the two interconnected components, which enables stability analysis from a graphical perspective. We notice that it is a sufficient condition but not a necessary one, namely, if~\eqref{eq:DW_sep} holds, then the system is stable, but if~\eqref{eq:DW_sep} is violated, we cannot conclude that the system is unstable. Instead, if the system is unstable, then we can conclude that~\eqref{eq:DW_sep} is certainly violated. We will later demonstrate that the conservatism of~\eqref{eq:DW_sep} is acceptable in the power system context. More importantly, we will show that it enables decentralized stability analysis. In ``\nameref{sb-IQC}'' we demonstrate the connections between IQC and the above DW shell separation condition, as well as the dissipativity condition.

% \subsection{Stability Conditions Based on the Projections of DW Shell}

Condition~\eqref{eq:DW_sep} requires checking the DW shell in the 3D space, and it can be simplified by considering the projections of DW shell onto lower-dimensional spaces. For instance, the numerical range, as defined in~\eqref{eq:WA}, can be thought of as the projection of DW shell onto the $x$-$y$ plane. Moreover, for a complex matrix $A \in \mathbb{C}^{n \times n}$, the projection of its DW shell onto the $x$-$z$ plane is referred to as the $x$-$z$ graph, defined as 
\begin{equation}\label{eq:PA}
P_{xz}(A)=\{(\Re(x{\hermconj}Ax),\|Ax\|^2): x\in \mathbb{C}^n, \|x\|=1\},
\end{equation}
which is a convex set embedded in $\mathbb{R} \times \mathbb{R}$. In ``\nameref{sb-DW-SRG},'' we show how the $x$-$z$ graph is related to well-known SRG.

\begin{pullquote}
The stability certificate of DW shell separation is a
special case of the integral quadratic constraints (IQCs).
\end{pullquote}

Consider two complex matrices $A, B \in \mathbb{C}^{n  \times n}$. Since the numerical range and the $x$-$z$ graph are both projections of DW shell, the following statement holds:
if the numerical range $W(A)$ is separated from $W(B)$, or the $x$-$z$ graph $P_{xz}(A)$ is separated from $P_{xz}(B)$, then we can conclude that the DW shells $DW(A)$ and $DW(B)$ are separated, as illustrated in Figure~\ref{Fig_sep_WA} and Figure~\ref{Fig_sep_PA}.

\begin{figure}[!]%[b]
\centering
\vskip-.25cm
\includegraphics[width=.35\textwidth]{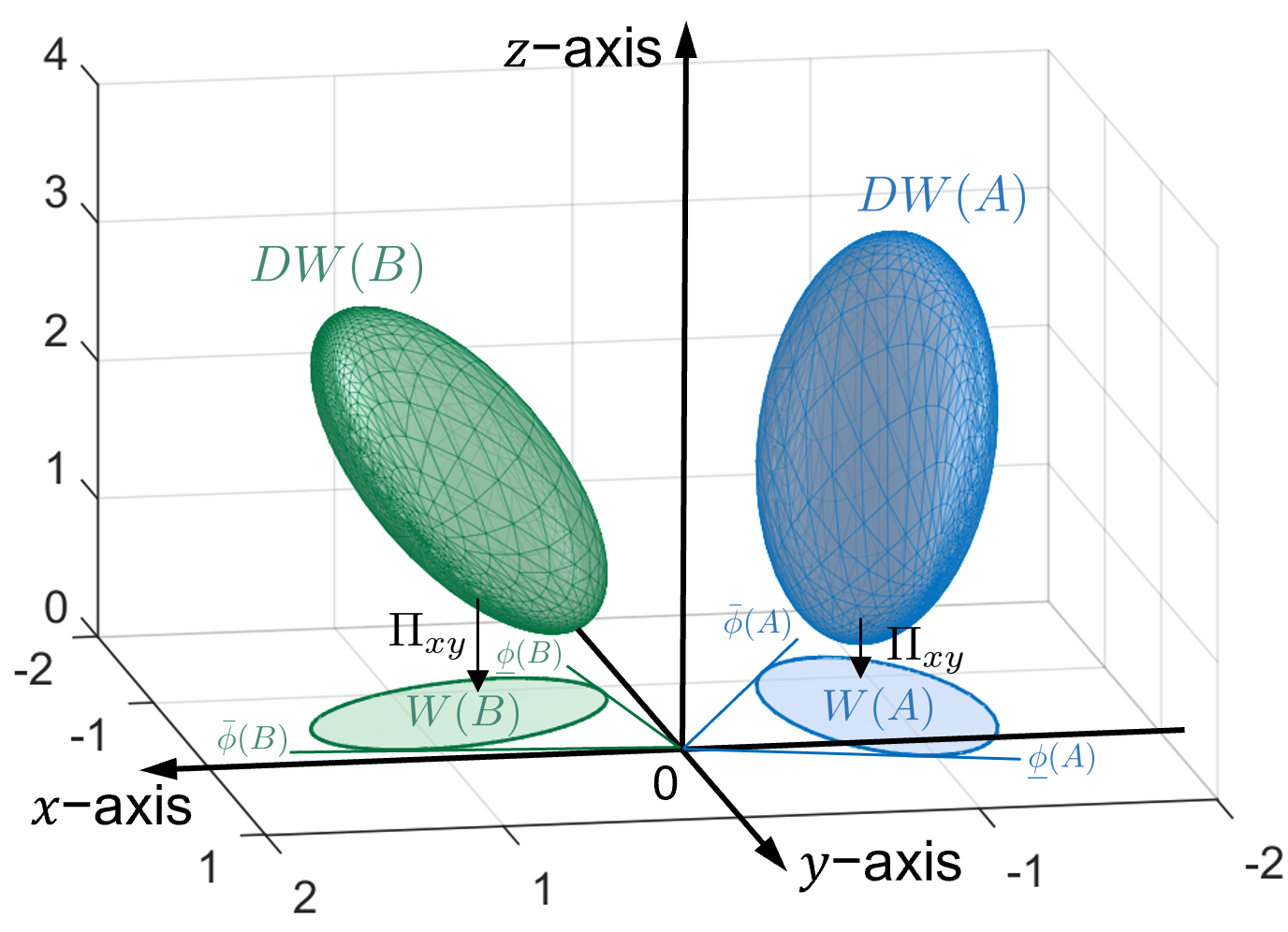}
\caption{The separation of numerical ranges implies the separation of DW shells.} \la{Fig_sep_WA}
\vspace{3mm}
\end{figure}

\begin{figure}[!]%[b]
\centering
\vskip-.25cm
\includegraphics[width=.35\textwidth]{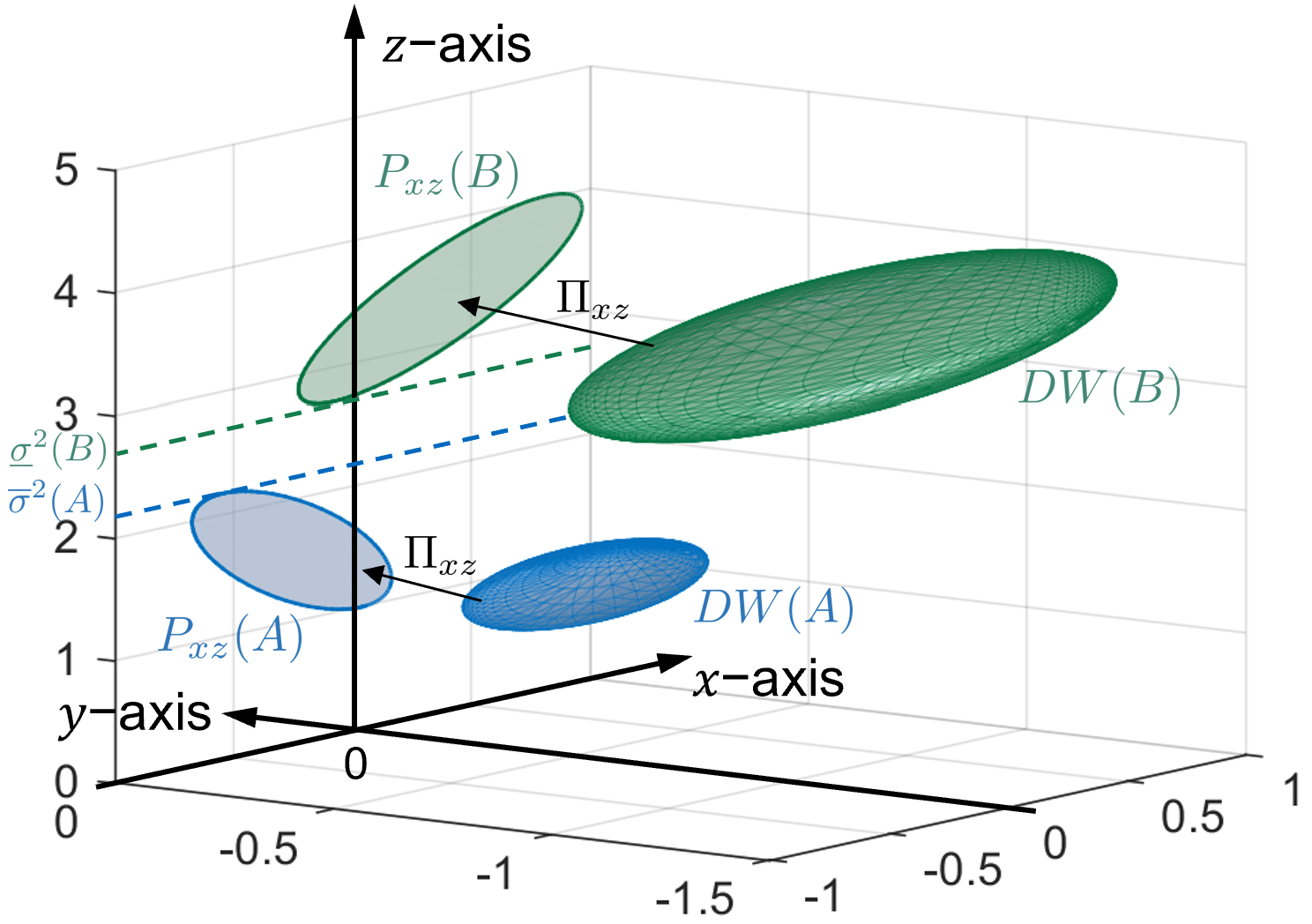}
\caption{The separation of $x$-$z$ graphs implies the separation of DW shells.} \la{Fig_sep_PA}
\vspace{0mm}
\end{figure}

\begin{highlightbox}
\textbf{Separation of Numerical Range and x-z Graph}

Consider the closed-loop system \( G(s)\#H(s) \) in Figure~\ref{fdbk}, where \( G\), \( H \in \mathcal{RH}_{\infty}^{m\times m}\). If, for each \( \omega \in [0,\infty) \), either the numerical range of \( G(j\omega) \) is separated from the numerical range of \( -\frac{1}{\tau} H^{-1}(j\omega) \) as
\begin{equation}\label{eq:W_sep}
    W(G(j\omega)) \cap W(-\tfrac{1}{\tau} H^{-1}(j\omega)) = \varnothing , \, \forall \tau \in (0,1] ,
\end{equation}
or the $x$-$z$ graph of \( G(j\omega) \) is separated from the $x$-$z$ graph of \( -\frac{1}{\tau} H^{-1}(j\omega) \) as
\begin{equation}\label{eq:P_sep}
    P_{xz}(G(j\omega)) \cap P_{xz}(-\tfrac{1}{\tau} H^{-1}(j\omega)) = \varnothing , \, \forall \tau \in (0,1] ,
\end{equation}
or the SRG of \( G(j\omega) \) is separated from the SRG of \( -\frac{1}{\tau} H^{-1}(j\omega) \) for $\tau \in (0,1]$, as shown in~\eqref{eq:SRG_sep},
then we can conclude that~\eqref{eq:DW_sep} holds, and moreover, the closed-loop system \( G(s)\#H(s) \) is stable.
\end{highlightbox}

The above result shows that the system stability can be analyzed by checking the numerical range or the $x$-$z$ graph, which for given $\omega$ and $\tau$ are both 2D graphs and are easier to check than the 3D DW shell, though the conditions in~\eqref{eq:W_sep} and~\eqref{eq:P_sep} are \textit{more conservative} than~\eqref{eq:DW_sep}.
In ``\nameref{sb-DW-SRG},'' we further demonstrate how one can obtain SRG by projecting the DW shell and show that starting from the condition of DW shell separation, one can also prove the SRG-based stability condition. 

\subsection{Small Phase and Small Gain}

We next illustrate how the matrix gain and phase are related to the DW shell and revisit stability conditions based on gain and phase. We adopt the concept of matrix phase as proposed in~\cite{chen2024phase, wang2023phases}, which is defined using the numerical range.

If the numerical range of a complex matrix $A$ does not include the origin, i.e., $0\notin W(A)$, then $A$ is said to be {sectorial}, and its sectorial decomposition is $A=T{\hermconj}DT$, where $T$ is nonsingular, and $D$ is a diagonal unitary matrix which has all diagonal entries lying on an arc of the unit circle with length smaller than $\pi$. 
Then, the {phases} of $A \in\mathbb{C}^{n\times n}$, denoted by 
\begin{equation}
    \overline\phi(A)=\phi_1(A) \geq \dots \geq \phi_n(A)=\underline\phi(A), 
    \tag{\bf Phases}
\end{equation}
are defined as the phases of the eigenvalues of $D$. 
The smallest phase $\underline{\phi}(A)$ and the largest phase
$\overline{\phi}(A)$ of a sectorial matrix $A$ are determined by the
two supporting rays of $W(A)$, as shown in Fig.~\ref{Fig_phase}, and the other phases of $A$ lie in between. Figure~\ref{Fig_DW_shell} shows the connections among DW shell, numerical range, and matrix phases graphically, and we will later demonstrate their connections in stability analysis. We revisit the so-called small-phase theorem in~\cite{chen2024phase} below.

\begin{figure}[!]%[b]
\centering
\vskip-.25cm
\includegraphics[width=.35\textwidth]{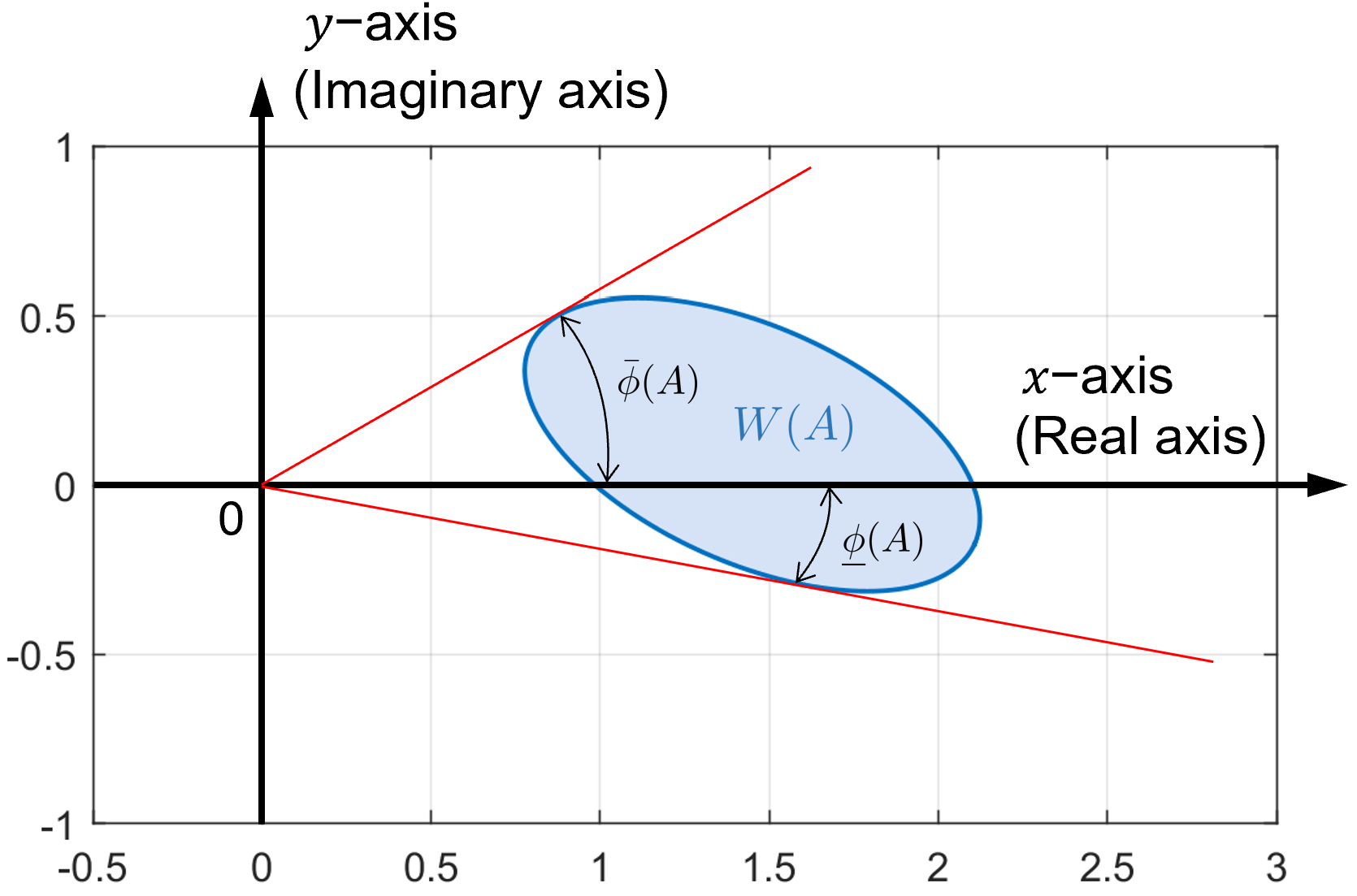}
\caption{Phases of a sectorial matrix $A$ are contained between $[\underline{\phi}(A),\,\overline{\phi}(A)]$, and the blue area $W(A)$ is the numerical range of $A$.} \la{Fig_phase}
\vspace{0mm}
\end{figure}

\begin{highlightbox}
\textbf{Small-Phase Theorem}

Consider \( G\), \( H \in \mathcal{RH}_{\infty}^{m\times m}\).
The closed-loop system \( G(s)\#H(s) \) in Figure~\ref{fdbk} is stable if \( G(j\omega) \) and \( H(j\omega) \) are sectorial, and for each \( \omega \in [0,\infty) \), 
    \begin{equation}\label{eq:SP}
    \hspace{-1mm}
    \overline{\phi}(G(j\omega)) 
    \hspace{-0.4mm} + \hspace{-0.4mm}
    \overline{\phi}(H(j\omega)) \!<\! \pi, \,\,
    \underline{\phi}(G(j\omega)) 
    \hspace{-0.4mm} + \hspace{-0.4mm}
    \underline{\phi}(H(j\omega)) \!>\! -\pi.
    \end{equation}
\end{highlightbox}

It was further shown in~\cite{huang2025geometric} that the small-phase condition in~\eqref{eq:SP} is a special case of the numerical range separation in~\eqref{eq:W_sep} when $G(\mathrm{j}\omega)$ and $H(\mathrm{j}\omega)$ are both sectorial. This implication can be interpreted graphically as follows. For each fixed $\omega$, let $A=G(\mathrm{j}\omega)$ and $B=-H^{-1}(\mathrm{j}\omega)$. The small-phase condition ensures the angular separation of $W(A)$ and $W(B)$, as shown in Fig.~\ref{Fig_sep_WA}. Since multiplication by the positive factor $1/\tau$ does not affect the matrix phases, the same numerical range separation holds when $B$ is replaced by $-\frac{1}{\tau}H^{-1}(\mathrm{j}\omega)$ for all $\tau\in(0,1]$, as required by~\eqref{eq:W_sep}. Since the numerical range is the projection of the DW shell onto the $x$-$y$ plane, \eqref{eq:W_sep} further implies~\eqref{eq:DW_sep}. We notice that the passivity condition can be thought of as restricting the phases of $G(j\omega)$ and $H(j\omega)$ within $(-\pi/2,\pi/2)$, that is,
\begin{equation}\label{eq:pas_phase}
\begin{array}{c}
    -\pi/2 < \underline{\phi}(G(j\omega)) \le \overline{\phi}(G(j\omega)) <\pi/2 \,,\\
    -\pi/2 < \underline{\phi}(H(j\omega)) \le \overline{\phi}(H(j\omega)) <\pi/2 \,,
\end{array}
\end{equation}
which is a special case of~\eqref{eq:SP}. This phase-based interpretation naturally leads to the following variant of the classical passivity theorem for LTI systems, in which \emph{strict} passivity is required for only one of the two interconnected systems~\cite{haberle2026decentralized}.
\begin{highlightbox}
{\textbf{Passivity Theorem}}

{The closed-loop system \( G(s)\#H(s) \) in Figure~\ref{fdbk} is stable if, for each \( \omega \in [0,\infty) \)}
    \begin{itemize}
        \item {$H(s)$ is strictly passive, and}
        \item {$G(s)$ is passive.}
    \end{itemize}
\end{highlightbox}

As a counterpart, the gains (magnitudes) of a complex matrix $A\in\mathbb{C}^{n\times n}$ are usually described by its singular values
\begin{equation}
    \overline{\sigma}(A) =\sigma_1(A) \geq \dots \geq \sigma_n(A)=\underline{\sigma}(A),
    \tag{\bf Gains}
\end{equation}
and the well-known small-gain theorem is as follows~\cite{skogestad2005multivariable}.

\begin{highlightbox}
\textbf{Small-Gain Theorem}

Consider $G, H \in \mathcal{RH}_{\infty}^{m\times m}$.
    The closed-loop system \( G(s)\#H(s) \) in Figure~\ref{fdbk} is stable if, for each \( \omega \in [0,\infty) \),
    \begin{equation}\label{eq:SG}
    \overline{\sigma}(G(j\omega))\,\overline{\sigma}(H(j\omega)) < 1.
    \end{equation}
\end{highlightbox}

It was shown in~\cite{huang2025geometric} that the small-gain condition in~\eqref{eq:SG} is a special case of the DW shell separation in~\eqref{eq:DW_sep}. At a given frequency $\omega$, let $A=G(\mathrm{j}\omega)$ and $B=-H^{-1}(\mathrm{j}\omega)$. Condition~\eqref{eq:SG} then becomes $\overline{\sigma}(A)<\underline{\sigma}(B)$. Since the extreme $z$ coordinates of the two DW shells are $\max\, z(DW(A))=\overline{\sigma}^{\,2}(A)$ and $\min \,z(DW(B))=\underline{\sigma}^{\,2}(B)$, the two shells can be separated by a plane of constant $z$, as shown in Fig.~\ref{Fig_sep_PA}. We then change $B$ from $-H^{-1}(\mathrm{j}\omega)$ to $-\frac{1}{\tau}H^{-1}(\mathrm{j}\omega)$ where $\tau\in(0,1]$, which scales its singular values by the factor $1/\tau$ and its $z$ coordinates by $1/\tau^2$. The illustrated separation still holds over the entire interval $\tau\in(0,1]$, indicating that the small-gain condition in~\eqref{eq:SG} implies~\eqref{eq:DW_sep}.

Note that the small-gain theorem can be combined with the small-phase theorem to reduce the conservativeness, that is, the closed-loop system is stable if, at any frequency, either the small-gain condition~\eqref{eq:SG} or the small-phase condition~\eqref{eq:SP} holds~\cite{zhao2022small}, which is less conservative than the original small-gain or small-phase conditions. 
Moreover, it was further demonstrated in~\cite{huang2024gain} that such a mixed gain-phase condition enables decentralized stability analysis of power systems. 

%*******************************************

\begin{sidebar}{}
\setcounter{sequation}{0}
\renewcommand{\thesequation}{S\arabic{sequation}}
\setcounter{stable}{0}
\renewcommand{\thestable}{S\arabic{stable}}
\setcounter{sfigure}{0}
\renewcommand{\thesfigure}{S\arabic{sfigure}}
\section{How are IQCs Connected to Dissipativity and DW Shell Separation?}
\label{sb-IQC}

Integral quadratic constraints (IQCs) are a popular concept for analyzing how structural perturbations affect system stability, and they can handle delay, saturation, and certain types of nonlinearities~\cite{SR_1}. In what follows, we provide self-contained derivations to demonstrate how IQCs are connected to dissipativity theory and the stability condition of DW shell separation. 

Consider the standard feedback interconnection shown in Figure~\ref{Fig_MDelta}, where $G$ is a linear time-invariant operator with the transfer function $G(s)$ in $\mathcal{RH}_{\infty}^{m\times m}$, and $\Delta$ is a bounded causal operator modeling an uncertainty, which can be nonlinear. The IQC theorem in~\cite{SR_1} shows that if, for every $\tau \in [0,1]$, the interconnection of $G$ and $\tau \Delta$ is well-posed, and moreover, 

i) if the IQC defined by $\Pi$ 
\begin{equation}\label{eq:IQC_Delta}
\int_{-\infty}^{\infty}
\begin{bmatrix} v(j\omega) \\  w(j\omega)\end{bmatrix}\hermconj
\Pi(j\omega,\tau)
\begin{bmatrix} v(j\omega) \\  w(j\omega)\end{bmatrix}d\omega
\geq 0 
\end{equation}
is satisfied by {$w = \tau \Delta \circ v$}, where $v(j\omega)$ and $w(j\omega)$ represent the harmonic spectrum of the signals $u$ and $w$ at the frequency $\omega$;

ii) and if there exists $\epsilon > 0$ such that
\begin{equation}\label{eq:IQC_G}
   \begin{bmatrix} G(j\omega) \\  I\end{bmatrix}\hermconj
\Pi(j\omega, \tau)
\begin{bmatrix} G(j\omega) \\  I\end{bmatrix}  \preceq -\epsilon I, \quad \forall \omega \in \mathbb{R} \,,
\end{equation}
then the feedback interconnection of $G$ and $\tau \Delta$ is stable. 

\sdbarfig{%
\centering
\includegraphics[width=.16\textwidth]{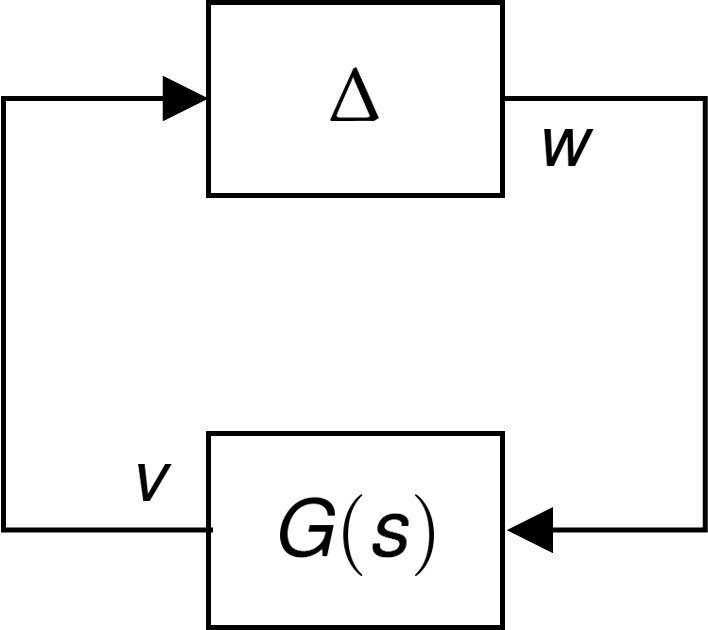}
}{A standard feedback system where $\Delta$ models the uncertainty and disturbance. \label{Fig_MDelta}}

To illustrate the connections with dissipativity and the DW shell separation, we next consider $\Delta$ to be linear time-invariant with associated transfer function $\Delta(s)$. Notice that in this case, the system in Figure~\ref{Fig_MDelta} is equivalent to the one in Figure~\ref{fdbk} considering $\Delta(s) = -H(s)$. Eq.~\eqref{eq:IQC_Delta} then becomes 
\begin{equation}\label{eq:IQC_Delta_s}
   \begin{bmatrix} I \\ \tau\Delta(j\omega) \end{bmatrix}\hermconj
\Pi(j\omega,\tau)
\begin{bmatrix} I \\ \tau\Delta(j\omega) \end{bmatrix}  \succeq 0, \quad \forall \omega \in \mathbb{R} \,,
\end{equation}
considering that $w(j\omega) = \tau \Delta(j\omega) v(j\omega)$. Notice that we drop the integral in~\eqref{eq:IQC_Delta} considering that the integral must hold for every valid input spectrum $v(j\omega)$, which requires the quadratic constraint~\eqref{eq:IQC_Delta_s} to hold at any frequency $\omega$.

For a fixed $\tau$ and a fixed $\omega$, we consider a special structure of $\Pi(j\omega,\tau)$ parameterized as
\begin{equation}\label{eq:Pi}
    \Pi(j\omega,\tau) = \begin{bmatrix}
        aI & (bI)\hermconj \\
        bI & -cI
    \end{bmatrix} \,,
\end{equation}
where $a \in \mathbb{R}$, $b \in \mathbb{C}$, $c \in \mathbb{R}$, and we drop their dependence on $\tau$ and $\omega$ for simplicity. By substituting~\eqref{eq:Pi} into~\eqref{eq:IQC_G} and~\eqref{eq:IQC_Delta_s}, we obtain 
\begin{equation}
    % aG^HG+bG  + b^HG^H - cI
    aG\hermconj(j\omega)G(j\omega) + bG(j\omega) + b\hermconj G\hermconj(j\omega)  - cI \preceq -\epsilon I \,,
\end{equation}
\begin{equation}
    aI+ b\tau \Delta\hermconj(j\omega) + b\hermconj \tau \Delta(j\omega) - c \tau^2 \Delta\hermconj(j\omega)\Delta(j\omega) \succeq 0 \,.
\end{equation}
Then, we multiply the right side of the above two equations by an arbitrary unit vector $x \in \mathbb{C}^m$ and the left side by $x\hermconj$, obtaining
\begin{equation}\label{eq:G_ineq}
    a\|G(j\omega)x\|^2 + 2\Re(bx\hermconj G(j\omega)x) - c \le -\epsilon \,,
\end{equation}
\begin{equation}\label{eq:D_ineq}
    a+ 2\Re(bx\hermconj \tau \Delta\hermconj(j\omega) x)  - c \|\tau \Delta\hermconj(j\omega) x\|^2 \ge 0 \,.
\end{equation}
Let $z_G = x\hermconj G(j\omega)x$, $\nu_G = \|G(j\omega)x\|^2$, $z_\Delta = x\hermconj \tau \Delta\hermconj(j\omega) x$, and $\nu_\Delta = \|\tau \Delta\hermconj(j\omega) x\|^2$. It then follows from the definition of DW shell in~\eqref{eq:DWA} that $(z_G,\nu_G) \in DW(G(j\omega))$, $(z_\Delta,\nu_\Delta) \in DW(\tau \Delta(j\omega))$, and further that $(\hat z_\Delta,\hat \nu_\Delta):=(\frac{z_\Delta\hermconj}{\nu_\Delta}, \frac{1}{\nu_\Delta}) \in DW(\frac{1}{\tau} \Delta^{-1}(j\omega))$~\cite{zhang2025phantom}. Moreover, Eqs.~\eqref{eq:G_ineq} and~\eqref{eq:D_ineq} can be rewritten as
\begin{equation}
    a \nu_G + 2\Re(bz_G) - c \le -\epsilon \,,
\end{equation}
\begin{equation}
    a \hat \nu_\Delta + 2\Re(b\hat z_\Delta) - c \ge 0 \,,
\end{equation}
and it can be deduced that for any $c-\epsilon < \hat c <c$, the hyperplane $H = \{(z,\nu)\in \mathbb{C} \times \mathbb{R}_{>0}: a \nu + 2\Re(bz) = \hat c, a \in \mathbb{R}, b \in \mathbb{C}\}$ separates $DW(G(j\omega))$ and $DW(\frac{1}{\tau} \Delta^{-1}(j\omega))$, as shown in Figure~\ref{Fig_DW_shell_IQC_separation}.

\sdbarfig{%
\centering
\includegraphics[width=.48\textwidth]{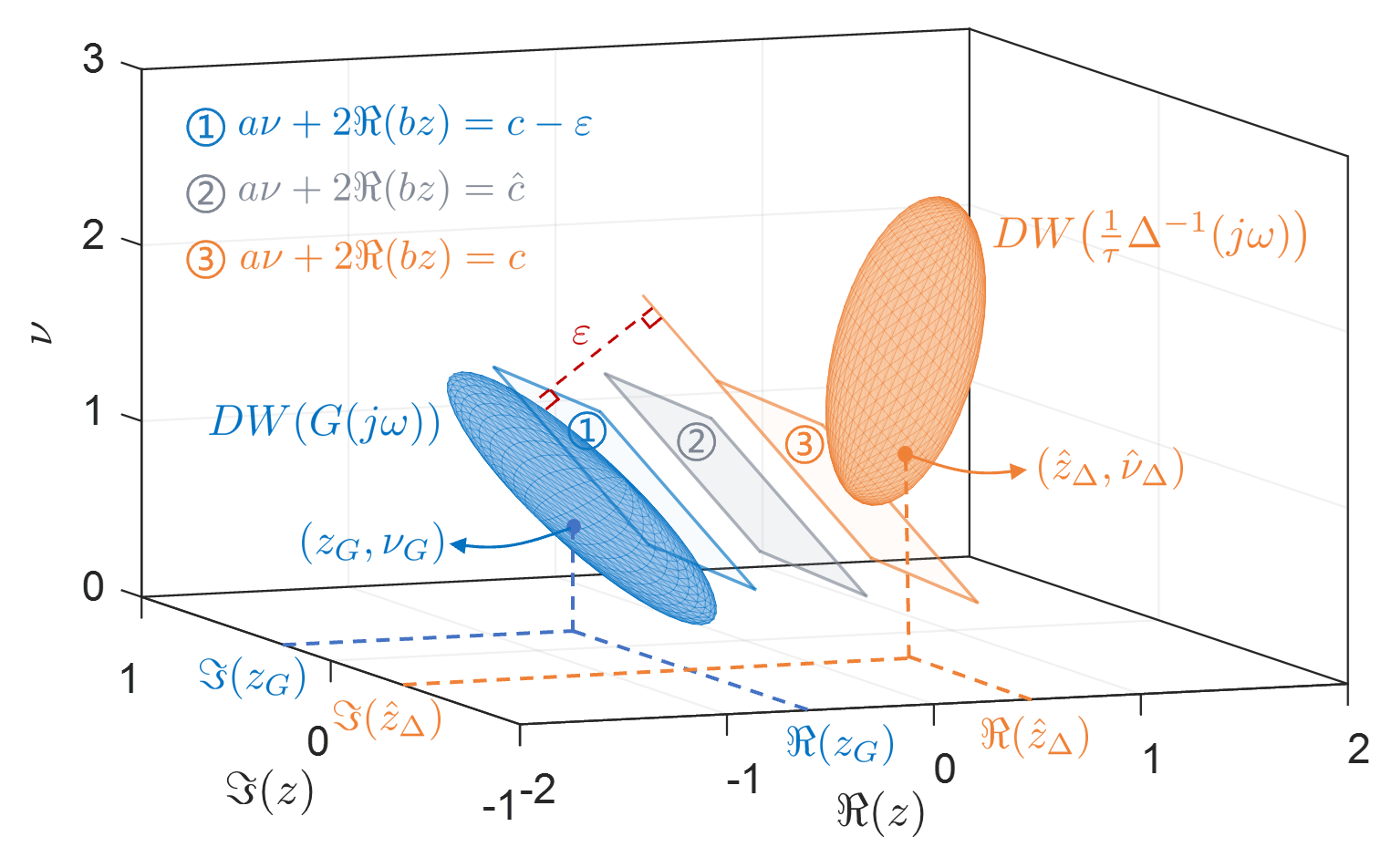}
}{Geometric interpretation of the IQC stability condition via DW shell separation. \label{Fig_DW_shell_IQC_separation}}

In summary, by noticing that $\Delta(s) = -H(s)$, one can derive the DW shell separation condition in~\eqref{eq:DW_sep} from the IQCs~\eqref{eq:IQC_Delta} and~\eqref{eq:IQC_G}, by considering a special form of $\Pi(j\omega, \tau)$ as in~\eqref{eq:Pi}. Hence, the stability certificate of DW shell separation is a special case of the IQC theory~\cite{hallinan2026decentralised,liu2026scalable}.

We next show how IQC, where we drop the integral and merely use the quadratic constraints as in~\eqref{eq:IQC_G} and~\eqref{eq:IQC_Delta_s}, is connected to dissipativity. As a first step, for any fixed $\tau$, we factor $\Pi(j\omega, \tau)$ into $\Pi(j\omega, \tau) = \Psi\hermconj(j\omega)M\Psi(j\omega)$, where $\Psi(j\omega)$ can be thought of as a filter and $M$ is a constant symmetric matrix. Then, we rewrite~\eqref{eq:IQC_G} and~\eqref{eq:IQC_Delta_s} as
\begin{equation}\label{eq:H_G}
    H_G\hermconj(j\omega) M H_G(j\omega) + \epsilon I \preceq 0, \quad \forall \omega \in \mathbb{R} \,,
\end{equation}
\begin{equation}\label{eq:H_Delta}
    -H_\Delta\hermconj(j\omega) M H_\Delta(j\omega) \preceq 0, \quad \forall \omega \in \mathbb{R} \,,
\end{equation}
where $H_G(j\omega) = \Psi(j\omega)\begin{bmatrix} G(j\omega) \\  I\end{bmatrix} $ is the transfer function matrix from $u_G = w$ to $y_G = \Psi(j\omega)\begin{bmatrix} v \\  w\end{bmatrix}$
and $H_\Delta(j\omega) = \Psi(j\omega)\begin{bmatrix} I \\ \tau\Delta(j\omega) \end{bmatrix}$ is the transfer function matrix from $u_\Delta = v$ to $y_\Delta = \Psi(j\omega)\begin{bmatrix} v \\  w\end{bmatrix}$. We notice that, by definition, $y_G =y_\Delta$ coincide.

\end{sidebar}

%*******************************************
\begin{sidebar}{}
\setcounter{sequation}{0}
\renewcommand{\thesequation}{S\arabic{sequation}}
\setcounter{stable}{0}
\renewcommand{\thestable}{S\arabic{stable}}
\setcounter{sfigure}{0}
\renewcommand{\thesfigure}{S\arabic{sfigure}}

Let $(A_G, B_G, C_G, D_G)$ be a minimal state-space realization of $H_G(j\omega)$ with state vector $x_G$, and let $(A_\Delta, B_\Delta, C_\Delta, D_\Delta)$ be a minimal state-space realization of $H_\Delta(j\omega)$ with state vector $x_\Delta$. Then, using the KYP lemma, it can be derived that~\eqref{eq:H_G} holds if and only if there exists a symmetric matrix $P_G \succ 0$ that satisfies the following LMI~\cite{SR_2}:
\begin{equation}\label{eq:LMI_P_G}
\begin{split}
    &\begin{bmatrix}
        A_G^\top P_G + P_G A_G & P_G B_G \\
        B_G^\top P_G & 0
    \end{bmatrix} + \begin{bmatrix} C_G & D_G \end{bmatrix}^\top M\begin{bmatrix} C_G & D_G \end{bmatrix} \\
    &+ \begin{bmatrix} 0 & 0 \\ 0 & \epsilon I \end{bmatrix} \preceq 0,
\end{split}
\end{equation}
and~\eqref{eq:H_Delta} holds if and only if there exists a symmetric matrix $P_\Delta \succ 0$ that satisfies the following LMI:
\begin{equation}
    \begin{bmatrix}
        A_\Delta^\top P_\Delta + P_\Delta A_\Delta & P_\Delta B_\Delta \\
        B_\Delta^\top P_\Delta & 0
    \end{bmatrix} - \begin{bmatrix} C_\Delta & D_\Delta \end{bmatrix}^\top M\begin{bmatrix} C_\Delta & D_\Delta \end{bmatrix} \preceq 0.
\end{equation}

We define the storage function of the system in~\eqref{eq:H_G} as $V_G(x_G) = x_G^\top P_G x_G$ and the the storage function of the system in~\eqref{eq:H_Delta} as $V_\Delta(x_\Delta) = x_\Delta^\top P_\Delta x_\Delta$. It follows from $\dot V_G(x_G) = x_G^\top (A_G^\top P_G + P_G A_G)x_G + u_G^\top B_G^\top P_G x_G + x_G^\top P_G B_G u_G$, $y_G = C_G x_G + D_G u_G$, and~\eqref{eq:LMI_P_G} that 
\begin{equation}\label{eq:VG_dot}
    \dot V_G(x_G) \le -y_G^\top M y_G  - \epsilon \|u_G\|^2 =: w_G(u_G, y_G)\,,
\end{equation}
where $w_G(u_G, y_G)$ is the supply rate.
Analogously, we obtain
\begin{equation}\label{eq:VDelta_dot}
    \dot V_\Delta(x_\Delta) \le y_\Delta^\top M y_\Delta  =: w_\Delta(u_\Delta, y_\Delta)\,,
\end{equation}
where $w_\Delta(u_\Delta, y_\Delta)$ is the supply rate. Let $V(x_G,x_\Delta) = V_G(x_G) + V_\Delta(x_\Delta)$ be the storage function of the closed-loop system, which, according to~\eqref{eq:VG_dot}, \eqref{eq:VDelta_dot}, and $y_G = y_\Delta$, satisfies
\begin{equation}
    \dot V(x_G,x_\Delta) \le w_G(u_G, y_G) + w_\Delta(u_\Delta, y_\Delta) = - \epsilon \|u_G\|^2 \le 0 \,,
\end{equation}
aligned with the dissipativity condition in~\eqref{eq:dissip}. That is to say, one can infer dissipativity and stability if the feedback system satisfies the IQCs in~\eqref{eq:IQC_G} and~\eqref{eq:IQC_Delta_s}.

{In closing, IQCs are a more general framework that allows one to derive dissipativity certificates and DW-shell separation, which again imply a range of other concepts: passivity, small gain, and SRG separation included. The price for this level of generality is that IQCs are also more unwieldy for design, less interpretable, and do not admit a graphical stability certificate.}

\end{sidebar}
%*******************************************

\subsection{Composite Graphical Stability Certificates}

Since the small-gain condition~\eqref{eq:SG}, the small-phase condition~\eqref{eq:SP}, the numerical range separation~\eqref{eq:W_sep}, the $x$-$z$ graph separation~\eqref{eq:P_sep}, and the SRG separation~\eqref{eq:SRG_sep} all imply the DW shell separation in~\eqref{eq:DW_sep}, we have the following stability claim. 

\begin{highlightbox}
\textbf{Graphical Stability Certificates}

Consider \( G\), \( H \in \mathcal{RH}_{\infty}^{m\times m}\).
The closed-loop system \( G(s)\#H(s) \) in Figure~\ref{fdbk} is stable if, for each \( \omega \in [0,\infty) \),
    \textbf{either}
     \begin{enumerate}

        \item \label{cond:passivity} the \textbf{passivity condition}~\eqref{eq:pas_phase} \textit{holds, \textbf{or}}
        \vspace{1mm}
    
        \item \label{cond:sp} the \textbf{small-phase condition}~\eqref{eq:SP} \textit{holds, \textbf{or}}
        \vspace{1mm}

        \item \label{cond:sg} the \textbf{small-gain condition}~\eqref{eq:SG} \textit{holds, \textbf{or}}
        \vspace{1mm}
    
        \item \label{cond:nr} the \textbf{numerical range separation}~\eqref{eq:W_sep} \textit{holds, \textbf{or}}
        \vspace{1mm}
    
        \item \label{cond:xz} the \textbf{x-z graph separation}~\eqref{eq:P_sep} \textit{holds, \textbf{or}}
        \vspace{1mm}

        \item \label{cond:SRG} the \textbf{SRG separation}~\eqref{eq:SRG_sep} \textit{holds, \textbf{or}}
        \vspace{1mm}
    
        \item \label{cond:dw} the \textbf{DW shell separation}~\eqref{eq:DW_sep} \textit{holds, \textbf{or}}
        \vspace{1mm}

        \item the sufficient version of GNC~\eqref{eq:suff_GNC}
        \textit{holds.}
\end{enumerate}

\end{highlightbox}

\begin{sidebar}{}
\setcounter{sequation}{0}
\renewcommand{\thesequation}{S\arabic{sequation}}
\setcounter{stable}{0}
\renewcommand{\thestable}{S\arabic{stable}}
\setcounter{sfigure}{0}
\renewcommand{\thesfigure}{S\arabic{sfigure}}
\section[Connection Between Passivity Index and x-z Graph]{Connection Between Passivity Index and x-z Graph}
\label{sb-xz-passivityindex}

We investigate the connection between the passivity index and the $x$-$z$ graph below. Consider the interconnected system again in Figure~\ref{fdbk}. The input feedforward passivity (IFP) index for $G(s)$ is defined in~\eqref{eq:IFPindex},
% \begin{equation}
%     \nu =  \min\limits_{\omega \in [0,\infty)} \frac{1}{2} \lambda_1(G(j\omega) + G^*(j\omega)) \,,
% \end{equation}
% where $\lambda_1$ denotes the minimum eigenvalue.
and the output feedback passivity (OFP) index for a minimum-phase $H(s)$ is defined in~\eqref{eq:OFPindex}. 
% \begin{equation}
%     \rho =  \frac{1}{2} \min_{\omega \in \mathbb{R}} \lambda_{\min}\bigl(H(j\omega)^{-1} + (H(j\omega)^{-1})^\mathsf{H}\bigr) \,.
% \end{equation}
Based on these definitions, we have the following classic stability condition: the closed-loop system \( G(s)\#H(s) \) in Figure~\ref{fdbk} is stable if $\nu + \rho >0$~\cite{SR_1}.

We further notice that 
\begin{equation*}
    v= \frac{1}{2} \lambda_{\min}\bigl(G(j\omega) + G(j\omega)\hermconj\bigr) = \min\limits_{\|x\| = 1} \Re(x{\hermconj}G(j\omega)x) \,,
\end{equation*}
which represents the exact location of the leftmost bounding vertical line of the $x$-$z$ graph of $G(j\omega)$, which is $x =v$ as shown in Figure~\ref{Fig_p_index_xz}~(a). Similarly, we have 
\begin{equation*}
    \rho =\frac{1}{2} \lambda_{\min}\bigl(H(j\omega)^{-1} + (H(j\omega)^{-1})^\mathsf{H}\bigr) = \min\limits_{\|x\| = 1} \Re(x{\hermconj}H^{-1}(j\omega)x) \,,
\end{equation*}
which represents the exact location of the leftmost bounding vertical line of the $x$-$z$ graph of $H^{-1}(j\omega)$. It also corresponds to the rightmost bounding vertical line of the $x$-$z$ graph of $-H^{-1}(j\omega)$ and of $-\tfrac{1}{\tau}H^{-1}(j\omega)$ under $\tau \in (0,1]$, which is $x = -\rho$.

Hence, if $\nu + \rho >0$, then there exists a vertical line that separates the $x$-$z$ graph of $G(j\omega)$ (with any $\omega \in [0,\infty)$) from the the $x$-$z$ graph of $-\tfrac{1}{\tau}H^{-1}(j\omega)$ (with any $\omega \in [0,\infty)$ and any $\tau \in (0,1]$), as shown in Figure~\ref{Fig_p_index_xz}, implying that the $x$-$z$ graph separation in~\eqref{eq:P_sep} holds for any $\omega \in [0,\infty)$ and the closed-loop system is stable. In other words, the stability condition based on the passivity index is a special case of the condition of $x$-$z$ graph separation, and it is therefore more conservative than the $x$-$z$ graph separation. For instance, Figure~\ref{Fig_p_index_xz} shows a case where the condition based on the passivity index is violated, but the system is stable since the $x$-$z$ graphs are separated.

\sdbarfig{%
\centering
\includegraphics[width=.43\textwidth]{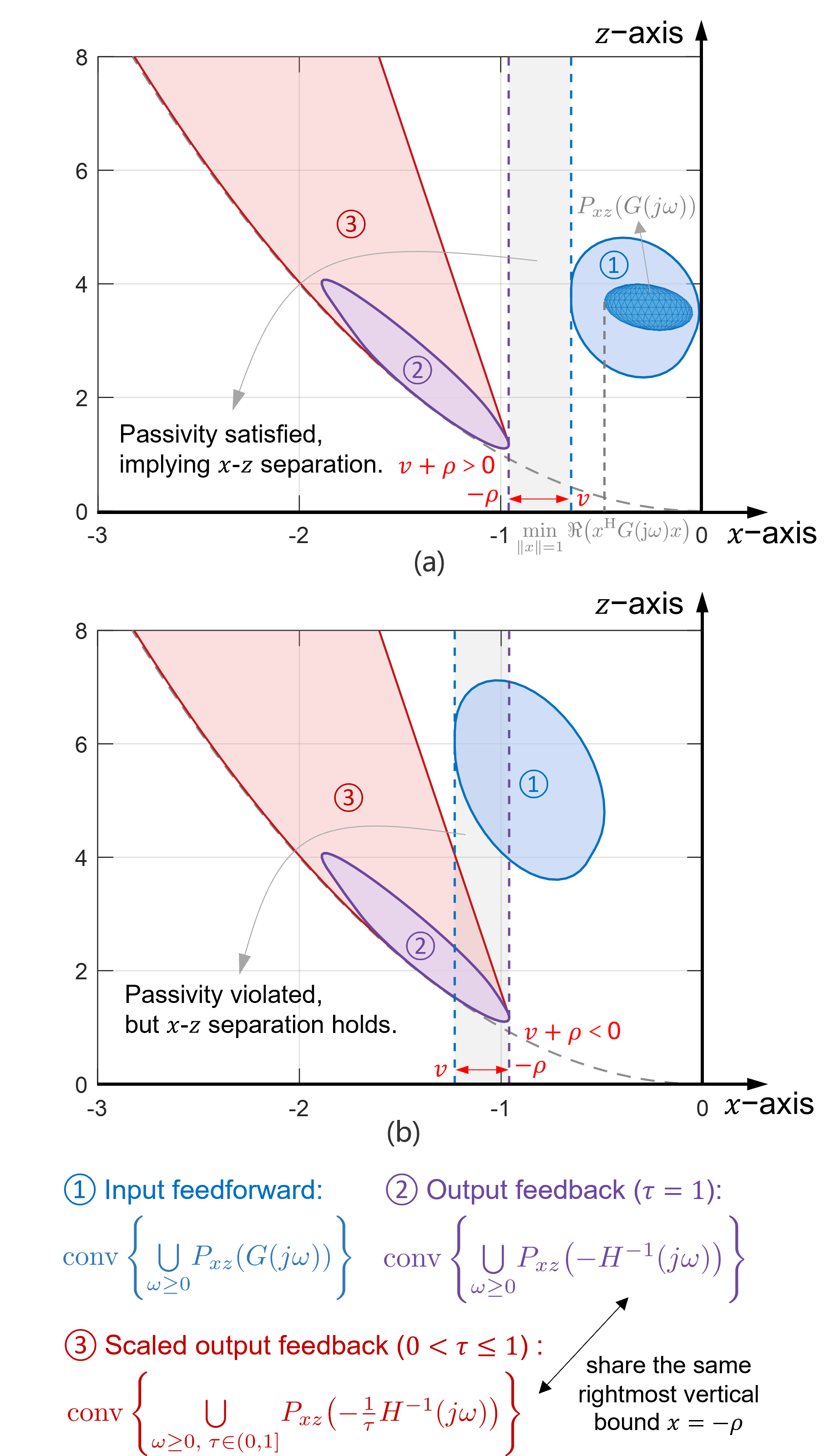}
}{Comparison of the stability conditions based on passivity index and the $x$-$z$ graph separation. \label{Fig_p_index_xz}}

\end{sidebar}

The above graphical stability certificates connect different paradigms and different approaches of stability analysis, and it shows that these stability conditions are complementary and can be checked in a frequency-wise manner. Moreover, the above stability conditions exhibit different levels of conservatism and require different computational efforts. For instance, the conditions of numerical range separation, $x$-$z$ graph separation, and SRG separation all require checking whether their 2D plots intersect, which are more computationally demanding, while less conservative, than the conditions of passivity, small gain, and small phase. The condition of DW shell separation is even less conservative, but it requires checking whether 3D plots intersect, which requires more computational effort. 

\begin{figure}[!t]%[b]
\centering
% \vskip-.25cm
\includegraphics[width=.48\textwidth]{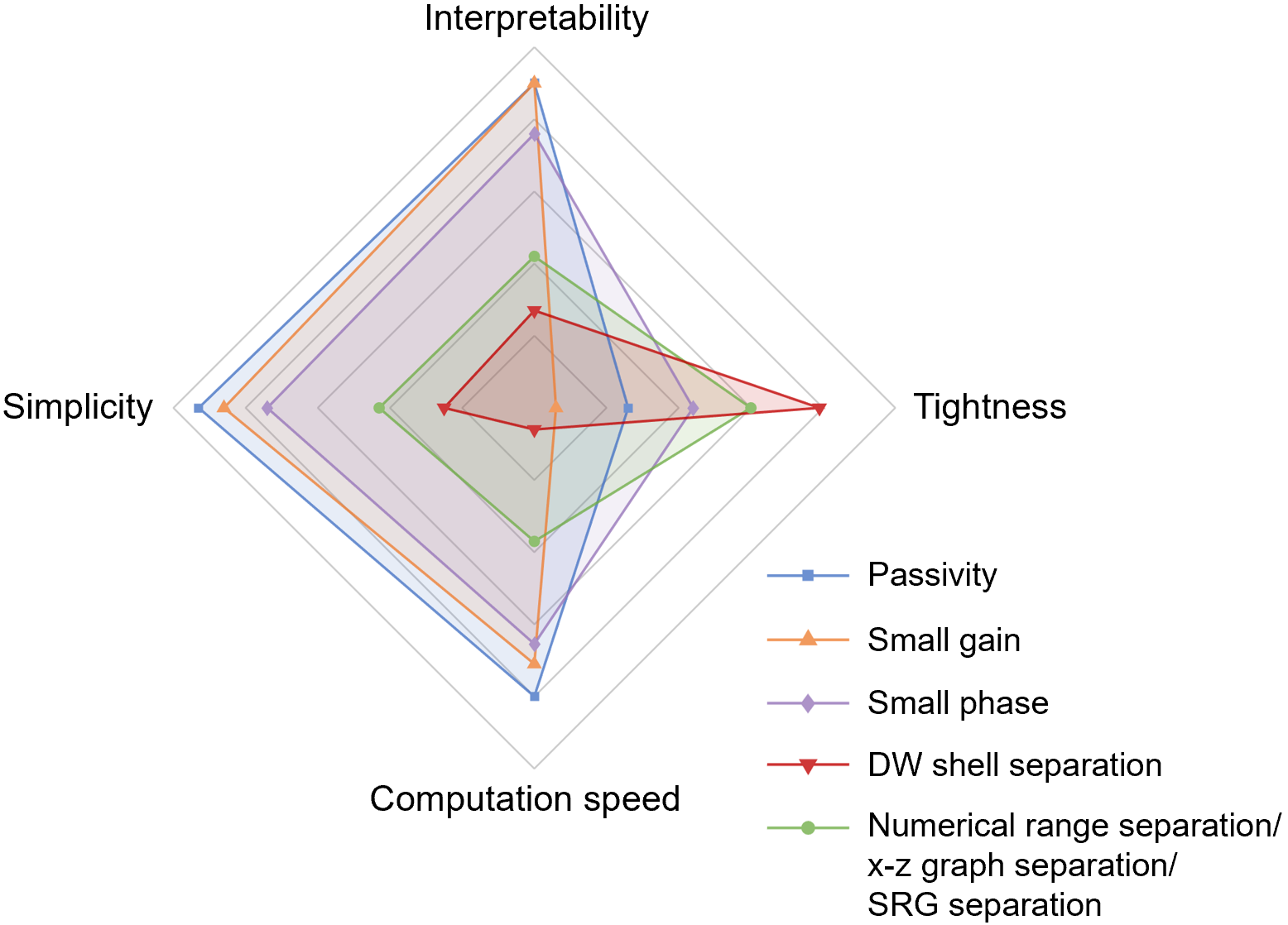}
\caption{A qualitative comparison of the stability certificates.} \la{Fig_compare_plot}
\vspace{0mm}
\end{figure}

Figure~\ref{Fig_compare_plot} compares the stability certificates in terms of interpretability, conservativeness, simplicity, and computational effort. It can be seen that the condition of DW shell separation has the highest computational effort but the least conservativeness; the passivity and small-gain conditions have the highest interpretability, while the passivity condition is generally less conservative than the small-gain condition; the small-phase theorem generalizes over passivity and is even less conservative.
In practical use, to balance the computational burden and the conservativeness of stability analysis, one can first check 1) the passivity condition, 2) the small-phase condition, and 3) the small-gain condition for the whole frequency range \( \omega \in [0,\infty) \). Then, at the frequencies where the above conditions are violated, check the 2D conditions: 1) the numerical range separation, and 2) the $x$-$z$ graph separation (or equivalently, the SRG separation). Finally, at the frequencies where none of the above conditions hold, check the 3D DW shell separation condition. Although one can also check the sufficient condition of GNC in~\eqref{eq:suff_GNC}, it involves computing the multiplication of $G(j\omega)$ and $H(j\omega)$, which impedes the understanding of their interaction and the further development of decentralized conditions.

\begin{figure}[!t]%[b]
\centering
% \vskip-.25cm
\includegraphics[width=.35\textwidth]{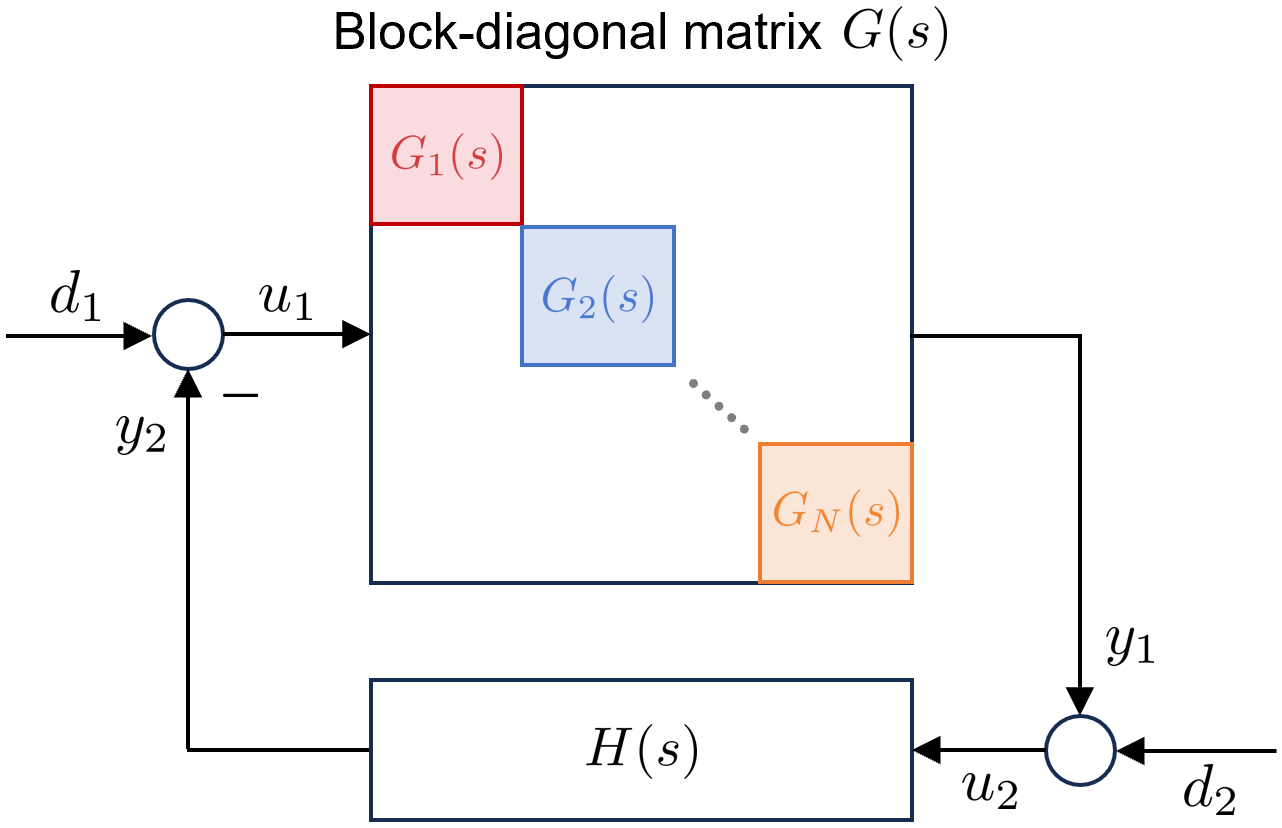}
\caption{The closed-loop diagram of a multi-agent system.} \la{Fig_blkdiag_G}
\vspace{0mm}
\end{figure}

\subsection{Graphical Decentralized Stability Certificates}
\label{subsec:geometric_decentralized_stability}

Consider now a multi-agent system, where $N$ agents interact with each other through an interconnection network. As illustrated in Figure~\ref{Fig_blkdiag_G}, the transfer function matrix $H(s)$ captures the dynamics of the network, and the block-diagonal transfer function matrix $G(s) = {\rm blkdiag}(G_1(s),G_2(s),...,G_N(s))$ captures the collective dynamics of the $N$ agents, where $G_i(s)$ is the transfer function (matrix) describing the dynamics of the $i$-th agent. We notice that the closed-loop system in Figure~\ref{Fig_blkdiag_G} is a special case of the one in Figure~\ref{fdbk}, where $G(s)$ has a block-diagonal structure. Hence, the previous stability conditions also apply to this system. However, instead of focusing on $G(s)$, we wish to develop decentralized certificates and reveal how each $G_i(s)$ should match the network characteristics $H(s)$ so that the overall interconnected system is stable.

We start with a special (though typical in power systems) case where $H^{-1}(j\omega)$ is constant and positive definite, that is, $H^{-1}(j\omega) = H^{-1} \succ 0$. For instance, $H^{-1}$ can be a grounded Laplacian matrix describing the network topology and line characteristics~\cite{dorfler2012kron}. In this case, the DW shell of $-H^{-1}$ is a polygon on the $x$-$z$ plane with $y=0$~\cite{li2009davis}. In particular, the vertices of the polygon lie exactly on the curve $z = x^2$, where the $i$-th vertex is $(-\lambda_i(H^{-1}),\lambda_i^2(H^{-1}))$ and $\lambda_i(H^{-1})$ is the $i$-th eigenvalue of $H^{-1}$.
The purple triangle in Figure~\ref{Fig_DW_H} is an example of $DW(-H^{-1})$ when $H \succ 0$ is a 3-by-3 matrix with three distinct eigenvalues. 

\begin{figure}[!]%[b]
\centering
% \vskip-.25cm
\includegraphics[width=.48\textwidth]{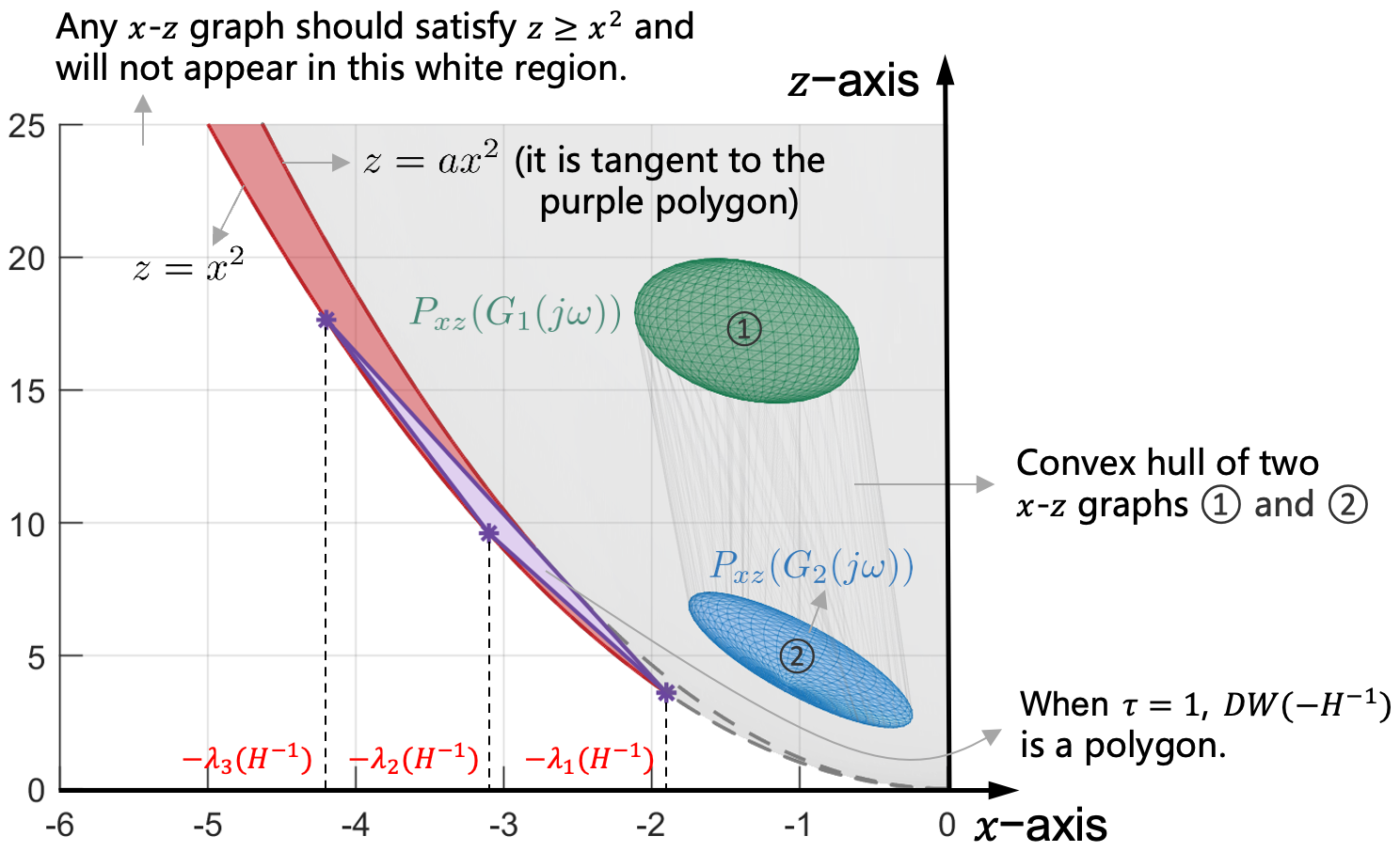}
\caption{Illustration of the DW shell homotogy of $-\tfrac{1}{\tau}H^{-1}$ with $\tau \in (0,1]$, i.e., the red area. The eigenvalues of a positive definite matrix $A \in \mathbb{R}^{n \times n}$ are organized as $\lambda_1(A) \le \lambda_2(A) \le \dots \le \lambda_n(A)$.} \la{Fig_DW_H}
\vspace{0mm}
\end{figure}

When further considering \( DW(-\frac{1}{\tau} H^{-1}) \) in the stability condition~\eqref{eq:DW_sep}, it becomes a DW shell homotopy since $\tau$ varies within the range of $\tau \in (0,1]$, as illustrated by the red area in Figure~\ref{Fig_DW_H}. 
This red area is expanded from the purple triangle (the case of $\tau = 1$), where, as $\tau$ approaches 0, the vertices slice along $z=x^2$ toward the upper-left direction, because the $i$-th vertex becomes $(-\frac{1}{\tau}\lambda_i(H^{-1}),\frac{1}{\tau^2}\lambda_i^2(H^{-1}))$. Hence, the lower bound of the red area is $z = x^2$, while
the upper bound of the red area is $z = ax^2$, where $a$ makes this upper bound tangent to the purple triangle.

We next investigate the DW shell of the block-diagonal $G(s)$. According to Property~2 of DW shell, the DW shell of $G(j\omega)$ is the convex hull of the DW shells of all its blocks, that is,
\[
DW(G(j\omega)) = {\rm conv}\{DW(G_1(j\omega)),...,DW(G_N(j\omega))\}\,.
\]
Further, since projection and convex hull commute, we have a similar result for the $x$-$z$ graph of $G(j\omega)$ as
\[
P_{xz}(G(j\omega)) = {\rm conv}\{P_{xz}(G_1(j\omega)),...,P_{xz}(G_N(j\omega))\}\,.
\]
Figure~\ref{Fig_DW_H} shows an example plot of the convex hull of two $x$-$z$ graphs for better illustration. We notice that any $x$-$z$ graph is always above $z=x^2$ because of Property~3 of DW shell. Moreover, we observe that the convex hull will not intersect the red area unless the individual $x$-$z$ graphs do, that is,
\begin{equation*}
    P_{xz}(G(j\omega)) \cap P_{xz}(-\tfrac{1}{\tau} H^{-1}) = \varnothing 
\end{equation*}
holds for all $\tau \in (0,1]$ if and only if
\begin{equation*}
    \forall i:P_{xz}(G_i(j\omega)) \cap P_{xz}(-\tfrac{1}{\tau} H^{-1}) = \varnothing \,.
\end{equation*}
This indicates that the condition of $x$-$z$ graph separation can be examined in a decentralized fashion, namely, agent by agent. The passivity, small-phase, and small-gain conditions also share similar decentralized compositionality properties, as shown in~\cite{huang2024gain}. Moreover, since SRG has a one-to-one correspondence to the $x$-$z$ graph, we can also derive a similar decentralized property. In ``\nameref{sb-xz-passivityindex},'' we further investigate how the condition of $x$-$z$ graph separation is related to the stability condition based on the passivity index.

\begin{figure}[!]%[b]
\centering
% \vskip-.25cm
\includegraphics[width=.43\textwidth]{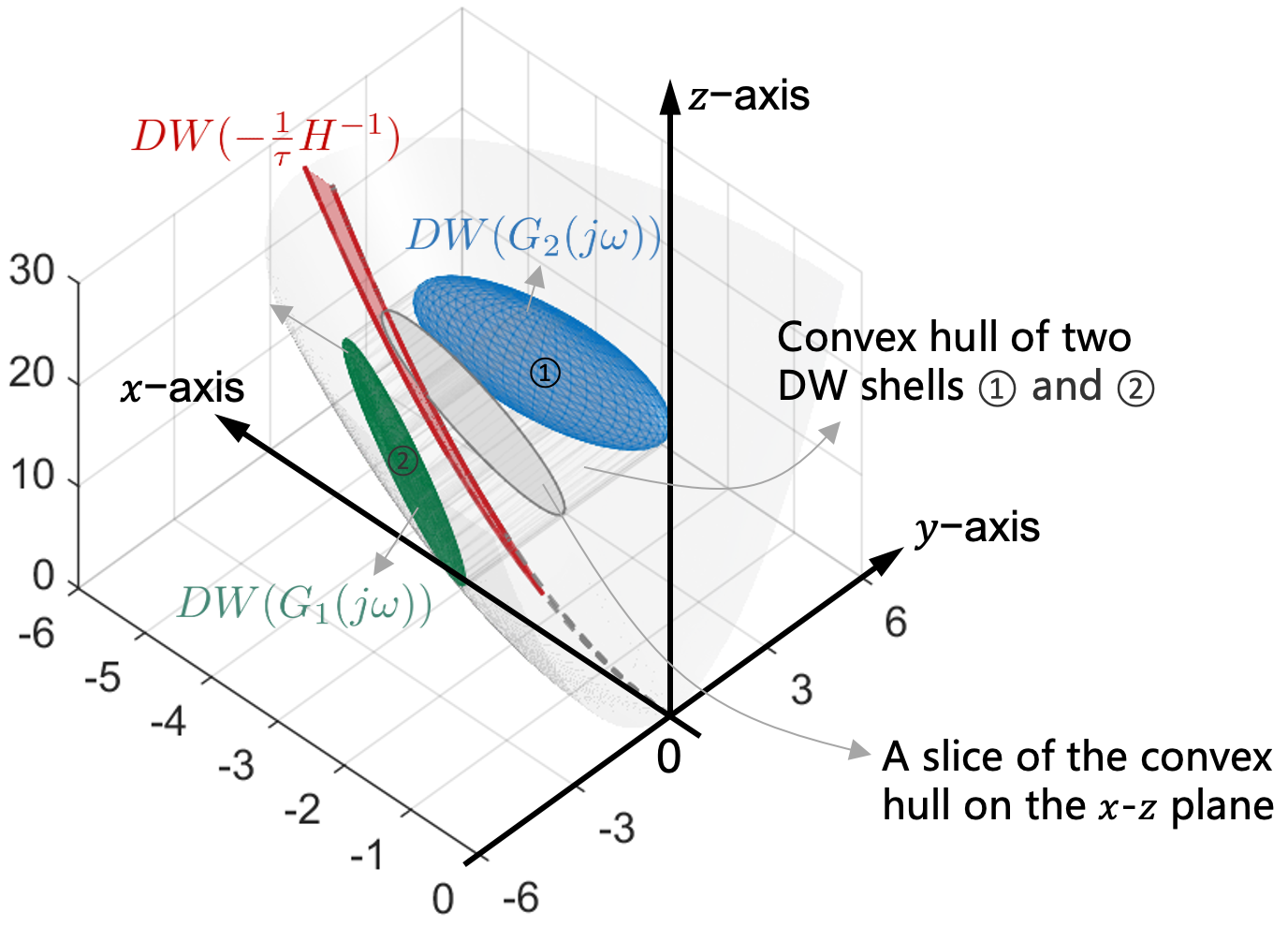}
\caption{The convex hull of the blue DW shell and the green DW shell intersects with the DW shell trajectory of $-\tfrac{1}{\tau}H^{-1}$, even though the two individual DW shells are separated from this DW shell trajectory.} \la{Fig_DW_intersect}
\vspace{0mm}
\end{figure}

However, for DW shell, such a decentralized property cannot be naturally obtained since the DW shell trajectory of $-\tfrac{1}{\tau}H^{-1}$ is a 2D plane figure in the 3D space, which may intersect with the convex hull even without intersecting the individual DW shells, as illustrated in Figure~\ref{Fig_DW_intersect}. To this end, we construct the following 3D geometry by expanding $P_{xz}(-\tfrac{1}{\tau} H^{-1})$ along the $y$-axis:
\begin{equation}\label{eq:DW_expd}
    \widehat{DW}(-\tfrac{1}{\tau} H^{-1}):=\{(x+jy,z):(x,z) \in P_{xz}(-\tfrac{1}{\tau} H^{-1}), x^2+y^2 \le z \} \,,
\end{equation}
where $\tau \in (0,1]$. We refer to $\widehat{DW}(-\tfrac{1}{\tau} H^{-1})$ as the \textit{expanded DW shell trajectory} of $-\tfrac{1}{\tau} H^{-1}$. Note that the projection of $\widehat{DW}(-\tfrac{1}{\tau} H^{-1})$ onto the $x$-$z$ plane is exactly $P_{xz}(-\tfrac{1}{\tau} H^{-1})$, that is, it shares the same $x$-$z$ graph with $-\tfrac{1}{\tau} H^{-1}$.

\begin{figure}[!]%[b]
\centering
% \vskip-.25cm
\includegraphics[width=.43\textwidth]{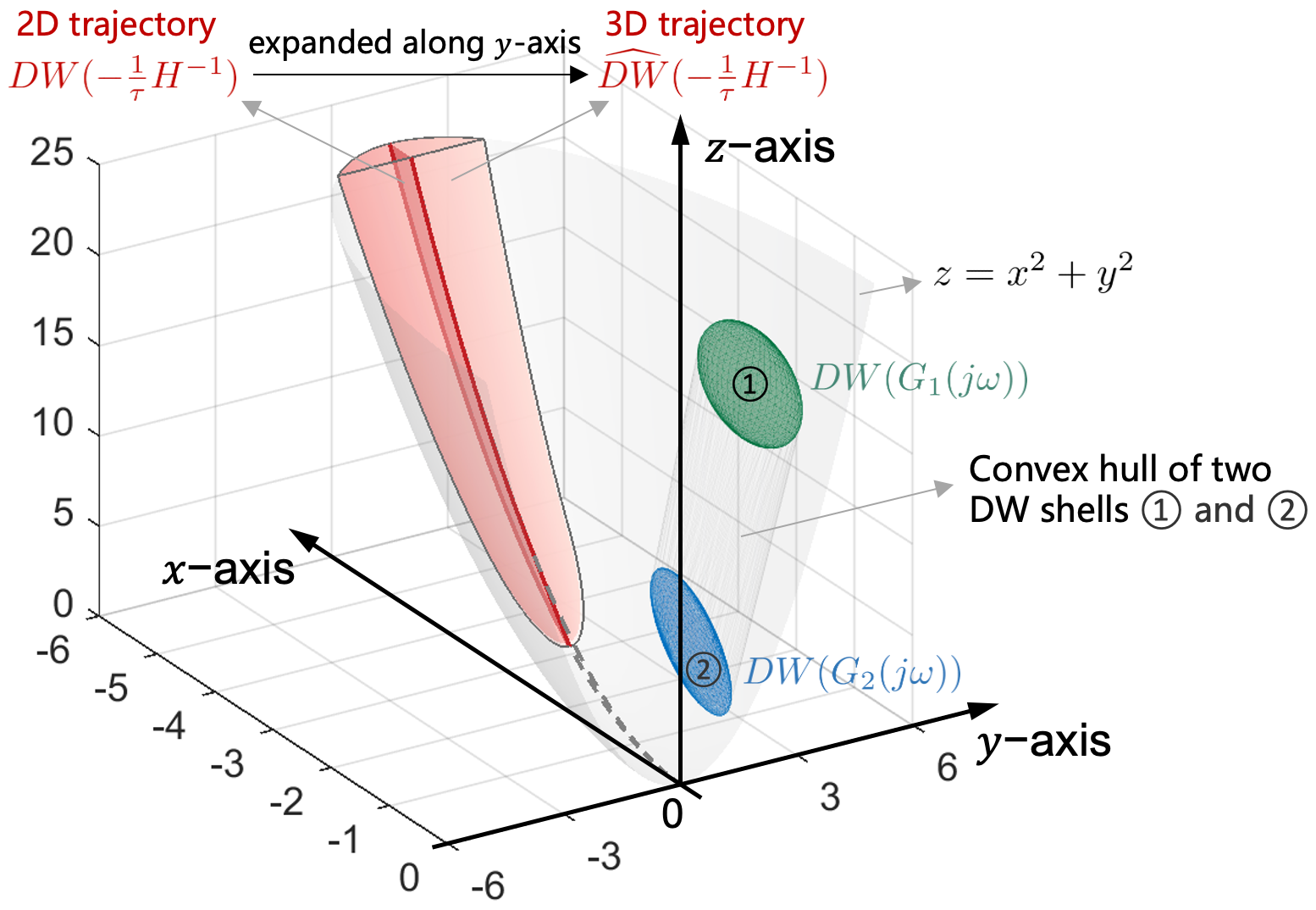}
\caption{Illustration of the expanded DW shell trajectory of $-\tfrac{1}{\tau} H^{-1}$.} \la{Fig_DW_expand}
\vspace{0mm}
\end{figure}

It can be seen from Figure~\ref{Fig_DW_expand} that inside the solid paraboloid $z\ge x^2+y^2$, $\widehat{DW}(-\tfrac{1}{\tau} H^{-1})$ is concave upward. Moreover, according to Property~3 of DW shell, any DW shell is embedded in $z\ge x^2+y^2$. Hence, we can deduce that the convex hull of multiple DW shells will not intersect with $\widehat{DW}(-\tfrac{1}{\tau} H^{-1})$ if and only if the individual DW shell does not intersect with $\widehat{DW}(-\tfrac{1}{\tau} H^{-1})$. To be specific, a decentralized condition is that 
\begin{equation*}
    DW(G(j\omega)) \cap \widehat{DW}(-\tfrac{1}{\tau} H^{-1}) = \varnothing 
\end{equation*}
holds for all $\tau \in (0,1]$ if and only if
\begin{equation*}
    \forall i:DW(G_i(j\omega)) \cap \widehat{DW}(-\tfrac{1}{\tau} H^{-1}) = \varnothing \,,
\end{equation*}
which implies
\begin{equation*}
    DW(G(j\omega)) \cap {DW}(-\tfrac{1}{\tau} H^{-1}) = \varnothing \,,
\end{equation*}
considering that 
\begin{equation}
    {DW}(-\tfrac{1}{\tau} H^{-1}) \subset \widehat{DW}(-\tfrac{1}{\tau} H^{-1})\,.
\end{equation}
Note that one can also expand the DW shell trajectory of $-\tfrac{1}{\tau} H^{-1}$ along other directions to possibly reduce the conservativeness of the separation condition.

\begin{figure}[!]%[b]
\centering
% \vskip-.25cm
\includegraphics[width=.43\textwidth]{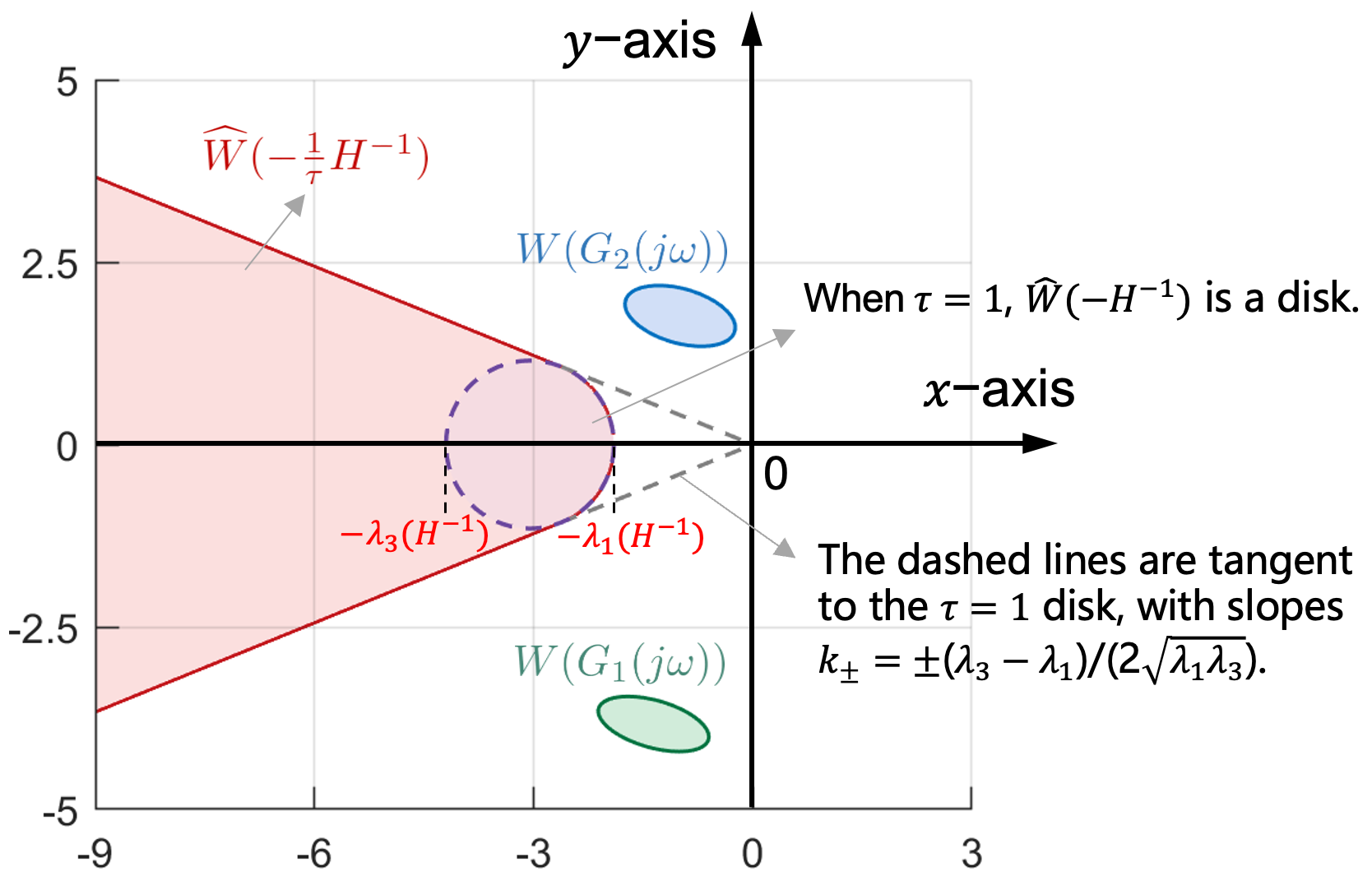}
\caption{Illustration of the expanded numerical range trajectory of $-\tfrac{1}{\tau} H^{-1}$.} \la{Fig_W_expand}
\vspace{0mm}
\end{figure}

We further define the \textit{expanded numerical range trajectory} as the projection of $\widehat{DW}(-\tfrac{1}{\tau} H^{-1})$ onto the $x$-$y$ plane, as illustrated in Figure~\ref{Fig_W_expand}, which is
\begin{equation}
    \widehat{W}(-\tfrac{1}{\tau} H^{-1}):= \{c: \exists z ~{\rm s.t.}~ (c,z) \in \widehat{DW}(-\tfrac{1}{\tau} H^{-1}) \} \,.
\end{equation}
A decentralized condition also exists for $\widehat{W}(-\tfrac{1}{\tau} H^{-1})$, analogous to the one above based on $\widehat{DW}(-\tfrac{1}{\tau} H^{-1})$.

Based on the above concepts, we are now ready to present the graphical and fully decentralized stability certificates below.

\begin{highlightbox}
\textbf{Graphical Decentralized Stability Certificates}

Consider \( G\), \( H \in \mathcal{RH}_{\infty}^{m\times m}\), where $G(s)$ is block-diagonal as $G(s) = {\rm blkdiag}(G_1(s),G_2(s),...,G_N(s))$ and $H(s) = H \succ 0$.
The closed-loop system \( G(s)\#H(s) \) in Figure~\ref{Fig_blkdiag_G} is stable if, for each \( \omega \in [0,\infty) \),
    \textbf{either}
     \begin{enumerate}

        \item 
        $\forall i:$ $G_i(j\omega)$  \textit{is \textbf{passive}, \textbf{or}}
        \vspace{-5mm}
        \begin{equation}\label{eq:pas}
        \vspace{-5mm}
            \,
        \end{equation}
        \vspace{1mm}

        \item the \textbf{decentralized phase condition}, i.e., 
        \begin{equation}\label{eq:DecPhase}
        \begin{cases}
        \textit{a) } \max\limits_i \, \overline{\phi}(G_i(j\omega)) < \pi, ~\min\limits_i \, \underline{\phi}(G_i(j\omega)) > -\pi, \\[2pt]
        \textit{b) } \max\limits_i \, \overline{\phi}(G_i(j\omega)) 
        - \min\limits_i \, \underline{\phi}(G_i(j\omega)) < \pi,
        \end{cases}
        \vspace{-2.5mm}
        \end{equation}
        \textit{holds, \textbf{or}}
        \vspace{1mm}

        \item 
        the \textbf{decentralized gain condition}, i.e.,
        \begin{equation}\label{eq:DecGain}
        \max_i \, \overline{\sigma}(G_i(j\omega)) 
        < \underline{\sigma}(H^{-1}) \,,
        \  \quad{\rm \textit{holds, \textbf{or}}}
        \end{equation}
        \vspace{-2mm}

        \item the \textbf{decentralized numerical range separation}, i.e.,
        \begin{equation}\label{eq:decentralized_W}
            \forall i:W(G_i(j\omega)) \cap \widehat{W}(-\tfrac{1}{\tau} H^{-1}) = \varnothing \,,\forall \tau \in (0,1],
        \end{equation}
        \textit{holds, \textbf{or}}
        \vspace{1mm}

        \item the \textbf{decentralized x-z graph separation}, i.e.,
        \begin{equation}\label{eq:decentralized_P}
            \forall i:P_{xz}(G_i(j\omega)) \cap P_{xz}(-\tfrac{1}{\tau} H^{-1}) = \varnothing \,,\forall \tau \in (0,1],
        \end{equation}
        \textit{holds, \textbf{or}}
        \vspace{1mm}

        \item the \textbf{decentralized SRG separation}, i.e.,
        \begin{equation}\label{eq:decentralized_SRG}
            \forall i:SRG(G_i(j\omega)) \cap SRG(-\tfrac{1}{\tau} H^{-1}) = \varnothing \,,\forall \tau \in (0,1],
            \vspace{-2.5mm}
        \end{equation}
        \textit{holds, \textbf{or}}
        \vspace{1mm}

        \item the \textbf{decentralized DW shell separation}, i.e.,
        \begin{equation}\label{eq:decentralized_DW}
            \forall i:DW(G_i(j\omega)) \cap \widehat{DW}(-\tfrac{1}{\tau} H^{-1}) = \varnothing \,,\forall \tau \in (0,1],
        \end{equation}
        \textit{holds.}

\end{enumerate}

\end{highlightbox}

The above condition of decentralized DW shell separation is derived by constructing an upper-bound surface in the 3D space that is concave upward, as shown in~\eqref{eq:DW_expd}. Such an upper-bound surface is not unique, and one may find a better upper surface to reduce the conservatism of the condition. 
In ``\nameref{sb-xz-general_condition},'' we further present a decentralized stability condition when $H(s)$ is a general MIMO transfer function matrix, without assuming that it is constant and positive definite.

\begin{sidebar}{}
\setcounter{sequation}{0}
\renewcommand{\thesequation}{S\arabic{sequation}}
\setcounter{stable}{0}
\renewcommand{\thestable}{S\arabic{stable}}
\setcounter{sfigure}{0}
\renewcommand{\thesfigure}{S\arabic{sfigure}}
\section{A Decentralized Stability Certificate for a Dynamic $H$ Block}
\label{sb-xz-general_condition}

We present a decentralized stability certificate when $H(s)$ is not a constant positive-definite matrix but a general transfer function matrix, that is, $H(s) \in \mathcal{RH}_{\infty}^{m\times m}$. In this case, the DW shell trajectory of $-\tfrac{1}{\tau} H^{-1}(j\omega)$ is no longer the red region in Figure~\ref{Fig_DW_H}, and the key to developing decentralized stability conditions is to find a region that contains this DW shell trajectory and has an upper bound that is concave upward.
Hence, we first construct an upper bound for this DW shell trajectory. 

\sdbarfig{%
\centering
\includegraphics[width=.43\textwidth]{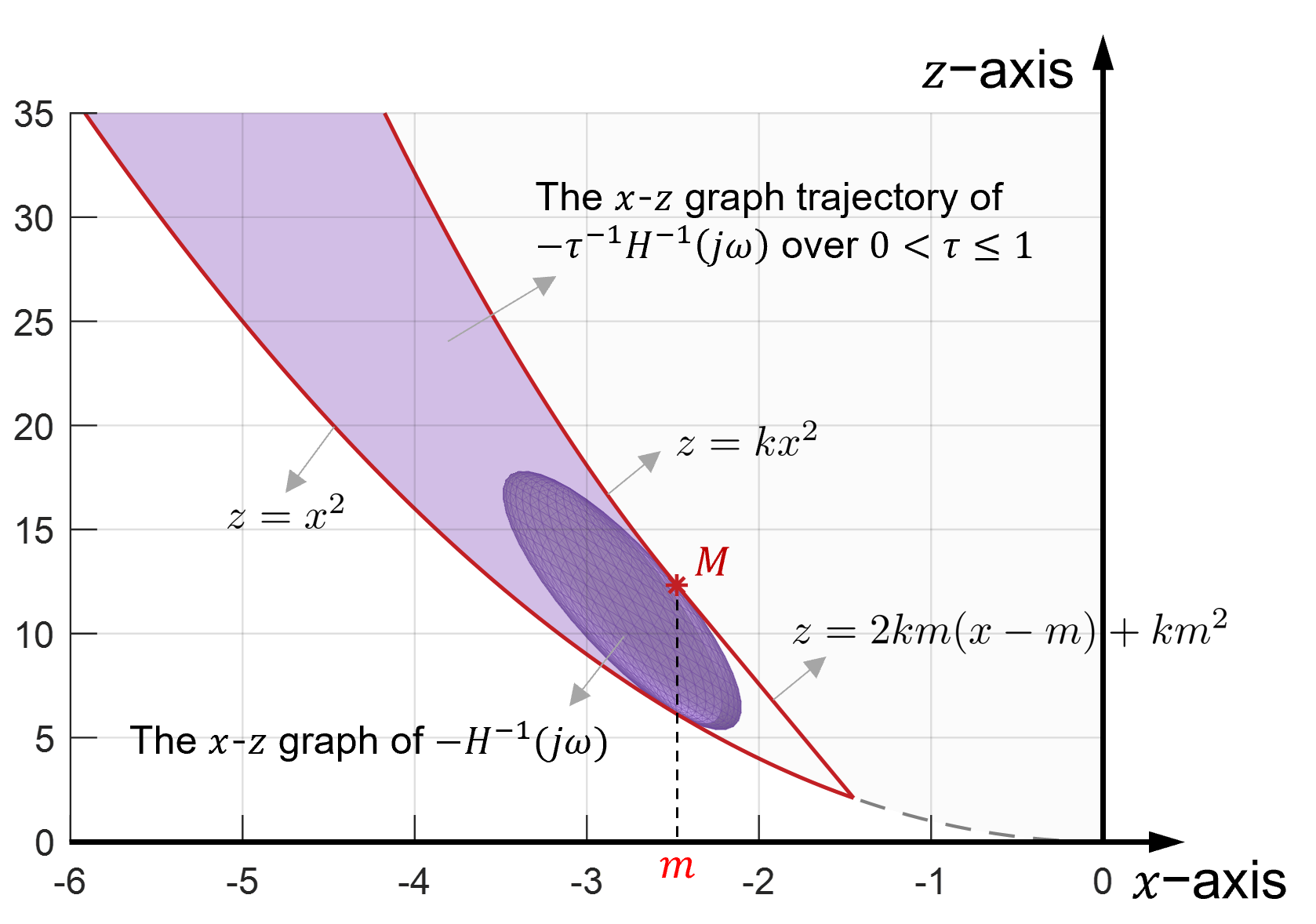}
}{Illustration of the expanded DW shell trajectory in the $x$-$z$ plane that contains the original DW shell trajectory of $-\tfrac{1}{\tau} H^{-1}(j\omega)$. \label{Fig_xz_Hs}}

Notice that according to~\cite{zhang2025phantom}, there exists a parabola $z = k x^2$ on the $x$-$z$ plane that upper bounds the $x$-$z$ graph of $-\tfrac{1}{\tau} H^{-1}(j\omega)$ over $\tau \in (0,1]$, where $k$ is selected so that $z = k x^2$ is tangent to the $x$-$z$ graph of $-H^{-1}(j\omega)$, as shown in Figure~\ref{Fig_xz_Hs}. We denote the intersection point as $M$, and its $x$ coordinate is $m$. Without loss of generality, we assume that $-H^{-1}(j\omega)$ is on the left half $x$-$z$ plane, similar to the case in Figure~\ref{Fig_DW_H}. We next draw a line $z = 2km(x-m) + km^2$ ($x\ge m$), which together with $z = k x^2$ (x<m), constitutes an upper bound that is concave upward. Based on this upper bound and the lower bound $z = x^2$ (i.e., the red lines in Figure~\ref{Fig_xz_Hs}), we construct the following expanded DW shell trajectory that contains the original DW shell trajectory of $-\tfrac{1}{\tau} H^{-1}(j\omega)$
\begin{equation}
\label{eq:expanded_DW_trajectory}
\begin{split}
    \widetilde{DW}(-\tfrac{1}{\tau} H^{-1}(j\omega)):= & \{(x+jy,z):x^2+y^2 \le z, \\
    & \,\, z\le kx^2 \quad {\rm if} \quad x<m, \\
    & \,\, z \le 2km(x-m) + km^2 \quad {\rm if} \quad x\ge m
    \} \,.
\end{split}
\end{equation}

The above expanded DW shell trajectory has an upper bound surface that is concave upward and thus can be used in decentralized stability analysis. We further define the projections of $\widetilde{DW}(-\tfrac{1}{\tau} H^{-1}(j\omega))$ onto the $x$-$y$ plane and $x$-$z$ plane as $\widetilde{W}(-\tfrac{1}{\tau} H^{-1}(j\omega))$ and $\widetilde{P}_{xz}(-\tfrac{1}{\tau} H^{-1}(j\omega))$, respectively. The SRG associated with $\widetilde{P}_{xz}(-\tfrac{1}{\tau} H^{-1}(j\omega))$ is denoted by $\widetilde{SRG}(-\tfrac{1}{\tau} H^{-1}(j\omega))$.
Based on these definitions, the generalized decentralized stability certificates with $H(s)$ are as follows. 

Consider \( G\), \( H \in \mathcal{RH}_{\infty}^{m\times m}\), where $G(s)$ is block-diagonal as $G(s) = {\rm blkdiag}(G_1(s),G_2(s),...,G_N(s))$ and the $x$-$z$ graphs of $-H^{-1}(j\omega)$ lie on the left half $x$-$z$ plane.
The closed-loop system \( G(s)\#H(s) \) in Figure~\ref{Fig_blkdiag_G} is stable if, for each \( \omega \in [0,\infty) \),
    \textbf{either}
     \begin{enumerate}

        \item $H(j\omega)$ \textit{is \textbf{strictly passive}} and $\forall i:$ $G_i(j\omega)$  \textit{is \textbf{passive}, \textbf{or}}
        \vspace{1mm}

        \item $H(j\omega)$ \textit{is \textbf{sectorial}} and the \textbf{decentralized phase condition}, i.e., 
        \begin{equation}\label{eq:DecPhase_g}
        \begin{cases}
        \textit{a) } \max\limits_i \, \overline{\phi}(G_i(j\omega)) < \pi - \overline{\phi}(H(j\omega)), \\[2pt]
        \textit{b) } \min\limits_i \, \underline{\phi}(G_i(j\omega)) > -\pi - \underline{\phi}(H(j\omega)), \\[2pt]
        \textit{c) } \max\limits_i \, \overline{\phi}(G_i(j\omega)) 
        - \min\limits_i \, \underline{\phi}(G_i(j\omega)) < \pi,
        \end{cases}
        \vspace{0mm}
        \end{equation}
        \textit{holds, \textbf{or}}
        \vspace{1mm}

        \item 
        the \textbf{decentralized gain condition}, i.e.,
        \begin{equation}\label{eq:DecGain_g}
        \max_i \, \overline{\sigma}(G_i(j\omega)) 
        < \underline{\sigma}(H^{-1}(j\omega)) \,,
        \  \quad{\rm \textit{holds, \textbf{or}}}
        \end{equation}
        \vspace{-2mm}

        \item the \textbf{decentralized numerical range separation}, i.e.,
        \begin{equation}\label{eq:decentralized_W_g}
            \forall i:W(G_i(j\omega)) \cap \widetilde{W}(-\tfrac{1}{\tau} H^{-1}(j\omega)) = \varnothing \,,
            \vspace{-2.5mm}
        \end{equation}
        \textit{holds, \textbf{or}}
        \vspace{1mm}

        \item the \textbf{decentralized x-z graph separation}, i.e.,
        \begin{equation}\label{eq:decentralized_P_g}
            \forall i:P_{xz}(G_i(j\omega)) \cap \widetilde{P}_{xz}(-\tfrac{1}{\tau} H^{-1}(j\omega)) = \varnothing \,,
        \end{equation}
        \textit{holds, \textbf{or}}
        \vspace{1mm}

        \item the \textbf{decentralized SRG separation}, i.e.,
        \begin{equation}\label{eq:decentralized_SRG_g}
            \forall i:SRG(G_i(j\omega)) \cap  \widetilde{SRG}(-\tfrac{1}{\tau} H^{-1}(j\omega)) = \varnothing \,,
            \vspace{-2.5mm}
        \end{equation}
        \textit{holds, \textbf{or}}
        \vspace{1mm}

        \item the \textbf{decentralized DW shell separation}, i.e.,
        \begin{equation}\label{eq:decentralized_DW_g}
            \forall i:DW(G_i(j\omega)) \cap \widetilde{DW}(-\tfrac{1}{\tau} H^{-1}(j\omega)) = \varnothing \,,
        \end{equation}
        \textit{holds.}

\end{enumerate}

The above condition is derived based on the constructed upper bounds of the $x$-$z$ graph trajectory and the DW shell trajectory over $\tau \in (0,1]$, and these upper bounds are concave upward, which enables decentralized stability analysis. Related work has developed systematic outer bounds for scaled graphs using regions defined by integral quadratic constraints (IQCs), thereby simplifying feedback stability analysis \cite{baronprada2026scaled}. We also notice that such upper bounds are not unique, and one may leverage the characteristics of $H(j\omega)$ to find other upper bounds that make the stability conditions less conservative.

\end{sidebar}

% \newpage

% (Linbin's part)

% connecting passivity to numerical range and DW shell, and further relate it to small gain and SRG. 

% will present a unified perspective based on DW shell

\section{Non-parametric Stability Certificates for PE-Dominated Power Systems}
\label{sec:nonpara}

% With the rapid development of renewable energy, energy storage systems, electric vehicles, etc., more and more power converters are integrated into power systems, as they serve as the interface of grid connection and energy conversion. 
PE-dominated power systems feature the interconnection of power networks and heterogeneous converters which serve as the interface of renewable energy sources, energy storage systems, smart loads, etc. 
In what follows, we present the system model and demonstrate how the graphical decentralized stability certificates developed in the previous section can be used for modular and scalable analysis of PE-dominated power systems.

\subsection{Modeling of Power Network and Converters}
\label{subsec:power_network_converter_modeling}

Consider a prototypical PE-dominated power system as illustrated in Figure~\ref{Fig_network}, where $N$ power converters are interconnected via the power network. In our setting, we allow the power converters to employ different control architectures and different control parameters. For instance, the converter may apply the widely-used GFL control, where a PLL is used to follow the grid frequency; one can also use GFM control which can actively establish the frequency and voltage for the power grid, e.g., by emulating the swing equation to generate the internal frequency and achieve synchronization~\cite{yang2020placing, wang2017unified1, li2022revisiting}.

\begin{figure}[!]%[b]
\centering
% \vskip-.25cm
\includegraphics[width=.45\textwidth]{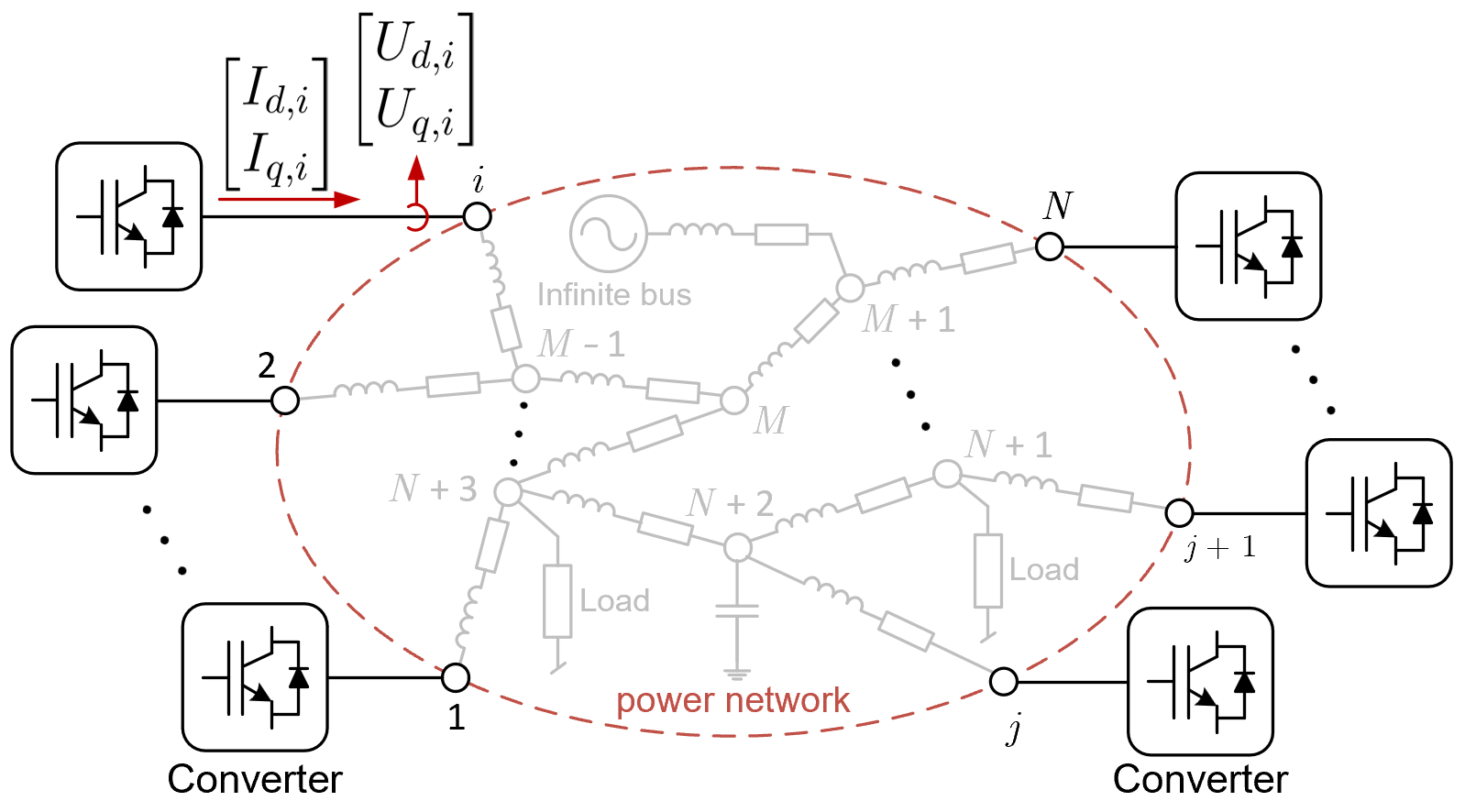}
\caption{Illustration of a PE-dominated power system where multiple converters are interconnected via the power network.} \la{Fig_network}
\vspace{0mm}
\end{figure}

The power network contains in total $M$ nodes, with $N$ converter nodes, denoted by Nodes~$1 \sim N$, and $M - N$ interior nodes ($M \ge N$) as Nodes~$N+1 \sim M$. The interior nodes may be connected to loads or shunt capacitors. 
We consider one common grounded node, denoted by Node~$M+1$. In small-signal dynamics, the infinite buses should also be considered as grounded since their voltage perturbations are zero. 
In a global $dq$ coordinate whose rotating speed is the nominal frequency $\omega_0$, 
the dynamic equation of the line between Node~$i$ and Node~$j$ ($i,j \in \{1, \dots,M+1\}$) can be expressed as
\begin{equation}\label{eq:edge_ij}
    \begin{bmatrix} \Delta I_{d,ij} \\ \Delta I_{q,ij} \end{bmatrix} = {Y}_{ij}(s) \left( \begin{bmatrix} \Delta U_{d,i} \\ \Delta U_{q,i} \end{bmatrix} - \begin{bmatrix} \Delta U_{d,j} \\ \Delta U_{q,j} \end{bmatrix} \right) \,,
\end{equation}
where $[ \Delta I_{d,ij} \; \Delta I_{q,ij} ]^\top$ is the perturbed current vector from Node~$i$ to Node~$j$ in the global $dq$ coordinate (with $\Delta$ meaning perturbations), $[ \Delta U_{d,i} \; \Delta U_{q,i} ]^\top$ is the perturbed voltage vector of Node~$i$ in the global $dq$ coordinate, and ${Y}_{ij}(s)$ is a $2 \times 2$ transfer function matrix encoding the dynamics of the line. For instance, if the line is composed of a resistor and an inductor (i.e., a typical power transmission line model), then we have 
\begin{equation}\label{eq:Y_ij}
    Y_{ij}(s) = B_{ij} \begin{bmatrix} s/\omega_0 + {\epsilon_{ij}} & -1 \\ 1 & s/\omega_0 + {\epsilon_{ij}} \end{bmatrix}^{-1} =: B_{ij} F_{\epsilon_{ij}}(s) \,,
\end{equation}
where $B_{ij} = \frac{1}{X_{ij}} \vspace{0.5mm}$ is the line susceptance ($X_{ij}$ is the reactance), {$\epsilon_{ij}$ is the R/X (resistance over reactance) ratio of the line~$ij$}, and $\omega_0$ is the nominal frequency of the system. If there is a load (resistor) at Node~$i$, then we have (with $j = M+1$)
\begin{equation}
    Y_{ij}(s) = \begin{bmatrix} R_{{\rm Load},i}^{-1} & 0 \\ 0 & R_{{\rm Load},i}^{-1} \end{bmatrix} \,,
\end{equation}
where $R_{{\rm Load},i}$ denotes the resistance of the load; as another example, if there is a shunt capacitor (e.g., in a Pi transmission line model) at Node~$i$, then we have (with $j = M+1$)
\begin{equation*}
    Y_{ij}(s) = \begin{bmatrix} sC_i & -\omega_0 C_i \\ \omega_0 C_i & sC_i \end{bmatrix} \,,
\end{equation*}
where $C_i$ is the shunt capacitor at Node~$i$.

Based on the dynamic equation of each line,
we can construct the $2M \times 2M$ grounded Laplacian (transfer function) matrix ${\bf Y}(s)$ where each $2 \times 2$ block is obtained as 
\begin{equation}
\begin{split}
    {\bf Y}_{ij}(s) &= - Y_{ij}(s), i \ne j, \\
    {\bf Y}_{ii}(s) &= \sum_{j = 1}^{M+1} Y_{ij}(s) .
\end{split}
\end{equation}
By eliminating the $M-N$ interior nodes through Kron reduction~\cite{dorfler2012kron}, we obtain a reduced network whose dynamics are captured by the following $2N \times 2N$ transfer function matrix
\begin{equation}\label{eq:Y_grid}
     \underbrace{ \begin{bmatrix} \Delta I_{d,1} \\ \Delta I_{q,1}  \\ \vdots \\ \Delta I_{d,N} \\ \Delta I_{q,N} \end{bmatrix}}_{=: {\bf I}} = \underbrace{[\;{\bf Y}_1(s) - {\bf Y}_2(s) {\bf Y}_4^{-1}(s) {\bf Y}_3(s)\; ] }_{=:{\bf Y}_{\rm grid}(s)}\underbrace{\begin{bmatrix} \Delta U_{d,1} \\ \Delta U_{q,1} \\  \vdots \\ \Delta U_{d,N} \\ \Delta U_{q,N} \end{bmatrix}}_{=:{\bf U}} ,
\end{equation}
where $[ \Delta I_{d,i} \; \Delta I_{q,i} ]^\top$ and $[\Delta U_{d,i} \; \Delta U_{q,i} ]^\top$ are respectively the perturbations of the injected current and terminal voltage at Node~$i$ in the global $dq$ coordinate, as illustrated in Figure~\ref{Fig_network}.
The matrices ${\bf Y}_1(s)$, ${\bf Y}_2(s)$, ${\bf Y}_3(s)$, and ${\bf Y}_4(s)$ are respectively $2N \times 2N$, $2N\times(2M-2N)$, $(2M-2N)\times 2N$, and $(2M-2N)\times (2M-2N)$ transfer function matrices defined by
\begin{equation*}
    {\bf Y}(s) =: \begin{bmatrix} {\bf Y}_1(s) & {\bf Y}_2(s) \\ {\bf Y}_3(s) & {\bf Y}_4(s) \end{bmatrix} .
\end{equation*}
% We rewrite~\eqref{eq:Y_grid} as
% \begin{equation}\label{eq:admittance_grid}
%     {\bf U} = \underbrace{{\bf Y}^{-1}_{\rm grid}(s)}_{=:{\bf Z}_{\rm grid}(s)} {\bf I} \,,
% \end{equation}
% which, together 

Since we consider a general expression of the line dynamics as in~\eqref{eq:edge_ij}, the Kron reduction process in~\eqref{eq:Y_grid}, which involves transfer function matrices, can be thought of as a generalization of the Kron reduction process of lossless networks. 
    In what follows, for the ease of illustration, we consider a power network with an identical R/X ratio $\epsilon$ so that $Y_{ij}(s) = B_{ij}{F}_\epsilon(s)$, where 
    \begin{equation}\label{eq:F_sigma}
    {F}_\epsilon(s) = \begin{bmatrix} s/\omega_0 + {\epsilon} & -1 \\ 1 & s/\omega_0 + {\epsilon} \end{bmatrix}^{-1}.
    \end{equation}
    Let ${\bf B} \!\in\! \mathbb{R}^{M \times M}$ be the grounded Laplacian matrix that encodes the network topology and susceptances, where ${\bf B}_{ij} = - B_{ij}$ if $i \ne j$, and ${\bf B}_{ii} = \sum_{j=1}^{M+1} B_{ij}$. By performing Kron reduction to $\bf B$, we obtain the reduced Laplacian matrix ${\bf B}_{\rm r} \in \mathbb{R}^{N \times N}$ which is calculated by ${\bf B}_{\rm r} = {\bf B}_1 - {\bf B}_2{\bf B}_4^{-1} {\bf B}_3$, where
    $$
        {\bf B}=: \begin{bmatrix} {\bf B}_1 \in \mathbb{R}^{N \times N} & {\bf B}_2 \in \mathbb{R}^{N \times (M-N)} \\ {\bf B}_3 \in \mathbb{R}^{(M-N) \times N} & {\bf B}_4 \in \mathbb{R}^{(M-N) \times (M-N)}\end{bmatrix}.
    $$
    Then, we have 
    \begin{equation}\label{eq:Ygrid_RX}
        {\bf Y}_{\rm grid}(s) = {\bf B}_{\rm r} \otimes {F}_\epsilon(s) \,,
    \end{equation}
    which is exactly a special case of~\eqref{eq:Y_grid}.
    Moreover, it can be seen that ${\bf Y}_{\rm grid}(s)$ in~\eqref{eq:Ygrid_RX} has a special structure as the Kronecker product of ${\bf B}_{\rm r}$ and ${F}_\epsilon(s)$, which allows simpler stability analysis compared with the general setting in~\eqref{eq:Y_grid} where the R/X ratios can be different across different lines.  

    At each node $i \in \{1,...,N\}$, a converter is installed, and its dynamics can be compactly described by 
    \begin{equation}\label{eq:admittance_model}
    \begin{bmatrix} \Delta I_{d,i} \\ \Delta I_{q,i} \end{bmatrix} = -S_i e^{J\theta_i} {\bf Y}_{{\rm C},i}(s) e^{-J\theta_i}\begin{bmatrix} \Delta U_{d,i} \\ \Delta U_{q,i} \end{bmatrix} \,,
    \end{equation}
    where ${\bf Y}_{{\rm C},i}(s)$ is a $2 \times 2$ transfer function matrix describing how a voltage perturbation causes the current fluctuation of the converter. This admittance is derived from the full small-signal converter model, which includes the dynamics of the filter and the control loops, such as the current control, power or voltage control, and synchronization control, as detailed in~\cite{xin2024many}. For system-level studies, however, reduced-order models retaining only the dominant converter dynamics are sometimes used. Moreover, $S_i$ is the capacity ratio of the $i$-th node’s rated capacity to the base capacity of per-unit calculation, $\theta_i$ is the steady-state angle difference between the global $dq$ coordinate and the $i$-th converter's local $dq$ coordinate, and $e^{J\theta_i} = \begin{bmatrix} \cos \theta_i & -\sin \theta_i \\ \sin \theta_i & \cos \theta_i \end{bmatrix} \vspace{1mm}$.

    We further extend~\eqref{eq:admittance_model} to capture the dynamics of all the $N$ converters as
    \begin{equation}\label{eq:admittance_devices}
    {\bf I} = -
    ({\bf S}\otimes I_2) \; e^{J \boldsymbol{\theta}} \; {\bf Y}_{\rm C}^N(s) \; e^{-J \boldsymbol{\theta}} \;
    {\bf U} \,,
\end{equation}
    where ${\bf S} \!=\! {\rm diag}\{S_1,\dots,S_N\} \!\!\in\!\! \mathbb{R}^{N \times N}$ is the capacity ratio matrix, $e^{J \boldsymbol{\theta}} \!\!=\!\! {\rm diag}\{e^{J\theta_1},\dots,e^{J\theta_N}\} \!\!\in\!\! \mathbb{R}^{2N \times 2N}$ is the steady-state coordinate transformation matrix,
and
${\bf Y}_{\rm C}^N(s)$ is a $2N \times 2N$ block-diagonal transfer matrix extended from ${\bf Y}_{{\rm C},i}(s)$ as
\begin{equation*}
    {\bf Y}_{\rm C}^N(s) = {\rm blkdiag}({\bf Y}_{{\rm C},1}(s),{\bf Y}_{{\rm C},2}(s),...,{\bf Y}_{{\rm C},N}(s)).
\end{equation*}

\begin{figure}[!t]
	\centering
	\includegraphics[width=2.9in]{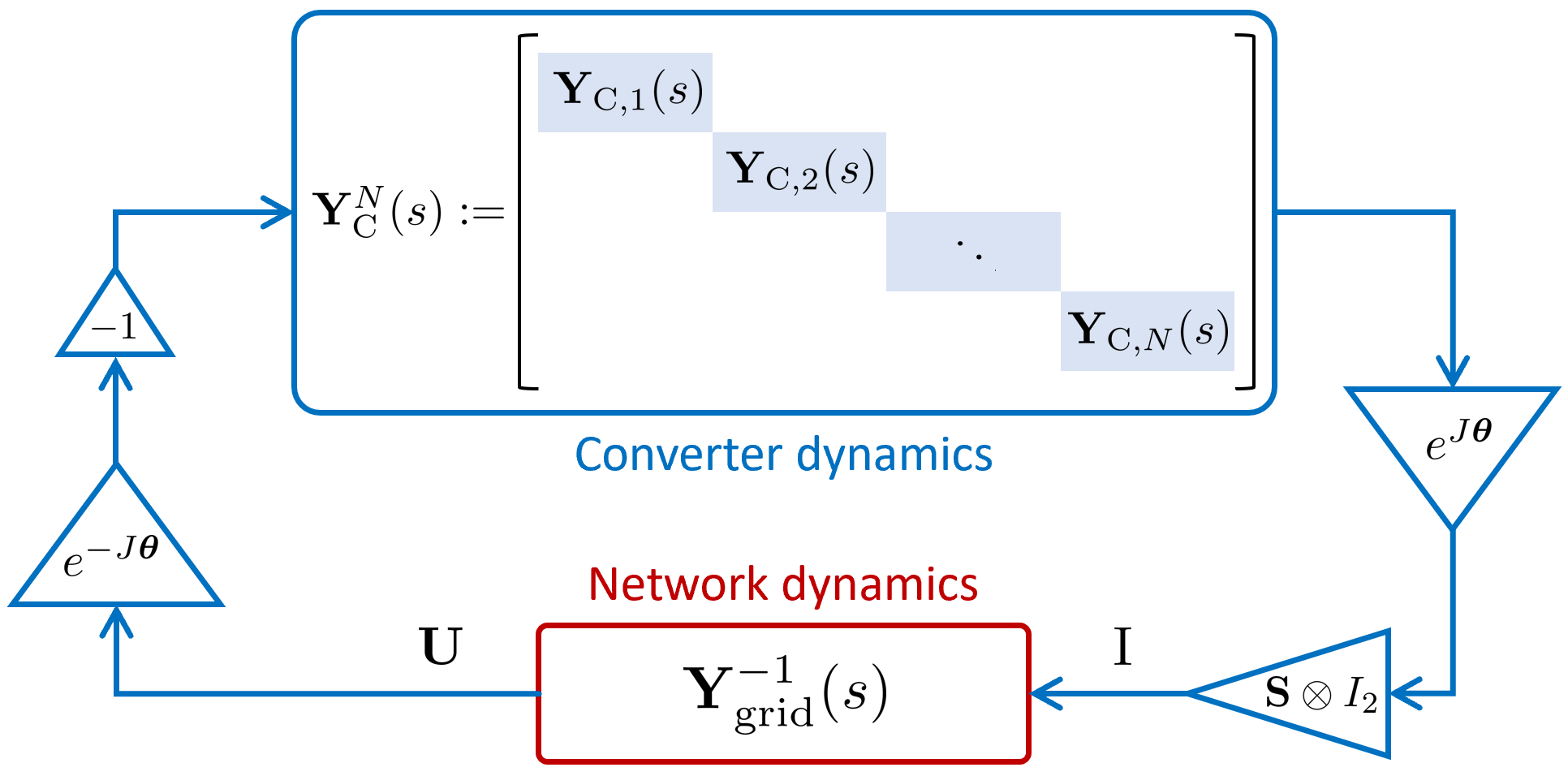}
	\vspace{0mm}
	%\DeclareGraphicsExtensions.
	\caption{Closed-loop model of a multi-converter system by combining~\eqref{eq:Y_grid} and~\eqref{eq:admittance_devices}.}
	\vspace{0mm}
	\label{Fig_closed_loop}
\end{figure}

\begin{figure}[!]%[b]
\centering
% \vskip-.25cm
\includegraphics[width=.48\textwidth]{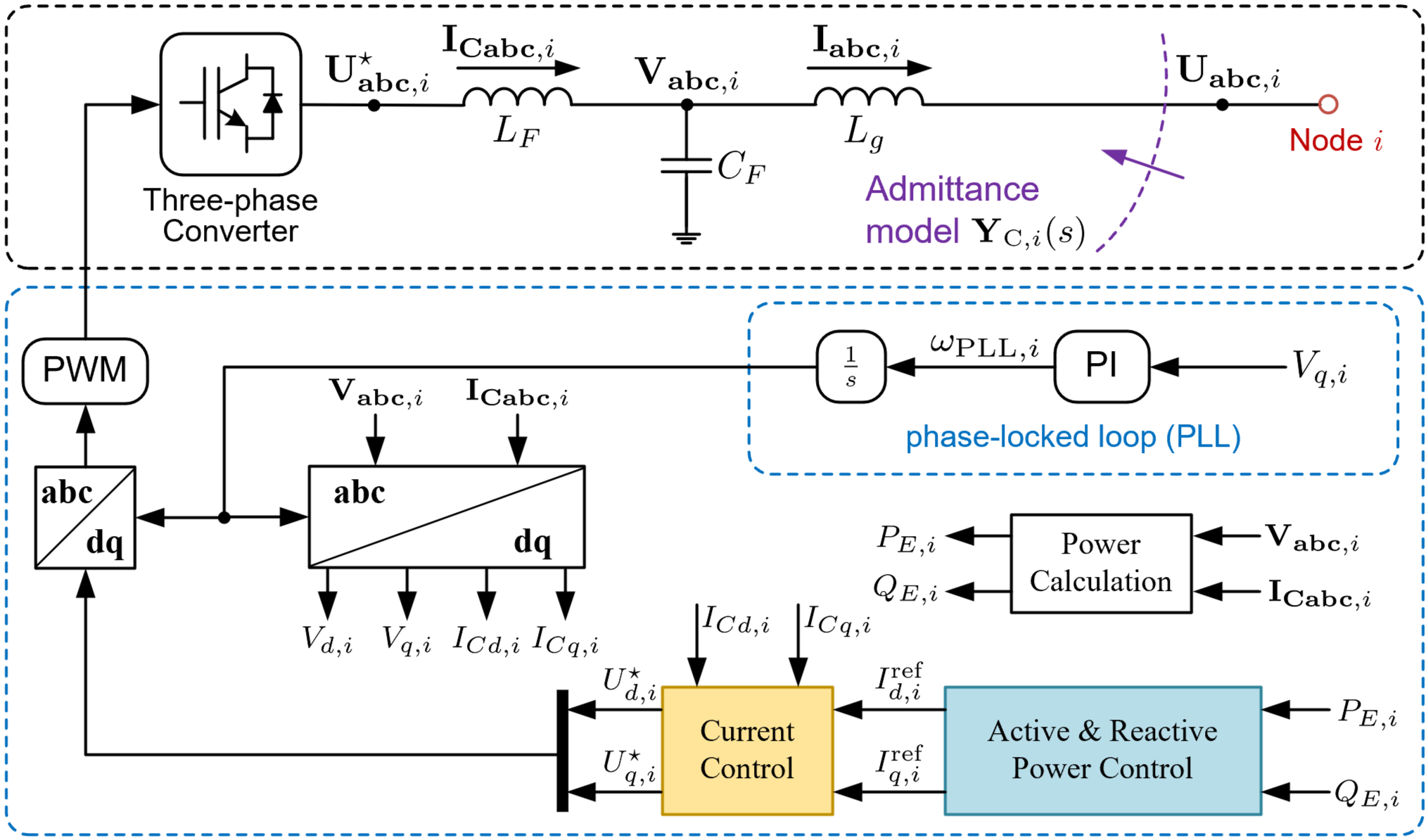}
\caption{A prototypical control diagram of GFL converters.} \la{Fig_GFL_converter}
\vspace{-3mm}
\end{figure}

Figure~\ref{Fig_GFL_converter} shows a GFL converter that is attached to Node~$i$ of the power network, where a PLL is used for synchronization and Park transformation, and the current and power control loops regulate output current and power based on feedback signals. It shows that the equivalent admittance ${\bf Y}_{{\rm C},i}(s)$ captures the dynamics of the converter seen from the terminal of the converter, where the internal dynamics of the control loops and LCL filters are hidden in the elements of ${\bf Y}_{{\rm C},i}(s)$. We refer to~\cite{xin2024many} for how ${\bf Y}_{{\rm C},i}(s)$ can be derived for GFL and GFM converters.

Based on~\eqref{eq:Y_grid} and~\eqref{eq:admittance_devices}, we obtain the closed-loop model of the multi-converter system, shown in Figure~\ref{Fig_closed_loop}, which enables decentralized stability analysis thanks to the block-diagonal structure of ${\bf Y}_{\rm C}^N(s)$.

\subsection{Loop Transformation for Decentralized Stability Analysis}
\label{subsec:power_system_decentralized_stability}

\begin{figure}[!t]
	\centering
	\includegraphics[width=2.9in]{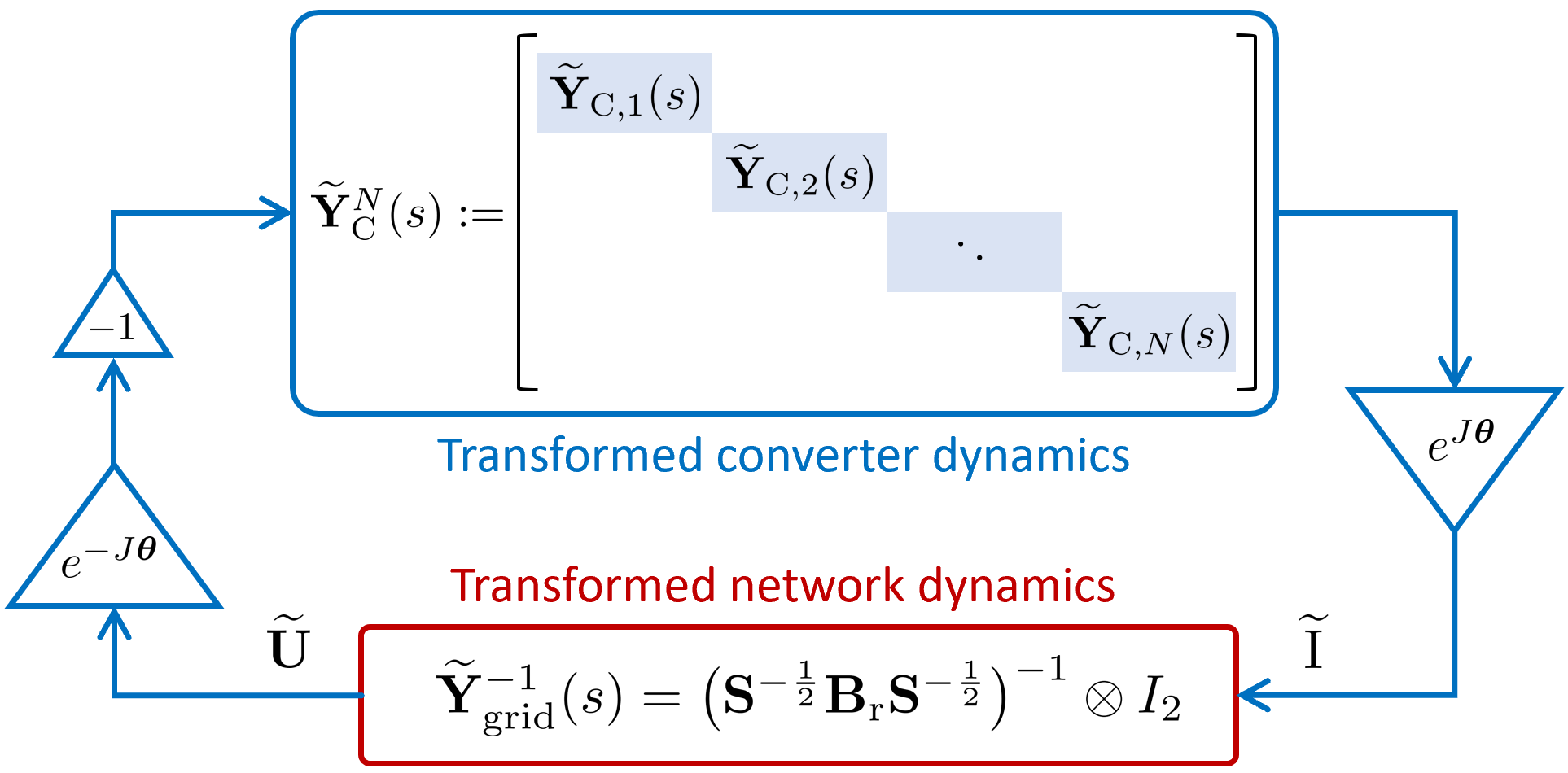}
	\vspace{0mm}
	\caption{Equivalent closed-loop model of the multi-converter system obtained through the loop transformation in~\eqref{eq:rescaled_sys} and~\eqref{eq:Y_matrices}.}
	\vspace{0mm}
	\label{Fig_eqclosed_loop}
\end{figure}

Without loss of generality, in what follows, we consider the power network dynamics in~\eqref{eq:Ygrid_RX} for non-parametric decentralized stability certificates. According to Figure~\ref{Fig_closed_loop}, the closed-loop dynamics of the multi-converter system can be described by $[({\bf S}\otimes I_2)e^{J \boldsymbol{\theta}} \; {\bf Y}_{\rm C}^N(s) \; e^{-J \boldsymbol{\theta}}]\; \# \; {\bf Y}^{-1}_{\rm grid}(s)$, where $\#$ denotes the feedback interconnection. By considering a multiplier similar to~\cite{huang2024gain}, this closed-loop system is equivalently represented as
\begin{equation}\label{eq:rescaled_sys}
        [e^{J \boldsymbol{\theta}} \; \widetilde{\bf Y}_{\rm C}^N(s) \; e^{-J \boldsymbol{\theta}} ]\; \# \; \widetilde{\bf Y}_{\rm grid}^{-1}(s),
\end{equation}
where we pushed the line dynamics in~\eqref{eq:F_sigma} to the converter subsystem, as shown in Figure~\ref{Fig_eqclosed_loop}:
\begin{equation}\label{eq:Y_matrices}
    \begin{cases}
    \begin{split}
        \widetilde{\bf Y}_{\rm grid}(s) =& ({\bf S}^{-\frac{1}{2}} \otimes I_2) {\bf Y}_{\rm grid}(s) ( {\bf S}^{-\frac{1}{2}} \otimes {{F}_\epsilon^{-1}(s)}) \\
        =& \,{\bf S}^{-\frac{1}{2}} {\bf B}_{\rm r} {\bf S}^{-\frac{1}{2}} \otimes I_2
    \end{split}\\
        \widetilde{\bf Y}_{\rm C}^N(s) = {\rm blkdiag}(\widetilde{\bf Y}_{{\rm C},1}(s),...,\widetilde{\bf Y}_{{\rm C},N}(s)) ,\\
        \widetilde{\bf Y}_{{\rm C},i}(s) = {\bf Y}_{{\rm C},i}(s){F}_\epsilon^{-1}(s),\; i \in \{1,..., N\}.
         
    \end{cases}
\end{equation}

It can then be observed that $\widetilde{\bf Y}_{\rm grid}^{-1}(s)$ is a constant and positive definite matrix, while $e^{J \boldsymbol{\theta}} \; \widetilde{\bf Y}_{\rm C}^N(s) \; e^{-J \boldsymbol{\theta}} $ is block-diagonal, with the $i$-th block being $e^{J\theta_i} \widetilde{\bf Y}_{{\rm C},i}(s) e^{-J\theta_i}$. Hence, the stability of the system in~\eqref{eq:rescaled_sys} can be analyzed using the geometric decentralized stability conditions in~\eqref{eq:pas}-\eqref{eq:decentralized_DW}. For instance, according to~\eqref{eq:decentralized_DW}, the system stability can be analyzed by checking if each converter's DW shell $DW(e^{J\theta_i} \widetilde{\bf Y}_{{\rm C},i}(j\omega) e^{-J\theta_i})$ intersects with the network's expanded DW shell trajectory $\widetilde{DW}(\widetilde{\bf Y}_{\rm grid}(j\omega))$. We further notice that $e^{J\theta_i}$ is a unitary matrix, so we have $DW(e^{J\theta_i} \widetilde{\bf Y}_{{\rm C},i}(j\omega) e^{-J\theta_i}) = DW(\widetilde{\bf Y}_{{\rm C},i}(j\omega))$ and analogously for other conditions, that is, we can ignore the steady-state coordinate transformation matrix and simply focus on 
\begin{equation}\label{eq:rescaled_sys_1}
        \widetilde{\bf Y}_{\rm C}^N(s) \; \# \; \widetilde{\bf Y}_{\rm grid}^{-1}(s).
\end{equation}

In short, the stability of the multi-converter power system in Figure~\ref{Fig_network}, whose closed-loop dynamics are equivalently described by~\eqref{eq:rescaled_sys_1}, can be analyzed using the graphical decentralized stability certificates in~\eqref{eq:pas}-\eqref{eq:decentralized_DW}, by considering $G(s) = \widetilde{\bf Y}_{\rm C}^N(s)$, $G_i(s) = \widetilde{\bf Y}_{{\rm C},i}(s)$, and $H^{-1} = \widetilde{\bf Y}_{\rm grid}(s) = {\bf S}^{-\frac{1}{2}} {\bf B}_{\rm r} {\bf S}^{-\frac{1}{2}} \otimes I_2\succ 0$. Here, the decentralized stability analysis is performed by comparing each $G_i(s)$ with $H^{-1}$ through DW shell, gain, phase, etc., where $H^{-1}$ reflects the network characteristics and $G_i(s)$ reflects the $i$-th converter's characteristics. In other words, if every converter satisfies any of the conditions in ~\eqref{eq:pas}-\eqref{eq:decentralized_DW}, then the system is guaranteed to be stable. This also motivates a design principle for the converters to enforce the overall system stability: given a power network $H^{-1}$, the controller of a converter should be designed such that $G_i(s)$ meets any of the conditions in ~\eqref{eq:pas}-\eqref{eq:decentralized_DW}, for instance, the decentralized DW shell separation condition.

We note that by applying the decentralized stability certificates to~\eqref{eq:rescaled_sys_1}, each converter is treated independently, and one may not be able to capture the interactions among different converters. For instance, it has been reported in~\cite{yang2020placing} that GFM converters can equivalently increase the power network strength and thus help stabilize GFL converters; however, such an effect could not be reflected by using~\eqref{eq:rescaled_sys_1} for decentralized stability analysis. To handle this problem, we can choose to partition the system in another way so that the GFM dynamics are fused into the network side, similar to~\cite{huang2024gain}. To be specific, instead of~\eqref{eq:rescaled_sys_1}, we focus on the following two subsystems
\begin{subequations}
\begin{empheq}[left={\empheqlbrace}]{align}
& {\rm subsystem~1:} \begin{bmatrix} 0 & \\ & \widetilde{\bf Y}_{\rm C}^{\rm GFM}(s) \end{bmatrix} \; \# \; \widetilde{\bf Y}^{-1}_{\rm grid}(s), \label{eq:subsys1} \\
& {\rm subsystem~2:} \; \widetilde{\bf Y}_{{\rm C}}^{\rm GFL}(s) \; \# \; \widetilde{\bf Y}^{-1}_{\rm gridC}(s), \label{eq:subsys2}
\end{empheq}
\end{subequations}
where $\widetilde{\bf Y}_{\rm C}^{\rm GFL}(s)$ is a $2N_L\times 2N_L$ block-diagonal matrix representing the dynamics of the $N_L$ GFL converters, $\widetilde{\bf Y}_{\rm C}^{\rm GFM}(s)$ is a block-diagonal matrix representing the dynamics of the $N_M$ GFM converters ($N_L+N_M = N$), and $\widetilde{\bf Y}_{\rm C}^N(s)= \begin{bmatrix} \widetilde{\bf Y}_{\rm C}^{\rm GFL}(s) & \\ & \widetilde{\bf Y}_{\rm C}^{\rm GFM}(s) \end{bmatrix}$. In subsystem~2, the GFL converters are interacting with an \textit{equivalent network} that includes the GFM dynamics as
\begin{equation}\label{eq:equil_network}
\widetilde{\bf Y}_{\rm gridC}(s) = \widetilde{\bf Y}_{\rm g}^1(s) - \widetilde{\bf Y}_{\rm g}^2(s) [\widetilde{\bf Y}_{\rm C}^{\rm GFM}(s) + \widetilde{\bf Y}_{\rm g}^4(s)]^{-1} \widetilde{\bf Y}_{\rm g}^3(s),
\end{equation}
where
\begin{equation}\label{eq:grid_partition}
\widetilde{\bf Y}_{\rm grid}(s) =:
    \begin{bmatrix}
    \widetilde{\bf Y}_{\rm g}^1(s) & \widetilde{\bf Y}_{\rm g}^2(s) \vspace{1mm} \\ 
    \widetilde{\bf Y}_{\rm g}^3(s) & \widetilde{\bf Y}_{\rm g}^4(s) \\
\end{bmatrix}.
\end{equation}
It was shown in~\cite{huang2024gain} that if both subsystems 1 and 2 are stable, then the original system in~\eqref{eq:rescaled_sys_1} is stable. Notice that subsystem~1 captures the GFM dynamics in the system, while subsystem~2 describes the interaction between the GFL converters and an equivalent network which contains the original network and the GFM dynamics. It was shown in~\cite{huang2024gain} and~\cite{luo2026impacts} that it is less conservative to use~\eqref{eq:subsys1} and~\eqref{eq:subsys2} for decentralized stability analysis than~\eqref{eq:rescaled_sys_1}, since~\eqref{eq:subsys2} and~\eqref{eq:equil_network} together describe how the GFM converters improve the network characteristics and stabilize the GFL converters, as will be shown below. In ``\nameref{sb-GFMindex},'' we further discuss how to simplify the computation in~\eqref{eq:equil_network} by focusing on the passivity index of GFM converters.

\begin{figure}[!]%[b]
\centering
% \vskip-.25cm
\includegraphics[width=.48\textwidth]{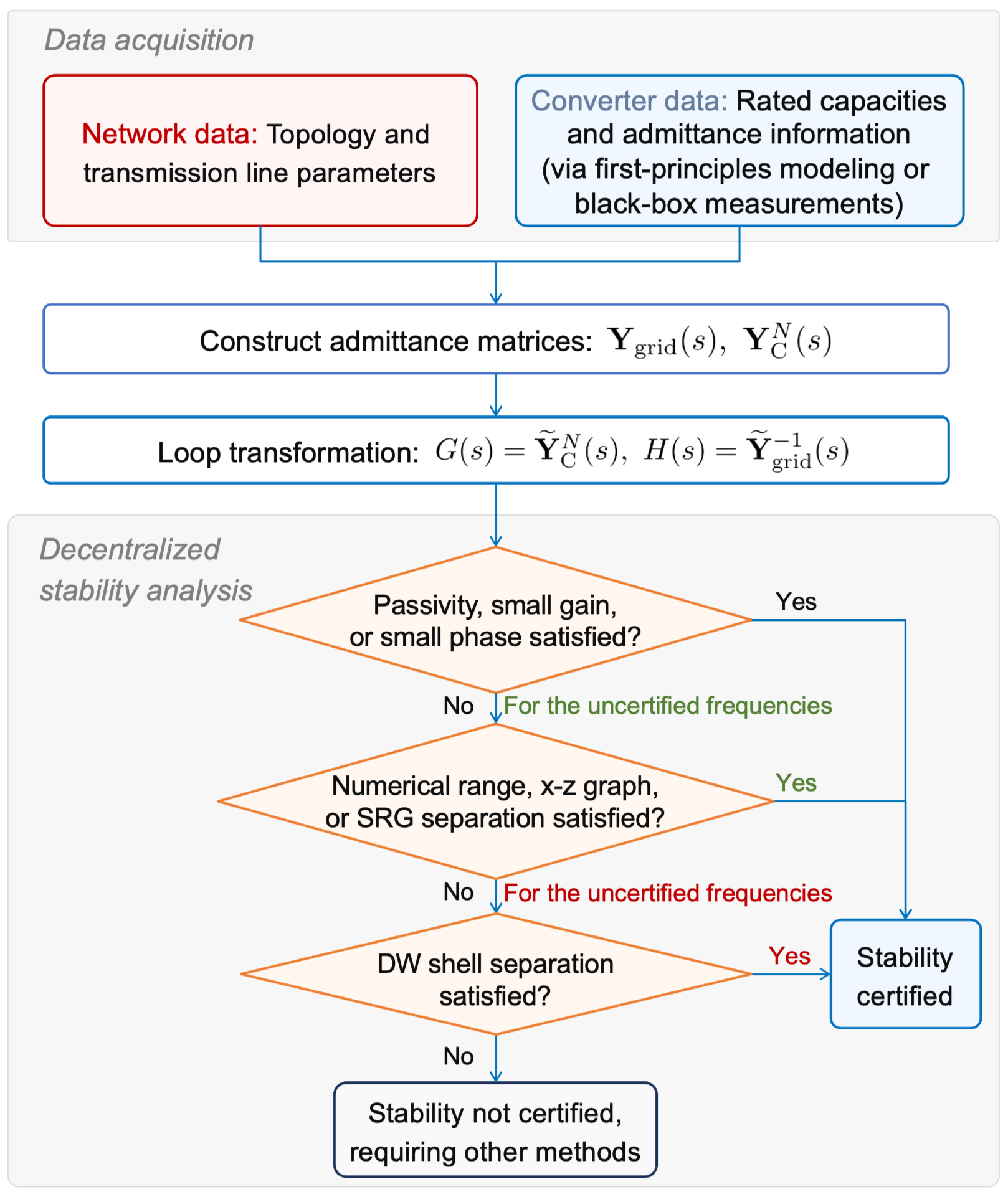}
\caption{A flowchart for the decentralized stability analysis process.} \la{Fig_flow}
\vspace{-3mm}
\end{figure}

\begin{sidebar}{}
\setcounter{sequation}{0}
\renewcommand{\thesequation}{S\arabic{sequation}}
\setcounter{stable}{0}
\renewcommand{\thestable}{S\arabic{stable}}
\setcounter{sfigure}{0}
\renewcommand{\thesfigure}{S\arabic{sfigure}}
\section[Passivity Index of GFM Converters]{Passivity Index of GFM Converters}
\label{sb-GFMindex}

The subsystem~2 in~\eqref{eq:subsys2} describes the interaction between the GFL converters and an equivalent network $\widetilde{\bf Y}_{\rm grid}(s)$, where the dynamics of GFM converters are fused into the equivalent network via~\eqref{eq:equil_network}. Notice that Eq.~\eqref{eq:equil_network} involves a Schur complement, which obtains the admittance matrix of the equivalent network $\widetilde{\bf Y}_{\rm gridC}(s)$ according to the admittance matrix of the original network $\widetilde{\bf Y}_{\rm grid}(s)$ and the admittance matrix of the GFM converters $\widetilde{\bf Y}_{\rm C}^{\rm GFM}(s)$. However, the expression~\eqref{eq:equil_network} is convoluted, and it is not straightforward to see how the GFM converters affect the equivalent network. To this end, Ref.~\cite{luo2026impacts} defines the passivity index of the $i$-th GFM converter at frequency $\omega$ as
\begin{equation}
    \nu_i(j\omega) = \lambda_{\min} (\mathcal H(\widetilde{\bf Y}_{{\rm C},i}^{\rm GFM}(j\omega))),
\end{equation}
where $\widetilde{\bf Y}_{{\rm C},i}^{\rm GFM}(j\omega)$ is the $i$-th block of $\widetilde{\bf Y}_{\rm C}^{\rm GFM}(s)$ and denotes the admittance matrix of the $i$-th GFM converter, $\mathcal{H}(M) = \frac{1}{2}(M+M{\hermconj})$ is the Hermitian part of matrix $M$. 

It was shown in~\cite{luo2026impacts} that the $x$-$z$ graph of the equivalent network is contained in an envelope
\begin{equation}
    \mathcal{E}_\alpha(j\omega) = \{(x,z) \, | \, x \le -\alpha(j\omega),\,z \ge x^2\}.
\end{equation}
The bound $\alpha$ above is obtained from the passivity indices of the GFM converters as
\begin{equation}\label{eq:GFM_alpha}
    \alpha(j\omega) := \lambda_{\min}[ Q_A - Q_B \left( Q_D+\Gamma(j\omega) \right)^{-1} Q_C ],
\end{equation}
where $\Gamma(j\omega)={\rm diag}\left(\nu_1(j\omega),\ldots,\nu_{N_M}(j\omega)\right)$, and we still consider the original network admittance matrix $\widetilde{\bf Y}_{\rm grid}(s)$ to be a constant matrix as in~\eqref{eq:Y_matrices} so that $\widetilde{\bf Y}_{\rm g}^1(s) = Q_A \otimes I_2$, $\widetilde{\bf Y}_{\rm g}^2(s) = Q_B \otimes I_2$, $\widetilde{\bf Y}_{\rm g}^3(s) = Q_C \otimes I_2$, and $\widetilde{\bf Y}_{\rm g}^4(s) = Q_D \otimes I_2$. 

Eq.~\eqref{eq:GFM_alpha} establishes a connection between the right bound of the $x$-$z$ graph envelope (i.e., $-\alpha(j\omega)$, as shown in Figure~\ref{Fig_GFM_index}) and the passivity indices of the GFM converters. Moreover, it implies that if $\nu_i(j\omega) \ge 0$ for all the GFM converters, then $\alpha(j\omega)$ is larger than $\lambda_{\min}(\widetilde{\bf Y}_{\rm grid}(s)) = \lambda_{\min}({\bf S}^{-\frac{1}{2}} {\bf B}_{\rm r} {\bf S}^{-\frac{1}{2}})$, which is known as the generalized short-circuit ratio (gSCR) of the original network that describes its strength~\cite{dong2018small, huang2024gain}. Similarly, the bound $\alpha(j\omega)$ serves as the strength of the equivalent network. Hence, an interpretation is that if $\nu_i(j\omega) \ge 0$ holds, the GFM converters equivalently increase the power network strength and stabilize the GFL converters. 

Figure~\ref{Fig_GFM_index} provides an example to illustrate this point, where the red region is the $x$-$z$ graph of the original network and its right bound is $-{\rm gSCR}$. It can be seen that the $x$-$z$ graph of the original network intersects with the $x$-$z$ graphs of a GFL converters, implying potential instability risk. However, when fusing the GFM dynamics into the network via~\eqref{eq:equil_network}, the $x$-$z$ graph envelope of the equivalent network is separated from the $x$-$z$ graphs of the GFL converter because $\alpha(j\omega) > {\rm gSCR}$, implying that the system is stable. One can conclude that a larger $\alpha(j\omega)$ is beneficial for the system stability from the view of $x$-$z$ graph separation, which is aligned with the interpretation that the network should be strong enough to support the operation of GFL converters. It can be deduced from~\eqref{eq:GFM_alpha} that a larger passivity index of any GFM converter leads to a larger $\alpha(j\omega)$, which further improves the system stability. Such an observation enables an explicit control design guideline for GFM converters, that is, the controller should be designed to increase the passivity index of a GFM converter as far as possible. We note that there are also other indices, such as the forming index in~\cite{SR_1}, to quantify the characteristics of GFM converters and provide guidance on their behavior design.

\sdbarfig{%
\centering
\includegraphics[width=.48\textwidth]{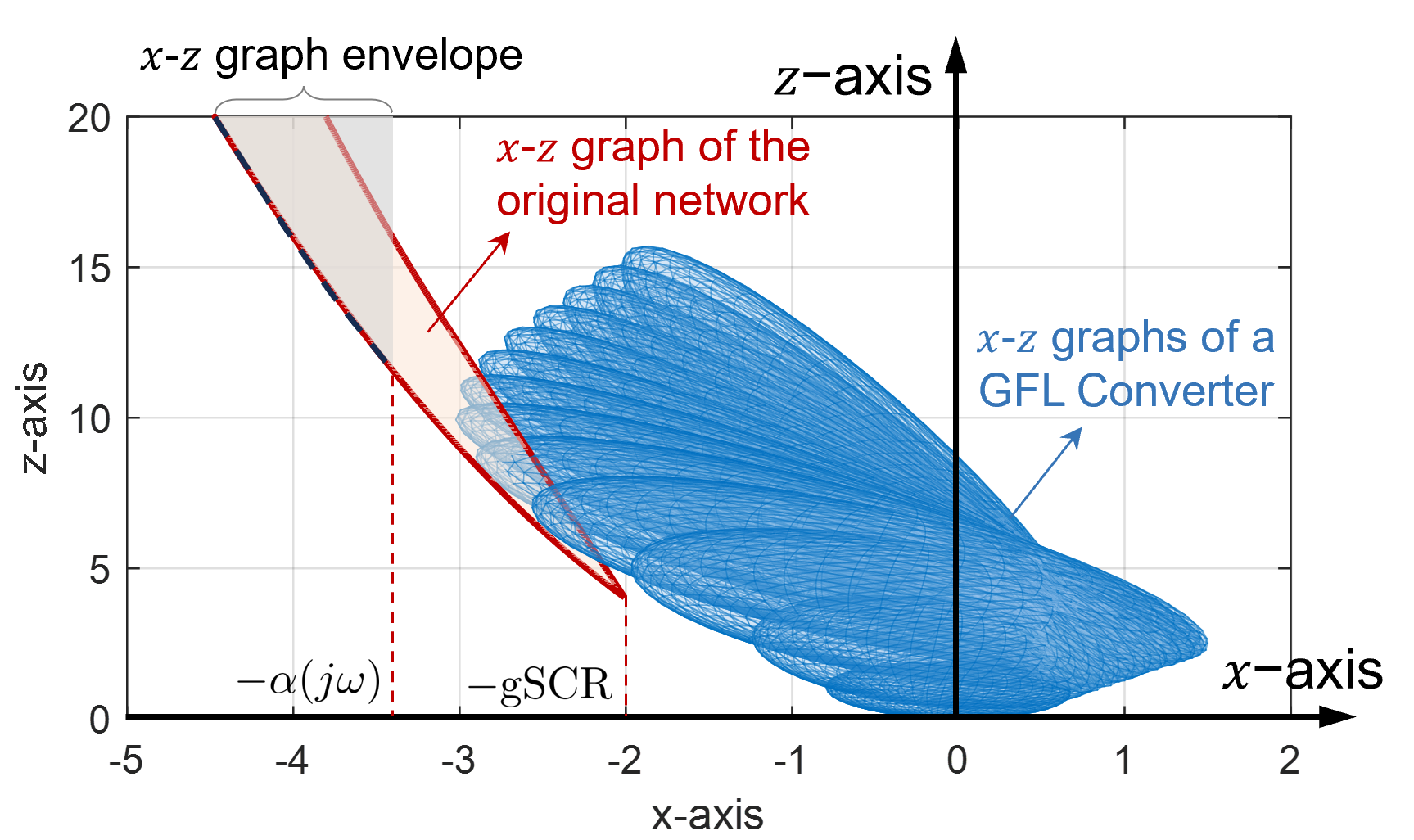}
}{Illustration of the $x$-$z$ graph envelope of the equivalent network. \label{Fig_GFM_index}}

% \begin{thebibliography}{SR}
% \bibitem[S1]{SR_1}
% J. Bao, P. L. Lee, ``Process Control: The Passive Systems Approach,'' \textit{London, U.K., Springer},
% 2007.
% \end{thebibliography}
\end{sidebar}

% We provide illustrative examples below. 

\subsection{Case Studies}
\label{subsec:case_studies}

In what follows, we provide two illustrative examples to show how the decentralized stability certificates enable modular and scalable analysis of PE-dominated power systems. For better illustration, Figure~\ref{Fig_flow} shows a flowchart that summarizes the decentralized stability analysis process of PE-dominated power systems based on the certificates in~\eqref{eq:pas}-\eqref{eq:decentralized_DW}.

\vspace{5.45mm}
\noindent \textit{\textbf{a) A 9-Bus Power System That Contains 3 GFL Converters}}

\begin{figure}[!t]
    \centering
    \includegraphics[width=0.9\linewidth]{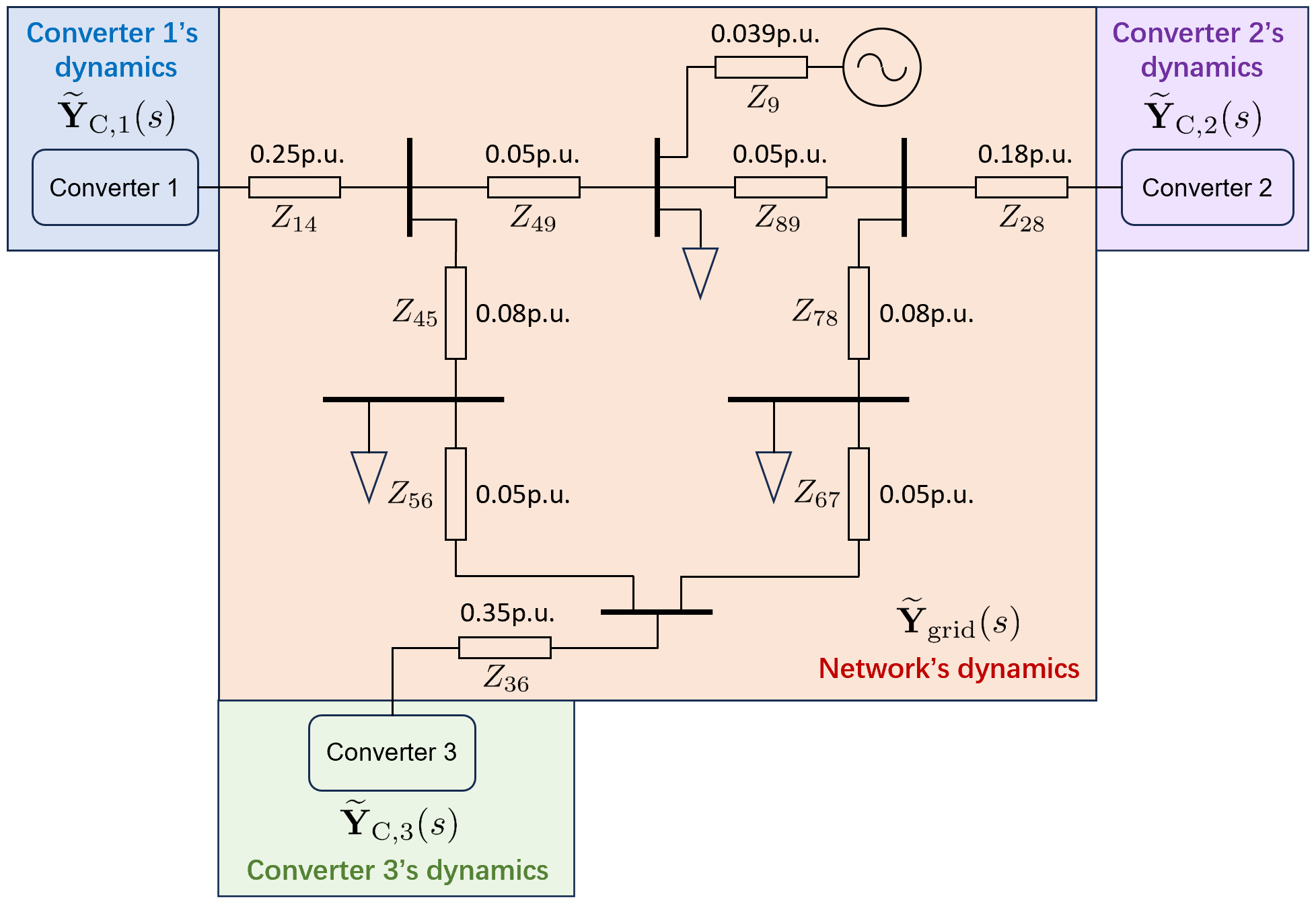}
    \caption{Topology of the three-converter 9-bus system.}
    \label{fig:9bus_topology}
\end{figure}

We consider a three-converter 9-bus system shown in Figure~\ref{fig:9bus_topology}, where all the converters operate in GFL mode. The per-unit line inductance parameters are indicated in Figure~\ref{fig:9bus_topology}, and all transmission lines have an identical \(R/X\) ratio of \(\epsilon=0.1\). The PLL bandwidths of the three converters are 30~rad/s, 20~rad/s, and 96.2~rad/s, respectively. According to~\eqref{eq:Y_matrices}, the converters' dynamics can be represented by the block-diagonal matrix $\widetilde{\bf Y}_{{\rm C}}^N(s)={\rm blkdiag}(\widetilde{\bf Y}_{{\rm C},1}(s),\widetilde{\bf Y}_{{\rm C},2}(s),\widetilde{\bf Y}_{{\rm C},3}(s))$, and the network dynamics are given by $\widetilde{\bf Y}_{\rm grid}(s)={\bf S}^{-\frac{1}{2}}{\bf B}_{\rm r}{\bf S}^{-\frac{1}{2}}\otimes I_2$, which is a positive-definite constant matrix. Following~\eqref{eq:rescaled_sys_1}, we let $G(s)=\widetilde{\bf Y}_{{\rm C}}^N(s)$ and $H(s)=\widetilde{\bf Y}_{\rm grid}^{-1}(s)$ and analyze the system stability using the geometric decentralized stability certificates.

Similar to~\cite{huang2024gain}, we refer to the network-dependent phase range $[-\pi-\underline{\phi}(\widetilde{\bf Y}_{\rm grid}(j\omega)),\,\pi-\overline{\phi}(\widetilde{\bf Y}_{\rm grid}(j\omega))]$ as the phase area of the network, and the phase area of the $i$-th converter is $[\underline{\phi}(\widetilde{\bf Y}_{{\rm C},i}(j\omega)),\,\overline{\phi}(\widetilde{\bf Y}_{{\rm C},i}(j\omega))]$. Since $\widetilde{\bf Y}_{\rm grid}(s)$ is positive definite, the phase area of the network reduces to $[-\pi,\pi]$. Hence, if the converters' phase areas are contained in the network's phase area, then the phase condition~\eqref{eq:DecPhase}, or the one in~\eqref{eq:DecPhase_g} for a more general case, is satisfied. The gain curves of the $i$-th converter and the network are respectively defined by $\overline{\sigma}(\widetilde{\bf Y}_{{\rm C,}i}(j\omega))$ and $\underline{\sigma}(\widetilde{\bf Y}_{\rm grid}(j\omega))$. If the converter's gain is smaller than the network's gain, then the small gain condition holds.

\begin{figure}[!t]
    \centering
    \includegraphics[width=0.85\linewidth]{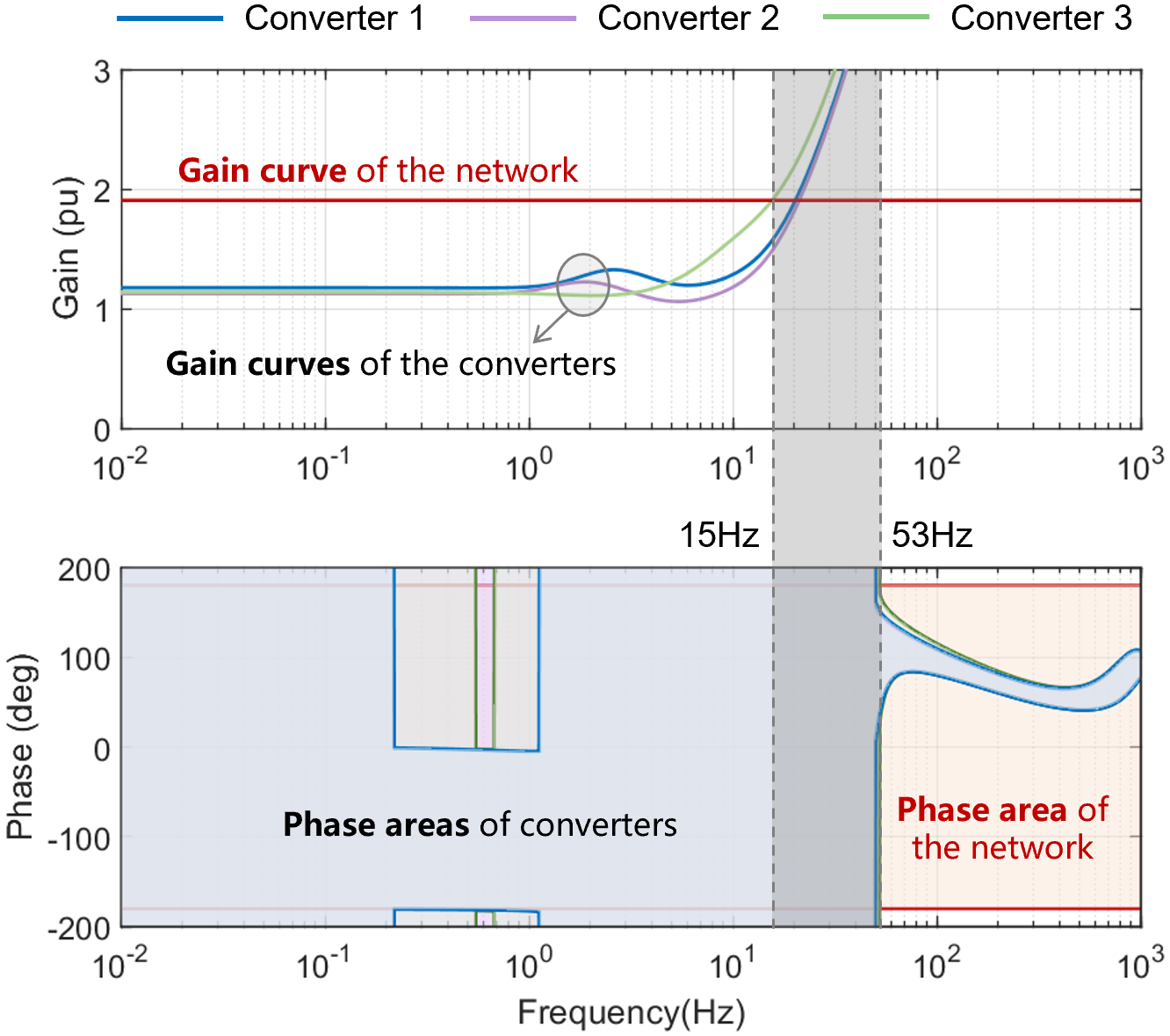}
\caption{Gain and phase of the three-converter system.}
    \label{fig:9bus_gain_phase}
\end{figure}

Figure~\ref{fig:9bus_gain_phase} plots the gain curves and phase areas of the three converters and the network, which are determined by $\widetilde{\bf Y}_{{\rm C},i}(j\omega)$ and $\widetilde{\bf Y}_{\rm grid}(j\omega)$, respectively. Below 15~Hz, the gains of all the converters are smaller than that of the network, satisfying the decentralized gain condition~\eqref{eq:DecGain}. We notice that the gain of the network is in fact $\underline{\sigma}(\widetilde{\bf Y}_{\rm grid}(j\omega))=\underline{\sigma}({\bf S}^{-\frac{1}{2}}{\bf B}_{\rm r}{\bf S}^{-\frac{1}{2}})={\rm gSCR}$, where gSCR denotes the so-called generalized short-circuit ratio in power systems~\cite{dong2018small,gao2026analysis}. Above 53~Hz, the phase areas of all the converters (the largest phase difference is less than $180^\circ$) are contained in the phase area of the network (i.e., $[-\pi,\pi]$), satisfying the decentralized phase condition~\eqref{eq:DecPhase}. Between 15~Hz and 53~Hz, neither the gain condition nor the phase condition holds, and the system stability can be further analyzed by, for instance, checking the decentralized $x$-$z$ graph separation condition~\eqref{eq:decentralized_P}, or equivalently, the decentralized SRG separation condition~\eqref{eq:decentralized_SRG}.

\begin{figure}[!t]
    \centering
    \includegraphics[width=1.0\linewidth]{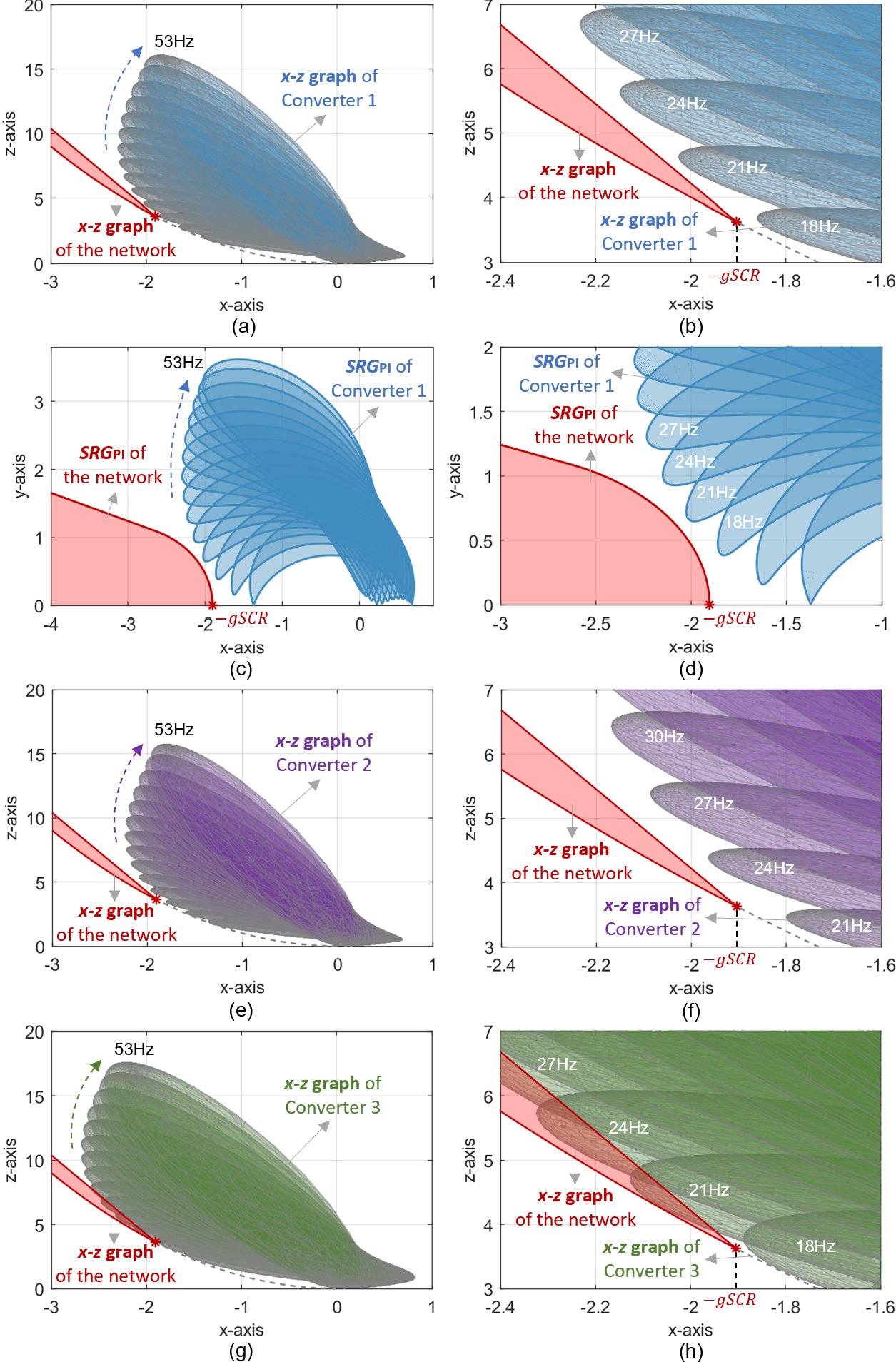}
\caption{Graphical stability analysis of the three-converter system.
Subplots (a), (e), and (g) show the $x$-$z$ graphs of Converters 1, 2,
and 3 together with the network, respectively, while subplot (c) shows the SRGs
of Converter 1 and the network. Subplots (b), (d), (f), and (h) are
the zoomed-in views of (a), (c), (e), and (g), respectively.}
    \label{fig:9bus_xz_srg}
\end{figure}

\begin{figure}[!t]
    \centering
    \includegraphics[width=1.0\linewidth]{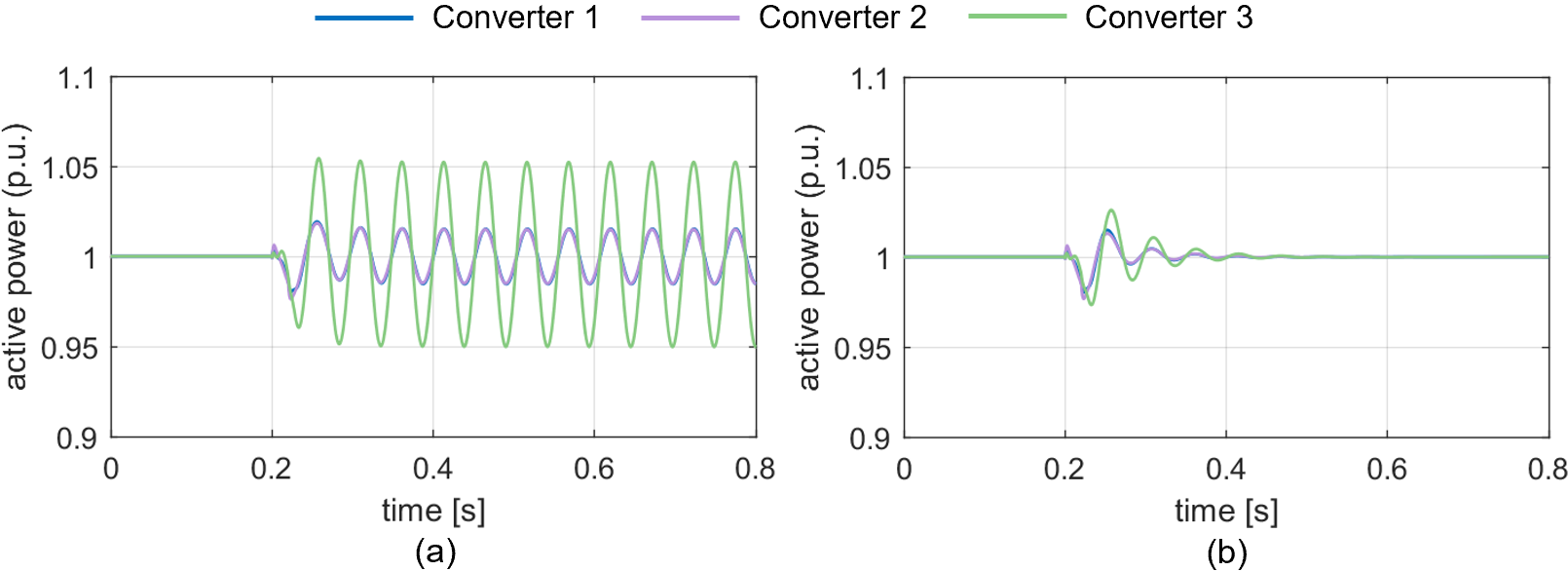}
\caption{Time-domain responses of the three-converter system under a small disturbance: (a) before and (b) after modifying the PLL bandwidth of Converter 3.}
    \label{fig:9bus_time_domain}
\end{figure}

Figure~\ref{fig:9bus_xz_srg} plots the $x$-$z$ graphs of the three converters and the network from 15~Hz to 53~Hz, where the $x$-$z$ graph trajectory of the network over $\tau\in(0,1]$ constitutes the red region; its upper boundary is concave upward, and its lower boundary corresponds to a part of the parabola $z=x^2$ with $x\in(-\infty,-{\rm gSCR}]$. It can be observed from Figure~\ref{fig:9bus_xz_srg}~(a) and (b) that the $x$-$z$ graphs of Converter~1 are separated from the red region, i.e., the $x$-$z$ graph trajectory of the network. Figure~\ref{fig:9bus_xz_srg}~(c) and (d) further show the corresponding $\mathrm{SRG}_{+}$ plots of Converter~1 and the network, which exhibit the same separation due to the bijective relation in~\eqref{eq:bijective} and therefore satisfy the decentralized SRG separation condition~\eqref{eq:decentralized_SRG}. Similarly, as shown in Figure~\ref{fig:9bus_xz_srg}~(e) and (f), the $x$-$z$ graphs of Converter~2 are also separated from the red region. However, in Figure~\ref{fig:9bus_xz_srg}~(g) and (h), the $x$-$z$ graphs of Converter~3 intersect with the red region, indicating that the decentralized $x$-$z$ graph separation condition~\eqref{eq:decentralized_P} is violated. Hence, the system stability cannot be guaranteed in this frequency range, suggesting a potential instability risk induced by Converter~3. This apparent risk is confirmed by the sustained oscillations in Figure~\ref{fig:9bus_time_domain}~(a).

To improve the stability, the PLL bandwidth of Converter~3 is then reduced from 96.2~rad/s to 20~rad/s, while Converters~1 and~2 remain unchanged. The resulting $x$-$z$ graphs of Converter~3 coincide with those of Converter~2 shown in Figure~\ref{fig:9bus_xz_srg}~(e) and (f), and are therefore separated from the red region. According to the decentralized $x$-$z$ graph separation condition~\eqref{eq:decentralized_P}, together with the gain and phase conditions satisfied outside this frequency range, the modified three-converter system is stable. This result is confirmed by the damped time-domain responses in Figure~\ref{fig:9bus_time_domain}~(b).

We remark that the above decentralized stability analysis is based on comparing the network characteristics with each converter's characteristics. The network characteristics are assumed to be obtained a priori, for instance, by the system operators since they have the power network data and model, and then such information is used to examine whether the converters meet the certificates in a decentralized manner. Such a process avoids building the high-dimensional closed-loop model and explicitly show how each converter's characteristics should match those of the power network. Moreover, it poses conditions, determined by the power network, for the converters to satisfy locally so that the global closed-loop stability is guaranteed, which can be included in the future grid code.

\vspace{4mm}
\noindent \textit{\textbf{b) A 68-Bus Power System That Contains GFL and GFM Converters}}

\begin{figure}[!t]
\centering
\includegraphics[width=0.9\linewidth]{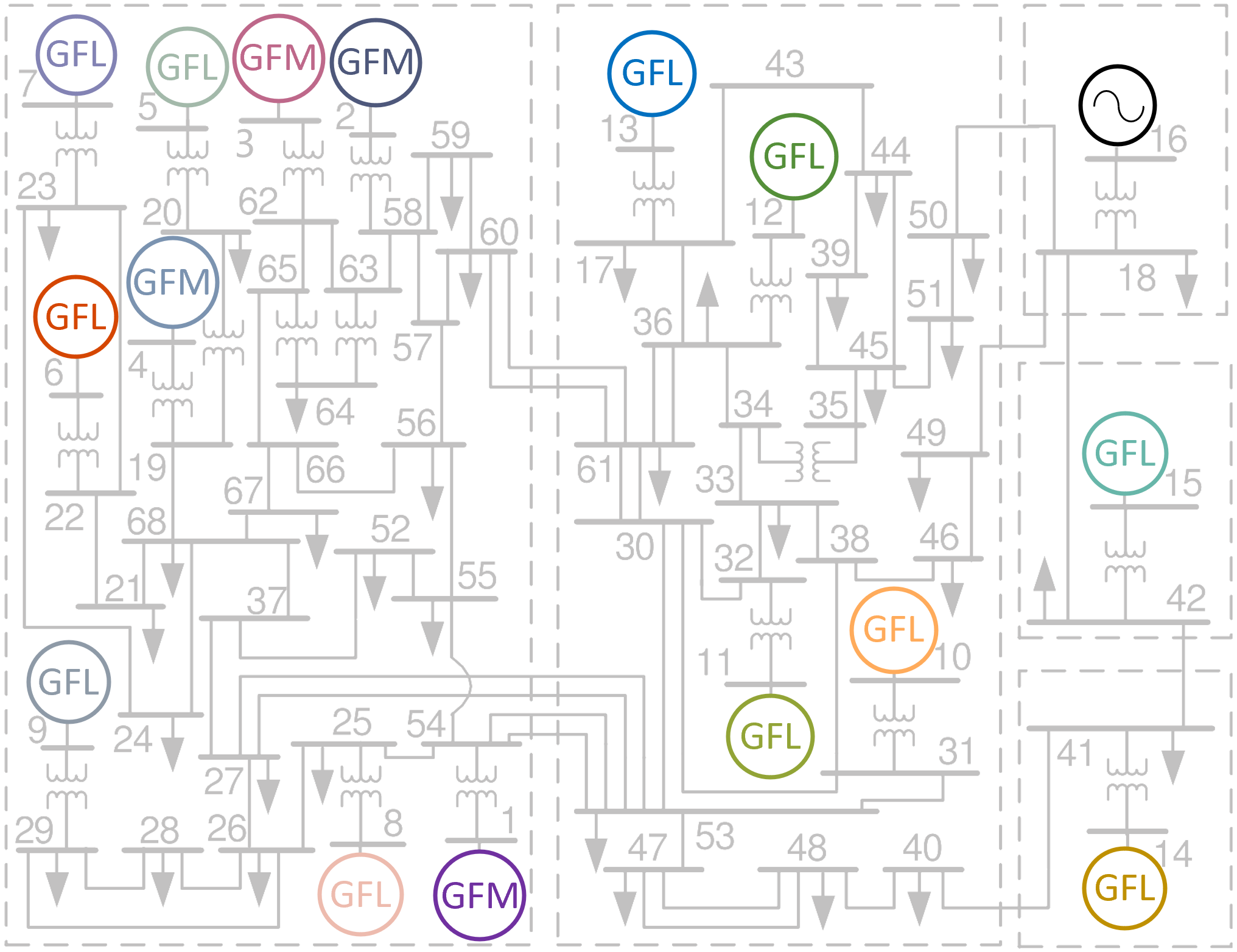}
\caption{Topology of the modified IEEE 68-bus system.}
\label{fig:68bus_topology}
\end{figure}

We further consider a modified IEEE 68-bus system as shown in Figure~\ref{fig:68bus_topology}, where four GFM converters are connected to Buses~1--4 (denoted by GFM converters~1--4), eleven GFL converters are connected to Buses~5--15 (denoted by GFL converters~5--15), and Bus~16 is modeled as an infinite bus to emulate a remote (weakly-connected) area. The transmission line parameters of the modified 68-bus system are adopted from~\cite{luo2026impacts}. We then apply the proposed approach to analyze this GFM-GFL hybrid system and demonstrate its applicability to power systems with diverse converter dynamics.

\begin{figure}[!t]
\centering
\includegraphics[width=1.0\linewidth]{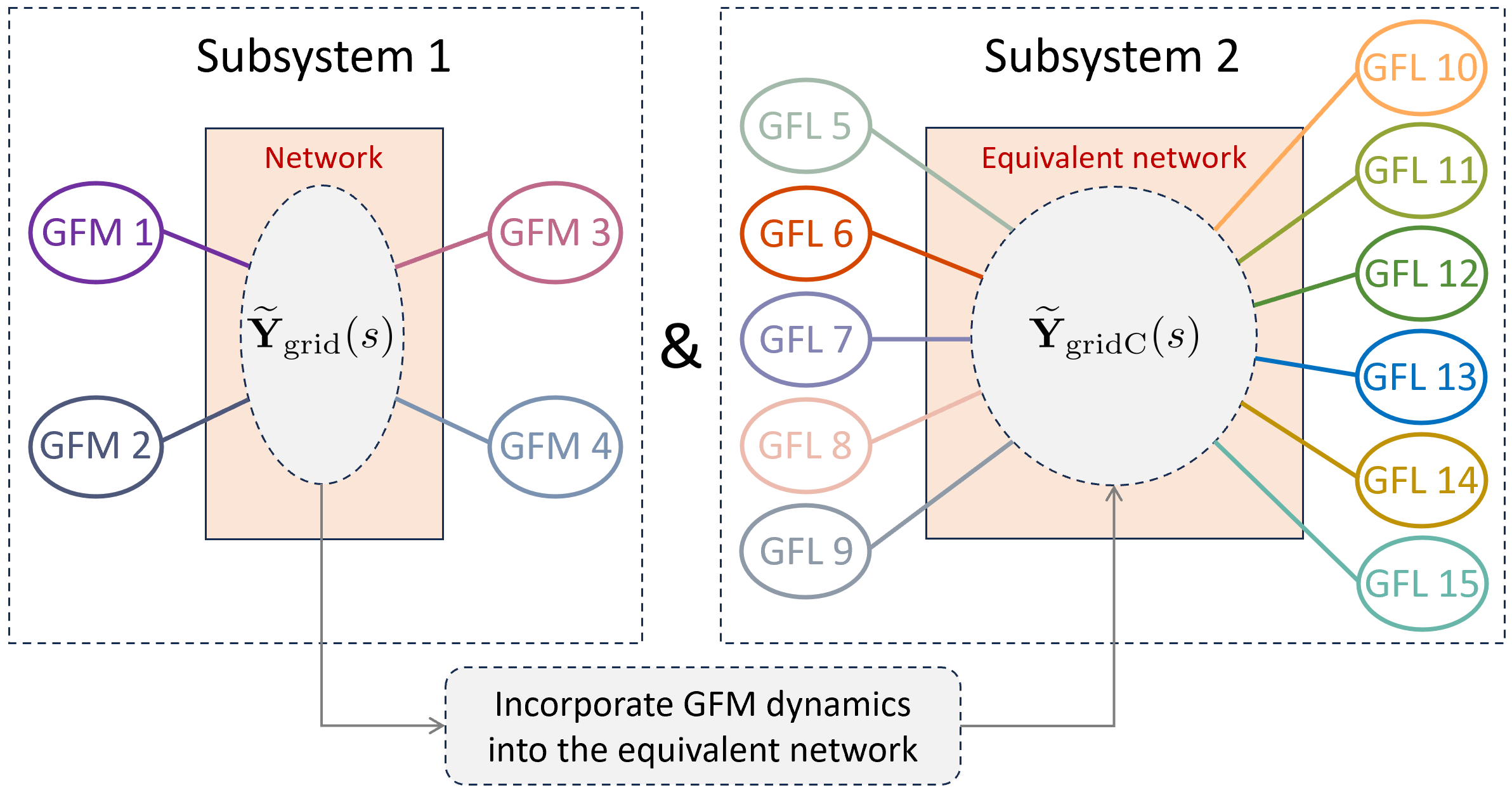}
\caption{Equivalent subsystems of the modified IEEE 68-bus system defined in~\eqref{eq:subsys1} and~\eqref{eq:subsys2}. The network in subsystem~1 and the equivalent network in subsystem~2 are related via~\eqref{eq:grid_partition} and~\eqref{eq:equil_network}.}
\label{fig:68bus_partition}
\end{figure}

Following the system partitioning introduced in~\eqref{eq:subsys1} and~\eqref{eq:subsys2}, the first step is to partition the system into two subsystems, as shown in Figure~\ref{fig:68bus_partition}. It can be seen that the four GFM converters are connected to the original network in subsystem~1, while the eleven GFL converters are connected to the equivalent network in subsystem~2. The equivalent network admittance \(\widetilde{\bf Y}_{\rm gridC}(s)\) is obtained from the partitioned original network and the GFM converter dynamics according to~\eqref{eq:grid_partition} and~\eqref{eq:equil_network}, which can be interpreted as incorporating the GFM dynamics into the network. According to Lemma~V.1 in~\cite{huang2024gain}, the original system is stable if both subsystems are stable.

\begin{figure}[!t]
\centering
\includegraphics[width=0.85\linewidth]{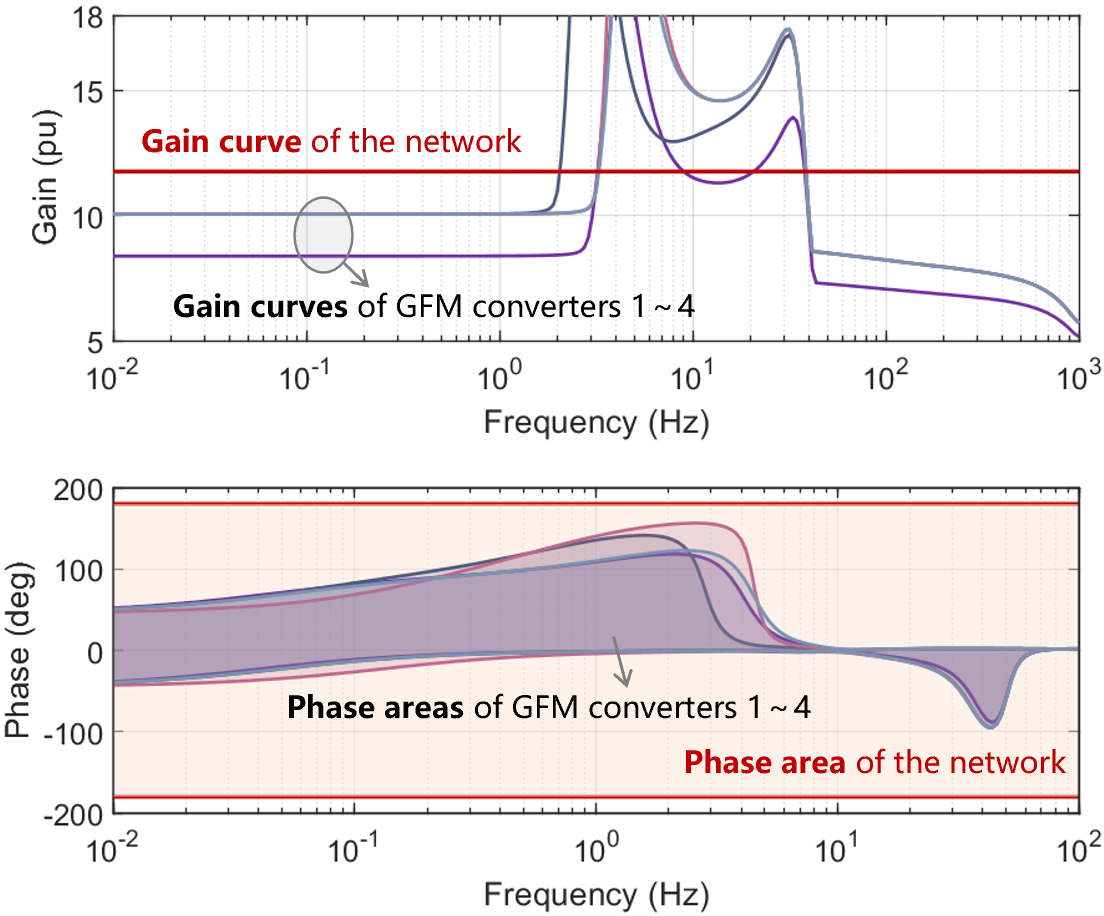}
\caption{Gain curves and phase areas of subsystem 1 of the modified 68-bus system, where the original network is focused.}
\label{fig:68bus_gfm_gain_phase}
\end{figure}

We first analyze the stability of subsystem~1. To consider heterogeneous GFM dynamics, the inertia, damping, and grid-side inductance triples $(J_i,D_i,L_{g,i})$ of GFM converters~1--4 are set to $(4,50,0.12)$, $(10,50,0.10)$, $(4,20,0.10)$, and $(4,50,0.10)$, respectively. Figure~\ref{fig:68bus_gfm_gain_phase} plots the gain curves and phase areas of the four GFM converters as well as the gain curve and phase area of the original network, where the GFL converter nodes are eliminated through Kron reduction as subsystem~1 does not contain GFL converters. It can be seen that the phase areas of all four GFM converters (the largest phase difference is less than $180^\circ$) are contained in the phase area of the network over the entire frequency range. Therefore, the decentralized phase condition~\eqref{eq:DecPhase} is satisfied, and subsystem~1 is stable.

\begin{figure}[!t]
\centering
\includegraphics[width=0.85\linewidth]{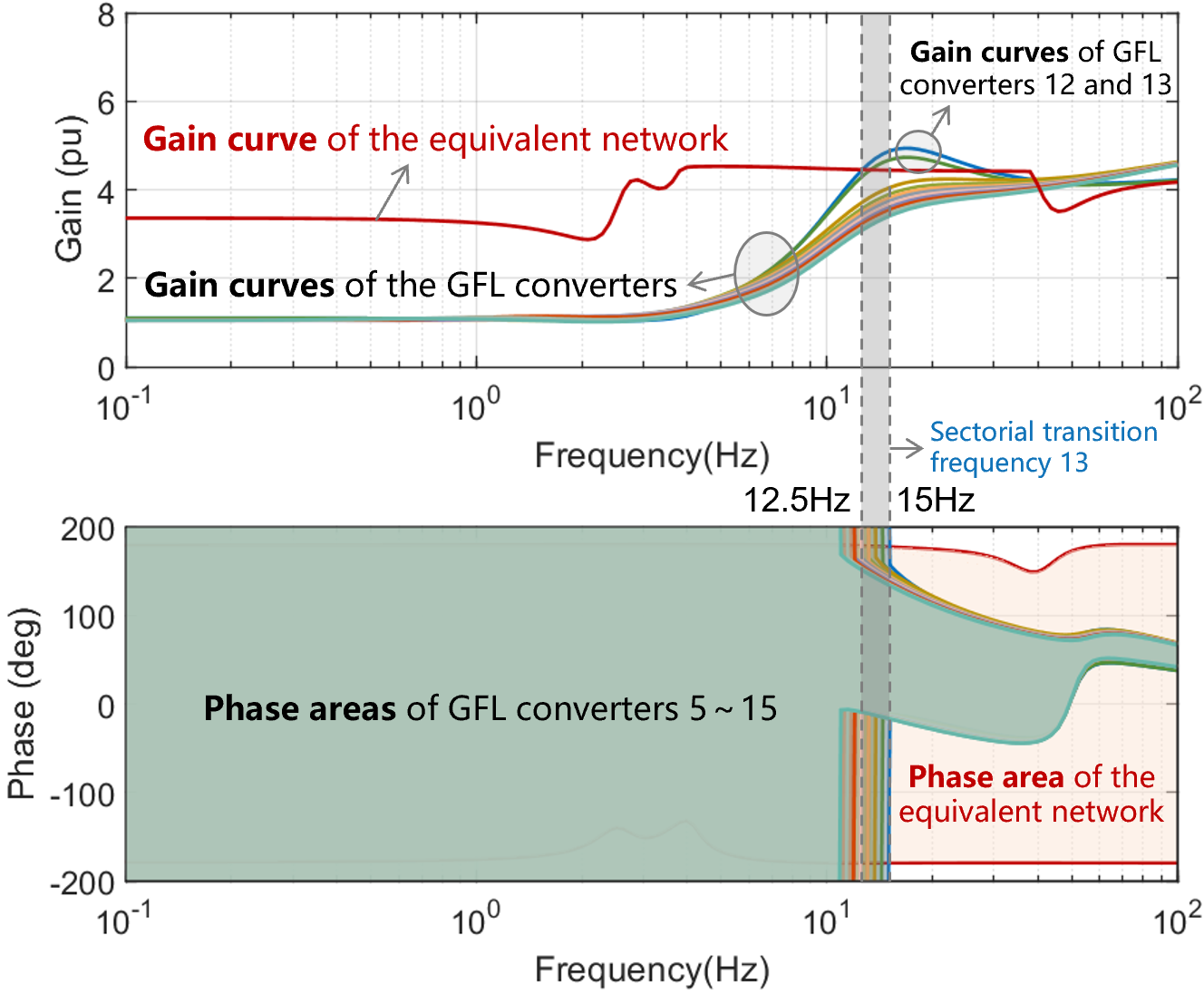}
\caption{Gain curves and phase areas of subsystem 2 of the modified 68-bus system, where the equivalent network is focused.}
\label{fig:68bus_gfl_gain_phase}
\end{figure}

We next analyze the stability of subsystem~2. The PLL bandwidths of GFL converters~5--15 are set to 10, 15, 20, 25, 30, 35, 40, 60, 70, 50, and 5~(rad/s), respectively. Figure~\ref{fig:68bus_gfl_gain_phase} plots the gain curves and phase areas of the eleven GFL converters as well as the gain curve and phase area of the equivalent network. The gain curve and phase area of the equivalent network vary with frequency because the GFM dynamics are incorporated into $\widetilde{\bf Y}_{\rm gridC}(s)$. Below 12.5~Hz, the gains of all GFL converters are smaller than that of the equivalent network, satisfying the decentralized gain condition~\eqref{eq:DecGain_g}. GFL converter~13 has the highest sectorial transition frequency of 15~Hz. Above this frequency, the phase areas of all GFL converters are contained in that of the equivalent network, and the union of their phase areas has a width of less than $180^\circ$, satisfying the decentralized phase condition~\eqref{eq:DecPhase_g}. Between 12.5~Hz and 15~Hz (the gray area in Figure~\ref{fig:68bus_gfl_gain_phase}), the decentralized phase condition is not satisfied, while only the gains of GFL converters~12 and~13 exceed that of the equivalent network. Hence, neither the decentralized gain condition nor the decentralized phase condition can certify the stability of subsystem~2 in this frequency range. We therefore examine the decentralized $x$-$z$ graph separation condition~\eqref{eq:decentralized_P_g} for these two converters.

\begin{figure}[!t]
\centering
\includegraphics[width=1.0\linewidth]{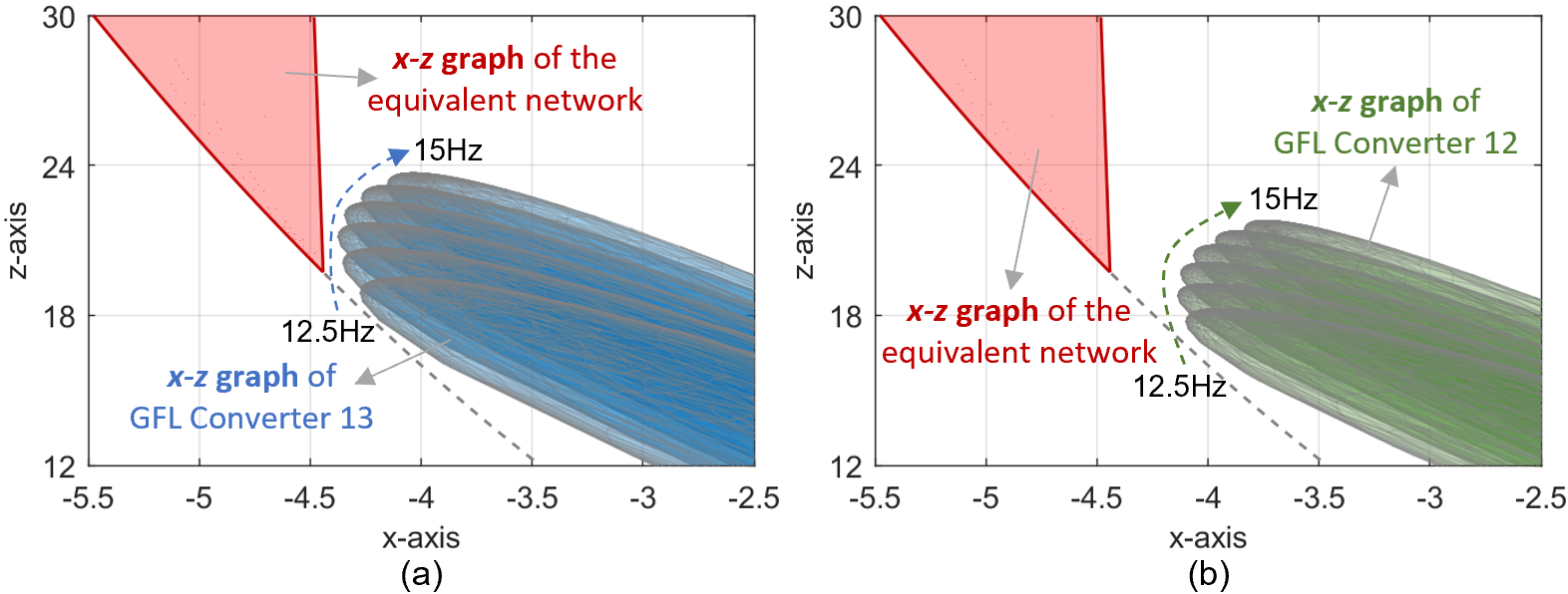}
\caption{$x$-$z$ graph analysis of the critical GFL converters and the equivalent network: (a) GFL Converter 13 and (b) GFL Converter 12.}
\label{fig:68bus_xz}
\end{figure}

Figure~\ref{fig:68bus_xz} plots the $x$-$z$ graphs of GFL converters~12 and~13 and the equivalent network between 12.5~Hz and 15~Hz. The red region represents the union of $\widetilde{P}_{xz}\bigl(-\tfrac{1}{\tau}\widetilde{\mathbf{Y}}_{\mathrm{gridC}}(j\omega)\bigr)$ over the frequency range of $[12.5~{\rm Hz}, 15~{\rm Hz}]$. It can be seen that the $x$-$z$ graphs of these two critical converters are separated from the red region. The remaining GFL converters satisfy the gain condition in~\eqref{eq:DecGain_g} in this frequency range, which also guarantees their $x$-$z$ graph separation from the equivalent network. Hence, the decentralized $x$-$z$ graph separation condition~\eqref{eq:decentralized_P_g} is satisfied between 12.5~Hz and 15~Hz. Together with the decentralized gain condition below 12.5~Hz and the decentralized phase condition above 15~Hz, this guarantees the stability of subsystem~2. Since subsystem~1 is also stable, the modified 68-bus system is stable.

\begin{figure}[!t]
\centering
\includegraphics[width=0.9\linewidth]{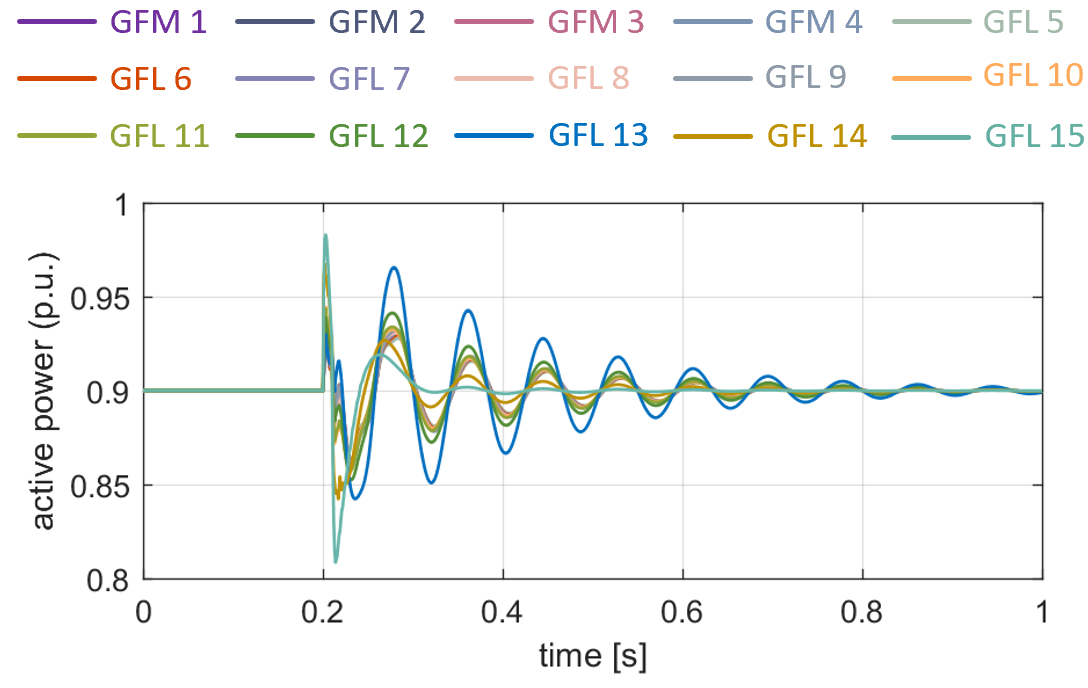}
\caption{Active-power responses of the modified 68-bus system.}
\label{fig:68bus_time_domain}
\end{figure}

Time-domain simulations are conducted to validate the above analysis, where a voltage dip of 0.1~pu occurs at the infinite bus (Bus~16) from 0.2~s to 0.21~s. As shown in Figure~\ref{fig:68bus_time_domain}, the active-power responses of all 15 converters converge to their steady-state values after the disturbance, consistent with the stability conclusion obtained from the proposed decentralized analysis.

% \newpage

% (Linbin's part)

% will present the modeling of PE-dominated power systems, both regarding the impedance modeling and the PQ modeling so that Verena can present the parametric results.

% will also present the non-parametric conditions based on DW shell and so on.

% \newpage

\section{Parametric Stability Certificates for Power Converters}
\label{sec:para}
The non-parametric stability certificates presented in the previous section are particularly useful for stability analysis when the complete dynamic models of the devices and the network are available. These models may be obtained either from first-principles modeling or from data-driven impedance measurements \cite{haberle2023mimo,abdalmoaty2025ultrafast}. However, while well suited for stability assessment, non-parametric certificates provide limited insight into the influence of individual control parameters and are less convenient for controller tuning. This motivates the development of parametric stability conditions, as discussed in the following.

In this section, we present examples of such conditions based on specific, albeit simplified, parameterizations of the network and device dynamics. These parameterizations allow us to derive explicit analytical decentralized stability certificates in terms of individual device parameters. The resulting conditions provide practical local tuning rules for converter control design and may also guide the development of future grid-code stability requirements.

%%%%%%%%%% check again %%%%%%%%%%%%%%%%%%%%%%
\subsection{Decentralized Small-Signal Stability Conditions}

To begin with, we notice that the closed-loop model in Figure~\ref{Fig_closed_loop} is not unique, as one can choose other input and output signals to derive the transfer function matrices. For instance, the voltage signals are chosen as the inputs, and the current signals are chosen as the outputs in~\eqref{eq:admittance_model}, which are popular choices in the modeling of GFL converters. To facilitate GFM converter modeling and align with grid-code requirements for active and reactive power support, one may alternatively choose the active and reactive power signals as inputs and the frequency and voltage magnitude as outputs.

\subsubsection{Small-Signal Modeling in Power Coordinates}

In power and voltage polar coordinates, the converter's dominant input--output dynamics are described by the filtered droop dynamics as follows,
\begin{equation}\label{eq:GFM_model}
    \begin{bmatrix} \Delta \omega_{i} \\ \Delta  |v|_{i} \end{bmatrix} = -
    \underbrace{\begin{bmatrix} \frac{d_{\mathrm{p},i}}{\tau_{\mathrm{p},i}s+1} & 0  \\ 0 & \frac{d_{\mathrm{q},i}}{\tau_{\mathrm{q},i}s+1} \end{bmatrix}}_{D_i(s)}
    \begin{bmatrix} \Delta P_i \\ \Delta Q_i \end{bmatrix} \,,
\end{equation}
where $\Delta P_i$ and $\Delta Q_i$ are respectively the perturbations of the active and reactive power of the $i$-th converter; $\Delta \omega_{i}$ and $\Delta |v|_{i}$ are respectively the perturbed frequency and voltage magnitude; {$d_{\mathrm{p},i}\in\mathbb{R}_{>0}$ and $d_{\mathrm{q},i}\in\mathbb{R}_{>0}$ are the active and reactive power droop gains, and $\tau_{\mathrm{p},i}\in\mathbb{R}_{\geq0}$ and $\tau_{\mathrm{q},i}\in\mathbb{R}_{\geq0}$ are the active and reactive power low-pass filter time constants.}

In fact, the input and output signals in~\eqref{eq:GFM_model} can be computed from the voltage and current signals in~\eqref{eq:admittance_model}. {More specifically, the time-domain bus voltages are modeled in polar coordinates as $U_{\mathrm{d},i}(t) \coloneqq |v|_i(t)\cos{\delta_i(t)}$ and $U_{\mathrm{q},i}(t) \coloneqq |v|_i(t)\sin{\delta_i(t)} $ with magnitude $|v|_i(t)$ and relative angle $\delta_i(t) = \theta_i(t) - \theta_0(t)$ in SI units, where $\theta_i(t)$ is the voltage angle at bus $i\in\{1,...,N \}$, and $\theta_0(t)$ the Park transformation angle of the global $\mathrm{dq}$ coordinates, which can be expressed as}
\begin{align}\label{eq:bus_voltage}
    \begin{split}
       \hspace{-2mm}|v|_i(t) \hspace{-0.7mm}= \hspace{-0.7mm}\sqrt{\hspace{-0.7mm}U_{\mathrm{d},i}(t)^2\hspace{-0.7mm}+\hspace{-0.7mm}U_{\mathrm{q},i}(t)^2},\quad \delta_i(t) \hspace{-0.7mm}= \arctan\hspace{-0.7mm}\left(\tfrac{U_{\mathrm{q},i}(t)}{U_{\mathrm{d},i}(t)}\right)\hspace{-0.5mm}.\hspace{-2mm}
    \end{split}
\end{align}
{We consider the active and reactive branch powers}
\begin{align}
\begin{split}
    P_{ij}(t) &= U_{\mathrm{d},i}(t)I_{\mathrm{d},ij}(t)+U_{\mathrm{q},i}(t)I_{\mathrm{q},ij}(t)\\
    Q_{ij}(t) &= -V_{\mathrm{d},i}(t)I_{\mathrm{q},ij}(t)+U_{\mathrm{q},i}(t)I_{\mathrm{d},ij}(t),
    \end{split}
\end{align}
{\color{black} and define the associated bus power injections $P_{i}(t)\coloneqq\textstyle\sum_{j\neq i}^nP_{ij}(t)$ and $Q_{i}(t)\coloneqq\textstyle\sum_{j\neq i}^nQ_{ij}(t)$, which are obtained as}
\begin{align}\label{eq:branch_power_injections}
\begin{split}
    \color{black}P_{i}(t) &\color{black}= U_{\mathrm{d},i}(t)I_{\mathrm{d},i}(t)+U_{\mathrm{q},i}(t)I_{\mathrm{q},i}(t)\\
    \color{black}Q_{i}(t) &\color{black}= -U_{\mathrm{d},i}(t)I_{\mathrm{q},i}(t)+U_{\mathrm{q},i}(t)I_{\mathrm{d},i}(t).
\end{split}
\end{align}
{\color{black} By linearizing \eqref{eq:bus_voltage} and \eqref{eq:branch_power_injections}, transforming them into the frequency domain, and performing some analytical computations, one can transform~\eqref{eq:GFM_model} into the input-output form in~\eqref{eq:admittance_model}, and vice versa (see~\cite{haberle2026decentralized} for details). Finally, if we extend \eqref{eq:GFM_model} to include the dynamics of all $N$ converters, we obtain the following $2n\times 2n$ transfer matrix in analogy to \eqref{eq:admittance_devices}, i.e.,} 
\begin{align}\label{eq:full_converter_polar}
    \color{black} \hspace{-1.8mm}-\hspace{-0.3mm}\underset{\eqqcolon\,{\tiny\begin{bmatrix}
       \Delta \omega(s)\\\Delta { |v|}(s)
   \end{bmatrix}}}{\underbrace{
   \begin{bmatrix}
       \Delta \omega_1(s)\\\Delta {|v|}_{1} (s)\\ \vdots\\ \Delta \omega_N(s)\\\Delta {|v|}_{N} (s)
   \end{bmatrix}}} \hspace{-1.4mm}= \hspace{-1.4mm} \underset{\eqqcolon\,D(s)}{\underbrace{\begin{bmatrix}
       D_1(s) \hspace{-0.8mm}& \hspace{-0.8mm}0_{2\times 2} \hspace{-0.8mm}& \hspace{-0.8mm} \dots  \hspace{-0.8mm}& \hspace{-0.8mm} 0_{2\times 2}\\0_{2\times2}  \hspace{-0.8mm}&  \hspace{-0.8mm}D_2(s) \hspace{-0.8mm} &  \hspace{-0.8mm}\dots  \hspace{-0.8mm}&  \hspace{-0.8mm}0_{2\times 2}\\
       \vdots  \hspace{-0.8mm}&  \hspace{-0.8mm}\vdots
        \hspace{-0.8mm}& \hspace{-0.8mm}\ddots \hspace{-0.8mm}& \hspace{-0.8mm}\vdots\\0_{2\times 2} \hspace{-0.8mm}& \hspace{-0.8mm}0_{2\times 2} \hspace{-0.8mm}& \hspace{-0.8mm}\dots \hspace{-0.8mm}& \hspace{-0.8mm}D_N(s)
   \end{bmatrix}}}
    \underset{\eqqcolon\,{\tiny\begin{bmatrix}
       \Delta { P}(s)\\\Delta { Q}(s)
   \end{bmatrix}}}{\underbrace{\begin{bmatrix}
       \Delta P_1(s)\\ \Delta Q_1(s)\\\vdots\\ \Delta P_N(s)\\ \Delta Q_N(s)
   \end{bmatrix}}} \hspace{-0.6mm}. \hspace{-1.5mm}
\end{align}

{\color{black} Notice that when writing the converter's dynamics in the form of~\eqref{eq:GFM_model}, one also needs to derive the network dynamics using power and polar coordinates, i.e., in analogy to \eqref{eq:Y_grid}, we obtain}
\begin{align}\label{eq:full_network_polar}
   \color{black}\hspace{-1mm} \underset{\eqqcolon\,{\tiny\begin{bmatrix}
       \Delta { P}_{\rm e}(s)\\\Delta { Q}_{\rm e}(s)
   \end{bmatrix}}}{\underbrace{\begin{bmatrix}
       \Delta P_{{\rm e},1}(s)\\ \Delta Q_{{\rm e},1}(s)\\\vdots\\ \Delta P_{{\rm e},N}(s)\\ \Delta Q_{{\rm e},N}(s)
   \end{bmatrix}}}
  \hspace{-0.5mm} = \hspace{-0.5mm}\underset{\eqqcolon \, {N}(s)}{\underbrace{
   \begin{bmatrix}
       N_{11}(s)&\cdots&N_{1N}(s)\\\vdots & \ddots& \vdots\\ N_{N1}(s)& \cdots & N_{NN}(s)
   \end{bmatrix}}}
   \underset{\eqqcolon\,{\tiny\begin{bmatrix}
       \Delta \delta(s)\\\Delta {|v|_\mathrm{n}}(s)
   \end{bmatrix}}}{\underbrace{
   \begin{bmatrix}
       \Delta \delta_1(s)\\\Delta {|v|}_{\mathrm{n},1} (s)\\ \vdots\\ \Delta \delta_N(s)\\\Delta {|v|}_{\mathrm{n},N} (s)
   \end{bmatrix}}}\hspace{-1mm} 
\end{align}
{\color{black} which for a resistive-inductive power transmission line model as in \eqref{eq:Y_ij} would be given by the $2\times2$ transfer matrix blocks $N_{ii}(s)$ and $N_{ij}(s)$ (see~\cite{haberle2026decentralized} for a detailed derivation)}
\begin{align}\label{eq:full_network_polar_matrix_blocks}
\begin{split}
    \color{black}N_{ii}(s) &= \color{black}\textstyle\sum_{j\neq i}^n B_{ij}\tfrac{|v|_{0,i}^2}{1+(\varepsilon+\tfrac{s}{\omega_0})^2}\begin{bmatrix}
        1&\varepsilon+\tfrac{s}{\omega_0}\\-(\varepsilon+\tfrac{s}{\omega_0}) & 1
    \end{bmatrix}\\
    &\color{black}\quad\quad\quad+\textstyle\sum_{j\neq i}^n B_{ij}\tfrac{|v|_{0,i}^2-|v|_{0,i}|v|_{0,j}}{1+\varepsilon^2}\begin{bmatrix}
        -1& \varepsilon \\ \varepsilon &1
    \end{bmatrix}\\
    \color{black}N_{ij}(s) &\color{black}= B_{ij}\tfrac{|v|_{0,i}|v|_{0,j}}{1+(\varepsilon+\tfrac{s}{\omega_0})^2}\begin{bmatrix}
        -1&-(\varepsilon+\tfrac{s}{\omega_0})\\(\varepsilon+\tfrac{s}{\omega_0})&-1
    \end{bmatrix},
\end{split}
\end{align}
{\color{black} where $\Delta {|v|}_{\mathrm{n},i} (s) \coloneqq \tfrac{\Delta |v|_i(s)}{|v|_{0,i}}$ is the voltage magnitude at bus $i$ normalized at the steady-state $|v|_{0,i}$. Moreover, to derive \eqref{eq:full_network_polar_matrix_blocks}, as in any power system small-signal model, we linearized at a small steady-state angle difference $\delta_{0,i}\approx \delta_{0,j}$, which is standard and justified since thermal limitations for transmission lines preclude large angle differences \cite{kundur2007power}. }

\begin{figure}[t!]
\vspace{-1mm}
    \centering
   \resizebox{0.37\textwidth}{!}{
\tikzstyle{roundnode}=[circle,draw=black!60,fill=black!5,scale=0.7]
\begin{tikzpicture}[scale=1,every node/.style={scale=0.8}]
\draw [rounded corners = 3,fill=backgroundblue!60] (-0.3,4.3) rectangle (1.3,3.5);
\node [scale=1.2] at (0.5,3.9) {$\mathcal{D}_0(s)$};
\draw [rounded corners = 3,fill=backgroundred!60] (-0.3,2.7) rectangle (1.3,1.9);
\node [scale=1.2] at (0.5,2.3) {$\mathcal{N}_0(s)$};
\draw [-latex](1.3,3.9) -- (2.5,3.9);
\draw (-0.3,2.3) -- (-1.1,2.3);

\draw[-latex] (1.8,2.3) -- (1.3,2.3);
\draw [-latex](2.6,2.3) -- (2,2.3); 
\draw [-latex](1.9,3.9) -- (1.9,2.4);
\node [roundnode] at (1.9,2.3) {};

\draw [-latex](-1,3.9) -- (-0.3,3.9);
\draw [-latex](-1.8,3.9) -- (-1.2,3.9); 
\draw [-latex](-1.1,2.3) node (v1) {} -- (-1.1,3.8);
\node [roundnode] at (-1.1,3.9) {};
\node at (-0.9,3.6) {$-$};
\node at (-1.8,4.35) {$\begin{bmatrix}\Delta p_\mathrm{d}\\ \Delta q_\mathrm{d}\end{bmatrix}$};
\node at (2.5,4.35) {$\begin{bmatrix} \Delta \omega\\\Delta |v|\end{bmatrix}$};
\node at (-0.65,4.35) {$\begin{bmatrix}\Delta p\\ \Delta q\end{bmatrix}$};
\node at (2.6,2.75) {$\begin{bmatrix} \Delta\omega_\mathrm{d} \\\Delta |v|_\mathrm{d}\end{bmatrix}$};
%\node at (-2.3,2.3) {$\begin{bmatrix}\Delta p_\mathrm{e}(s)\\ \Delta q_\mathrm{e}(s) \end{bmatrix}$};
\node at (0.5,3.3) {device dynamics};
\node at (0.5,1.7) {network dynamics};
\node at (-1.05,4.35) {$-$};
\draw[-latex] (-1.1,2.3) -- (-1.8,2.3);
\node at (-1.8,2.75) {$\begin{bmatrix}\Delta p_\mathrm{e}\\ \Delta q_\mathrm{e}\end{bmatrix}$};
\end{tikzpicture}
}
    \vspace{-9mm}
    \caption{\color{black} Closed-loop power system dynamics where $\Delta P_\mathrm{d}$ and  $\Delta Q_\mathrm{d}$ are the active and reactive power disturbances, $\Delta \omega_\mathrm{d}$ is the frequency disturbance, and $\Delta |{v}|_{\mathrm{d}}$ is the voltage magnitude disturbance.}
    \vspace{-4mm}
        \label{fig:power_sys_model_orig}
\end{figure}
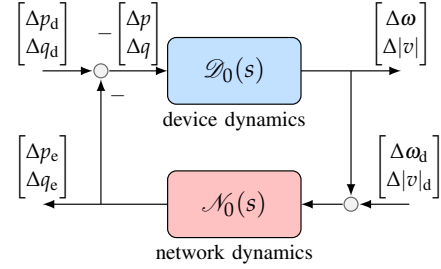

{\color{black} In these coordinates, the closed-loop model of the multi-converter system can be constructed by combining \eqref{eq:full_converter_polar} and \eqref{eq:full_network_polar} as illustrated in Figure \ref{fig:power_sys_model_orig}. Notice that we focus on the stability of the bus frequency and voltage magnitudes, given by $\Delta \omega_i(s)=\Delta \delta_i(s)s$ and $\Delta |v|_i(s)$ for $i\in\{1,...,N\}$ and consider these quantities as interconnection signals between the device and network dynamics in Figure \ref{fig:power_sys_model_orig}. Accordingly, the interconnected subsystems are defined as:}
\begin{align}\label{eq:N_0_D_0}
\begin{split}
   \color{black} \mathcal{D}_0(s) &\color{black}\coloneqq D(s)\\
    \color{black}\mathcal{N}_0(s) &\color{black}\coloneqq N(s)\cdot\mathrm{diag}(\tfrac{1}{s},\tfrac{1}{|v|_{0,1}},\dots,\tfrac{1}{s},\tfrac{1}{|v|_{0,N}}),
\end{split}
\end{align}
which is based on the power and polar coordinates to study the stability of multi-converter systems \eqref{eq:N_0_D_0} \cite{haberle2026decentralized,siahaan2024decentralized,watson2020control,Niehues2026,gross2022compensating}. One motivation for this choice is that it naturally supports parametric models of GFM devices, such as the model in \eqref{eq:GFM_model}. A convenient simplification is to neglect the inner-loop converter dynamics, which are typically manufacturer-specific, and instead derive stability conditions for the outer-loop controllers. These conditions can then serve as explicit tuning rules for control design. Moreover, common GFM outer-loop controllers, most notably droop control and its variants, are naturally formulated in power and polar coordinates, as illustrated in \eqref{eq:GFM_model}. This makes these coordinates particularly suitable for the parametric stability analysis of multi-converter systems.

Accordingly, the stability of the multi-converter system in Figure~\ref{fig:power_sys_model_orig}, described by the network and device dynamics in \eqref{eq:N_0_D_0}, can be analyzed using variants of the geometric decentralized stability certificates in \eqref{eq:pas}--\eqref{eq:decentralized_DW}. Since deriving explicit parametric stability conditions is analytically challenging, existing approaches \cite{haberle2026decentralized,siahaan2024decentralized,watson2020control,Niehues2026} have primarily focused on the comparatively simple passivity-, phase-, and gain-based conditions in \eqref{eq:pas}--\eqref{eq:DecGain}, which are more amenable to analytical expressions in terms of converter and network parameters. However, directly applying these conditions, particularly classical passivity requirements, can be overly conservative or even infeasible for the considered network models. This motivates the use of loop-shifting techniques, which enable less restrictive stability certificates by redistributing passivity properties between the device and network dynamics, as detailed in the following.

\begin{figure}[t!]
    \centering
    \vspace{-1mm}
    \begin{subfigure}{0.22\textwidth}
        \centering
        \resizebox{1.05\textwidth}{!}{
\tikzstyle{roundnode}=[circle,draw=black!60,fill=black!5,scale=0.7]
\begin{tikzpicture}[scale=1,every node/.style={scale=1}]
\draw [rounded corners = 3,fill=backgroundblue!60] (-0.3,4.3) rectangle (1.3,3.5);
\node [scale=1.2] at (0.5,3.9) {$\mathcal{D}(s)$};
\draw [rounded corners = 3,fill=backgroundred!60] (-0.3,2.7) rectangle (1.3,1.9);
\node [scale=1.2] at (0.5,2.3) {$\mathcal{N}(s)$};
\draw [-latex](1.3,3.9) -- (2.5,3.9);
\draw (-0.3,2.3) -- (-1.1,2.3);

\draw[-latex] (1.8,2.3) -- (1.3,2.3);
\draw [-latex](2.6,2.3) -- (2,2.3); 
\draw [-latex](1.9,3.9) -- (1.9,2.4);
\node [roundnode] at (1.9,2.3) {};

\draw [-latex](-1,3.9) -- (-0.3,3.9);
\draw [-latex](-1.8,3.9) -- (-1.2,3.9); 
\draw [-latex](-1.1,2.3) node (v1) {} -- (-1.1,3.8);
\node [roundnode] at (-1.1,3.9) {};
\node at (-0.9,3.6) {$-$};
\node at (-1.8,4.5) {$\begin{bmatrix}\Delta P_\mathrm{d}\\ \Delta Q_\mathrm{d}\end{bmatrix}$};
\node at (2.5,4.5) {$\begin{bmatrix} \Delta \omega\\\Delta |\tilde{v}|_\mathrm{n}\end{bmatrix}$};
\node at (-0.67,4.5) {$\begin{bmatrix}\Delta P\\ \Delta Q\end{bmatrix}$};
\node at (2.6,2.9) {$\begin{bmatrix} \Delta\omega_\mathrm{d} \\\Delta |\tilde{v}|_\mathrm{nd}\end{bmatrix}$};
%\node at (-2.3,2.3) {$\begin{bmatrix}\Delta P_\mathrm{e}(s)\\ \Delta Q_\mathrm{e}(s) \end{bmatrix}$};
\node[scale=1.2] at (0.5,3.2) {device dynamics};
\node[scale=1.2] at (0.5,1.6) {network dynamics};
\node at (-1.12,4.5) {$-$};
\draw[-latex](-1.1,2.3) -- (-1.8,2.3);
\node at (-1.8,2.9) {$\begin{bmatrix}\Delta P_\mathrm{e}\\ \Delta Q_\mathrm{e}\end{bmatrix}$};
\end{tikzpicture}
}
        \vspace{-8.5mm}
        \caption{\color{black} Original feedback interconnection.}
        \label{fig:power_sys_model}
    \end{subfigure}
    \hspace{0.05cm}
    \begin{subfigure}{0.24\textwidth}
        \centering
        \resizebox{1.05\textwidth}{!}{
\tikzstyle{roundnode}=[circle,draw=black!60,fill=black!5,scale=0.75]
\begin{tikzpicture}[scale=1,every node/.style={scale=1}]
\draw [rounded corners = 3, dashed,fill=backgroundred!20] (2,3.7) rectangle (-1.1,1.6);
\draw [rounded corners = 3, dashed, fill=backgroundblue!20] (2,4.4) rectangle (-1.1,6.5);
\draw [rounded corners = 3,fill=backgroundblue!60] (-0.3,6.2) rectangle (1.3,5.4);
\draw [rounded corners = 3,fill=backgroundred!60] (-0.3,2.5) rectangle (1.3,1.7);
\node [scale=1.2] at (0.5,5.8) {$\mathcal{D}(s)$};
\node [backgroundblue!100,scale=1.2] at (1.65,6.15) {$\mathcal{D}'$};
\node [scale=1.2] at (0.5,2.1) {$\mathcal{N}(s)$};
\node [backgroundred!100,scale=1.2] at (1.65,3.35) {$\mathcal{N}'$};
\draw [-latex](1.3,5.8) -- (3,5.8);
\draw (-0.9,2.1) -- (-1.4,2.1);

\draw[-latex] (2.2,2.1) -- (1.3,2.1);

\draw [-latex](2.3,5.8) -- (2.3,2.2);

\draw [-latex](-1.3,5.8) -- (-0.9,5.8);
\draw [-latex](-2.1,5.8) -- (-1.5,5.8); 
\draw [-latex](-1.4,2.1) node (v2) {} -- (-1.4,5.7);
\node [roundnode] at (-1.4,5.8) {};
\node at (-1.25,5.5) {$-$};

\draw [rounded corners = 3, fill = black!20] (-0.3,3.4) rectangle (1.3,2.6);
\node [scale=1.2] at (0.5,3) {$\Gamma(s)$};
\draw [rounded corners = 3, fill = black!20] (-0.3,5.3) rectangle (1.3,4.5);
\node [scale=1.2] at (0.5,4.9) {$\Gamma(s)$};
\draw [-latex](1.8,2.1) -- (1.8,3) -- (1.3,3);
\draw [-latex](-0.3,2.1) -- (-0.7,2.1); 
\draw [-latex](-0.3,3) -- (-0.8,3) -- (-0.8,2.2);
\node [roundnode] at (-0.8,2.1) {}; 

\node[roundnode] at (-0.8,5.8) {}; 
\draw [-latex](-0.7,5.8) -- (-0.3,5.8);
\draw [-latex](-0.3,4.9) -- (-0.8,4.9) -- (-0.8,5.7);
\draw[-latex] (1.8,5.8) -- (1.8,4.9) -- (1.3,4.9);

\node at (-2.1,6.4) {$\begin{bmatrix}\Delta P_\mathrm{d}\\ \Delta Q_\mathrm{d}\end{bmatrix}$};
\node at (3,6.4) {$\begin{bmatrix} \Delta \omega\\\Delta |\tilde{v}|_\mathrm{n}\end{bmatrix}$};
\node[scale=1.2] at (0.5,4.1) {device dynamics};
\node[scale=1.2] at (0.5,1.3) {network dynamics};
\node [roundnode] (v1) at (2.3,2.1) {};

\draw [-latex](3,2.1) --  (2.4,2.1);
\draw [-latex](-1.4,2.1) -- (-2.1,2.1);
\node at (-2.1,2.7) {$\begin{bmatrix}\Delta P_\mathrm{e}\\ \Delta Q_\mathrm{e}\end{bmatrix}$};
\node at (3,2.7) {$\begin{bmatrix} \Delta\omega_\mathrm{d} \\\Delta |\tilde{v}|_\mathrm{nd}\end{bmatrix}$};
\end{tikzpicture}
}
        \vspace{-8.5mm}
        \caption{\color{black} Loop-shifted interconnection with $\Gamma$.}
        \label{fig:loop_shifted_ND}
    \end{subfigure}
    \caption{\footnotesize {\color{black} Closed-loop feedback interconnections used for stability analysis.}}
    \label{fig:loop_shifting_ND}
    \vspace{-4mm}
\end{figure}
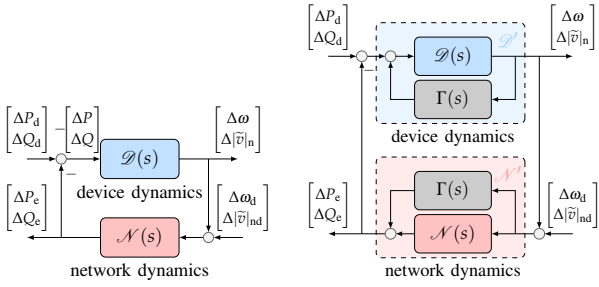

{\color{black} A recurring feature of parametric stability analyses in power and polar coordinates is the use of voltage-magnitude derivatives (recall that frequency is the derivative of the voltage angle) together with loop-shifting techniques related to the passivity-index framework introduced earlier. The voltage-magnitude derivative is introduced to preserve the symmetry of the network model. In particular, we use the angle and normalized-voltage derivatives}
\[
\Delta \omega_i(s)\coloneqq s\Delta\delta_i(s),
\qquad
\Delta|\tilde{v}|_{\mathrm{n},i}(s)
\coloneqq s\Delta |v|_{\mathrm{n},i}(s)
\]
{\color{black} as interconnection signals. We then analyze the closed-loop system shown in Figure~\ref{fig:power_sys_model}, where}
\begin{align}
\label{eq:N_D}
\begin{split}
    \mathcal{D}(s)
    &\coloneqq
    D(s)\,
    \mathrm{diag}\left(
    1,\frac{s}{|v|_{0,1}},
    \ldots,
    1,\frac{s}{|v|_{0,N}}
    \right),\\
    \mathcal{N}(s)
    &\coloneqq
    N(s)\,
    \mathrm{diag}\left(
    \frac{1}{s},\frac{1}{s},
    \ldots,
    \frac{1}{s},\frac{1}{s}
    \right).
\end{split}
\end{align}

{\color{black} The basic idea underlying loop shifting is that an excess of passivity in one subsystem can compensate for a passivity deficit in the other, while preserving the stability of their feedback interconnection. In the present setting, the subsystems $\mathcal{D}(s)$ and $\mathcal{N}(s)$ in Figure~\ref{fig:power_sys_model} do not directly satisfy the assumptions of the Passivity Theorem. In particular, $\mathcal{N}(s)$ is generally not passive \cite{dey2022passivity,dey2022passivity2}. We therefore employ dynamic loop-shifting techniques and consider the loop-shifted interconnection
$\mathcal{D}'\#\mathcal{N}'$ shown in Figure~\ref{fig:loop_shifted_ND}, where}
\begin{align}
\label{eq:N_dash_D_dash}
\begin{split}
    \mathcal{D}'(s)
    &\coloneqq
    \mathcal{D}(s)
    \bigl(I-\Gamma(s)\mathcal{D}(s)\bigr)^{-1},\\
    \mathcal{N}'(s)
    &\coloneqq
    \mathcal{N}(s)+\Gamma(s).
\end{split}
\end{align}
{\color{black} The loop-shifting term $\Gamma(s)$ is an analytical tool used to establish suitable passivity properties. It does not correspond to any physical modification of the converter hardware, network impedances, or the actual interconnected system.}

% {\color{black}The following subsection presents illustrative examples of parametric stability conditions obtained using this loop-shifting framework.}

{\color{black} The objective of this subsection is to guarantee internal stability of the original feedback interconnection in Figure \ref{fig:power_sys_model} by imposing conditions independently on the device models ${D}_i(s)$ in \eqref{eq:GFM_model} constituting $\mathcal{D}_0(s)$ in \eqref{eq:N_0_D_0}. The resulting conditions depend on the adopted representation of the network dynamics \eqref{eq:full_network_polar} and, in particular, on the level of simplification applied to the full model in \eqref{eq:full_network_polar_matrix_blocks}. The analysis in \cite{haberle2026decentralized} provides explicit parametric stability certificates for several such network representations.}

{\color{black} The derivation follows three main steps. First, a suitable network model is selected. Second, the loop-shifting term $\Gamma(s)$ is chosen such that the shifted network dynamics $\mathcal{N}'(s)$ satisfy the required passivity or small-phase property. Finally, strict passivity of the shifted device dynamics $\mathcal{D}'(s)$ is enforced through local algebraic conditions on the converter-control parameters in \eqref{eq:GFM_model}. This procedure reveals how the complexity and conservativeness of the resulting certificates depend on the fidelity of the network model.}

\begin{highlightbox}
\color{black}
\textbf{Network Model Simplifications}
The linearized network model in \eqref{eq:full_network_polar_matrix_blocks} admits three nested levels of approximation, as summarized below; see
\cite{haberle2026decentralized} for detailed derivations.

\medskip
\textbf{Dynamic Small-Signal Network Model}\\
The most detailed representation retains both the electromagnetic line dynamics and the nonuniform steady-state voltage magnitudes. For a dominantly inductive transmission network with $\varepsilon\ll 1$, the transfer-matrix blocks, characterized by
$s\neq 0$ and $|v|_{0,i}\neq |v|_{0,j}$, are
\begin{align}
\label{eq:full_network_polar_matrix_blocks_level0}
\begin{split}
N_{ii}(s)
&=
\sum_{j\neq i}^{N}
B_{ij}
\frac{|v|_{0,i}^{2}}
     {1+\left(\varepsilon+\frac{s}{\omega_0}\right)^{2}}
\begin{bmatrix}
1 & \frac{s}{\omega_0}\\[1mm]
-\frac{s}{\omega_0} & 1
\end{bmatrix}
\\
&\quad+
\sum_{j\neq i}^{N}
B_{ij}
\frac{|v|_{0,i}^{2}-|v|_{0,i}|v|_{0,j}}
     {1+\varepsilon^{2}}
\begin{bmatrix}
-1 & 0\\
0 & 1
\end{bmatrix},
\\
N_{ij}(s)
&=
B_{ij}
\frac{|v|_{0,i}|v|_{0,j}}
     {1+\left(\varepsilon+\frac{s}{\omega_0}\right)^{2}}
\begin{bmatrix}
-1 & -\frac{s}{\omega_0}\\[1mm]
\frac{s}{\omega_0} & -1
\end{bmatrix}.
\end{split}
\end{align}

\medskip
\textbf{Quasi-Stationary Network Model}\\
Neglecting the electromagnetic line dynamics by setting $s=0$ in
\eqref{eq:full_network_polar_matrix_blocks_level0}, while retaining the nonuniform steady-state voltage magnitudes, gives the
\emph{quasi-stationary} network model. A closely related representation is employed in
\cite{Niehues2026,siahaan2024decentralized}. Its transfer-matrix blocks are
\begin{align}
\label{eq:full_network_polar_matrix_blocks_level1}
\begin{split}
N_{ii}(s)
&=
\sum_{j\neq i}^{N}
B_{ij}\frac{1}{1+\varepsilon^{2}}
\begin{bmatrix}
|v|_{0,i}|v|_{0,j} & 0\\
0 & 2|v|_{0,i}^{2}-|v|_{0,i}|v|_{0,j}
\end{bmatrix},
\\
N_{ij}(s)
&=
B_{ij}
\frac{|v|_{0,i}|v|_{0,j}}
     {1+\varepsilon^{2}}
\begin{bmatrix}
-1 & 0\\
0 & -1
\end{bmatrix}.
\end{split}
\end{align}

\medskip
\textbf{Zero-Power-Flow Network Model}\\
A further simplification is obtained by assuming identical steady-state voltage magnitudes,
\[
|v|_{0,i}=|v|_{0,j}=|v|_0.
\]
Together with the quasi-stationary approximation, this yields the
\emph{zero-power-flow} network model similarly used in \cite{watson2020control}:
\begin{align}
\label{eq:full_network_polar_matrix_blocks_level2}
\begin{split}
N_{ii}(s)
&=
\sum_{j\neq i}^{N}
B_{ij}
\frac{|v|_0^{2}}{1+\varepsilon^{2}}
\begin{bmatrix}
1 & 0\\
0 & 1
\end{bmatrix},
\\
N_{ij}(s)
&=
B_{ij}
\frac{|v|_0^{2}}{1+\varepsilon^{2}}
\begin{bmatrix}
-1 & 0\\
0 & -1
\end{bmatrix}.
\end{split}
\end{align}

Finally, it is important to recognize that each network model simplification introduces additional assumptions that limit the range of dynamics captured by the resulting stability analysis. In particular, neglecting the network dynamics, as in the quasi-stationary models, may exclude certain oscillatory modes and stability phenomena that are captured by the full dynamic small-signal network model in \eqref{eq:full_network_polar_matrix_blocks_level0}. Hence, the choice of network model involves a trade-off between analytical tractability and the fidelity of the resulting stability assessment.
\end{highlightbox}

\color{black} These network representations form a hierarchy in which successive approximations remove dynamic or operating-point-dependent terms. As a result, the required loop shift $\Gamma(s)$ becomes progressively simpler and the corresponding converter-side conditions less restrictive.

\subsubsection{Loop-Shifting of the Network Dynamics}
For the dynamic network model in
\eqref{eq:full_network_polar_matrix_blocks_level0}, consider the block-diagonal loop-shifting term
\[
\Gamma(s)
=
\operatorname{diag}\bigl(
\Gamma_1(s),\ldots,\Gamma_N(s)
\bigr),
\qquad
\Gamma_i(s)
=
\begin{bmatrix}
\Gamma_i^{\mathrm{p}}(s) & 0\\
0 & \Gamma_i^{\mathrm{q}}(s)
\end{bmatrix}.
\]
An admissible choice derived in \cite{haberle2026decentralized}, which ensures the required network-side passivity or small-phase property, is
{
\begin{align*}
\Gamma_i^{\mathrm{p}}(s)
&=
\tfrac{1}{s}
\left(
\tfrac{\gamma_{1,i}^{\mathrm{p}}\omega_0^2s^2}
     {\omega_0^2+(\varepsilon\omega_0+s)^2}
+
\tfrac{\gamma_{2,i}^{\mathrm{p}}\omega_0^2}
     {\omega_0^2+(\varepsilon\omega_0+s)^2}
+
\gamma_{3,i}^{\mathrm{p}}
\right),
\\
\Gamma_i^{\mathrm{q}}(s)
&=
\tfrac{1}{s}
\left(
\tfrac{\gamma_{1,i}^{\mathrm{q}}\omega_0^2s^2}
     {\omega_0^2+(\varepsilon\omega_0+s)^2}
+
\tfrac{\gamma_{2,i}^{\mathrm{q}}\omega_0^2}
     {\omega_0^2+(\varepsilon\omega_0+s)^2}
+
\gamma_{3,i}^{\mathrm{q}}
\right),
\end{align*}
}
with
{
\begin{align*}
\gamma_{1,i}^{\mathrm{p}}
&=
\gamma_{1,i}^{\mathrm{q}}
=
\tfrac{2|v|_{\max}^{2}}{\omega_0^{2}}
\textstyle\sum_{j\neq i}^{n}B_{ij},
\qquad
\gamma_{2,i}^{\mathrm{p}}
=
\gamma_{2,i}^{\mathrm{q}}
=
-3|v|_{\max}^{2}
\textstyle\sum_{j\neq i}^{n}B_{ij},
\\
\gamma_{3,i}^{\mathrm{p}}
&=
-\tfrac{\gamma_{2,i}^{\mathrm{p}}}{1+\varepsilon^{2}},
\qquad
\gamma_{3,i}^{\mathrm{q}}
=
-\tfrac{\gamma_{2,i}^{\mathrm{q}}}{1+\varepsilon^{2}}
+
\widetilde{\gamma}_{3,i}^{\mathrm{q}},
\qquad
\widetilde{\gamma}_{3,i}^{\mathrm{q}}
\geq
\tfrac{0.8}{1+\varepsilon^{2}}
\textstyle\sum_{j\neq i}^{n}B_{ij}.
\end{align*}
}

The loop shift simplifies for the reduced network models. For the quasi-stationary model in
\eqref{eq:full_network_polar_matrix_blocks_level1}, it suffices to choose
\begin{align*}
\Gamma_i^{\mathrm{p}}(s)&=0,
&
\Gamma_i^{\mathrm{q}}(s)
&=
\tfrac{\widetilde{\gamma}_{3,i}^{\mathrm{q}}}{s}.
\end{align*}
For the zero-power-flow model in
\eqref{eq:full_network_polar_matrix_blocks_level2}, no loop shift is required:
\begin{align*}
\Gamma_i^{\mathrm{p}}(s)&=0,
&
\Gamma_i^{\mathrm{q}}(s)&=0.
\end{align*}

\subsubsection{Parametric Device-Level Conditions}
Once $\Gamma(s)$ has been selected, the remaining stability verification is decentralized because the loop-shifted device dynamics retain their block-diagonal structure:
\begin{align*}
\mathcal{D}'(s)
&=
\operatorname{diag}\bigl(
\mathcal{D}'_1(s),\ldots,\mathcal{D}'_N(s)
\bigr),
\\
\mathcal{D}'_i(s)
&=
\mathcal{D}_i(s)
\bigl(
I-\Gamma_i(s)\mathcal{D}_i(s)
\bigr)^{-1} =\operatorname{diag}\bigl(\mathcal{D}'_{\mathrm{p},i}(s),  \mathcal{D}'_{\mathrm{q},i}(s)\bigr)
\end{align*}
Thus, strict passivity can be verified independently for the loop-shifted active- and reactive-power control. Their shifted transfer functions are
\begin{align*}
\mathcal{D}'_{\mathrm{p},i}(s)
&=
\tfrac{d_{\mathrm{p},i}}
{\tau_{\mathrm{p},i}s
 +1
 -\Gamma_i^{\mathrm{p}}(s)d_{\mathrm{p},i}},
\qquad
\mathcal{D}'_{\mathrm{q},i}(s)
=
\tfrac{d_{\mathrm{q},i}s/|v|_{0,i}}
{\tau_{\mathrm{q},i}s
 +1
 -\Gamma_i^{\mathrm{q}}(s)
  d_{\mathrm{q},i}s/|v|_{0,i}}.
\end{align*}
Applying the scalar strict-passivity conditions to these two transfer functions yields explicit inequalities for the active-power--frequency and reactive-power--voltage droop controllers.

In particular, such \emph{decentralized stability conditions} can serve as \emph{local tuning rules} for the dynamic droop control \eqref{eq:GFM_model} of each converter $i \in \{1, \dots, N\}$. The stability conditions for each converter $i$ depend on
\begin{itemize}
    \item its own tunable local droop control parameters, namely, the droop gains $d_{\mathrm{p},i},\, d_{\mathrm{q},i}$ and the time constants $\tau_{\mathrm{p},i},\, \tau_{\mathrm{q},i}$,
    \item certain network parameters, including the susceptances of the neighboring lines $B_{ij}$, the resistance-to-inductance ratio $\varepsilon$, and the maximum steady-state bus voltage magnitude $|v|_{\mathrm{max}}$. These parameters are either locally accessible or globally agreed upon, e.g., in grid codes.
\end{itemize}
Crucially, the conditions are local and entirely independent of the control parameters of other converters, making them directly applicable for scalable and decentralized stability assessment, as well as for local controller tuning and grid-code design. Beyond that, the decentralized stability certificates support \emph{heterogeneous} device-level controllers by allowing each converter to be designed with independent droop gains and time constants, without requiring uniform parameters across the system, as is the case in, e.g., \cite{gross2022compensating}.

\begin{figure}[t!]
    \centering
\begin{subfigure}{0.2\textwidth}
    \centering
    \vspace{-1mm}
\scalebox{0.35}{\includegraphics[]{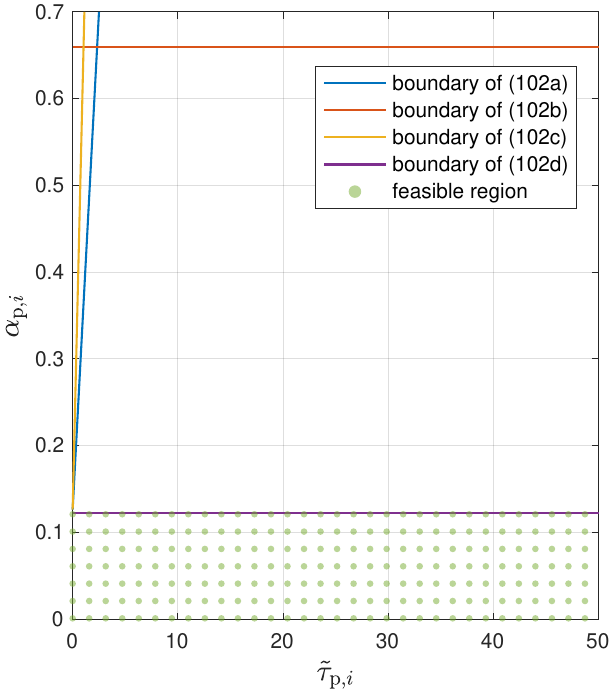}}
\vspace{-5mm}
    \caption{2D plot with $\varepsilon = 0.3$.}
    \vspace{-1mm}
    \label{fig:2d_p}
\end{subfigure}
\hspace{0.2cm}
\begin{subfigure}{0.26\textwidth}
    \centering
    \vspace{-1mm}
\scalebox{0.37}{\includegraphics[]{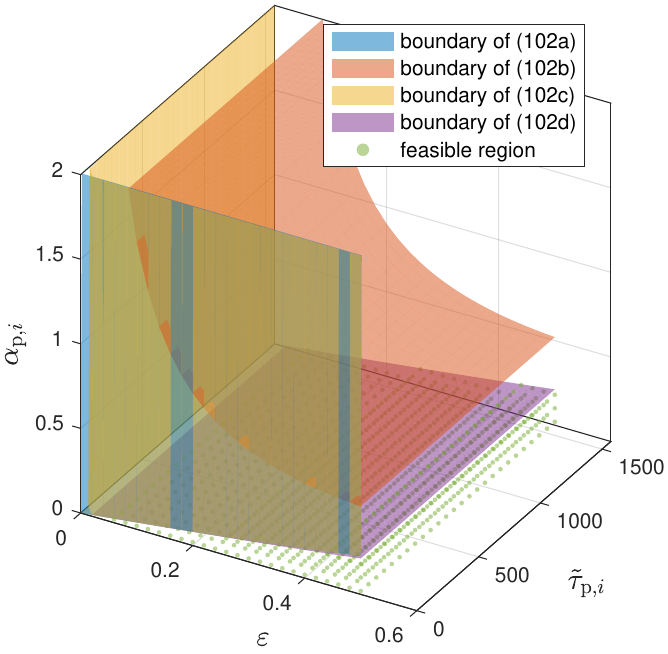}}
\vspace{-1mm}
    \caption{3D plot with variable $\varepsilon$.}
    \vspace{-1mm}
    \label{fig:3d_p}
\end{subfigure}
\caption{Graphical illustration of the stability conditions in \eqref{eq:p_conditions}, where $\alpha_{\mathrm{p},i}\hspace{-0.5mm}=\hspace{-0.5mm}\tfrac{d_{\mathrm{p},i}}{\omega_0}|v|_{\mathrm{max}}^2\sum_{j\ne i}^NB_{ij}$ and  $\tilde{\tau}_{\mathrm{p},i}\hspace{-0.5mm}=\hspace{-0.5mm}\tau_{\mathrm{p},i}\omega_0$ (in rad).}
\label{fig:p_conditions_3D}
\vspace{-2mm}
\end{figure}

\begin{figure}[t!]
    \centering
\begin{subfigure}{0.2\textwidth}
    \centering
\scalebox{0.35}{\includegraphics[]{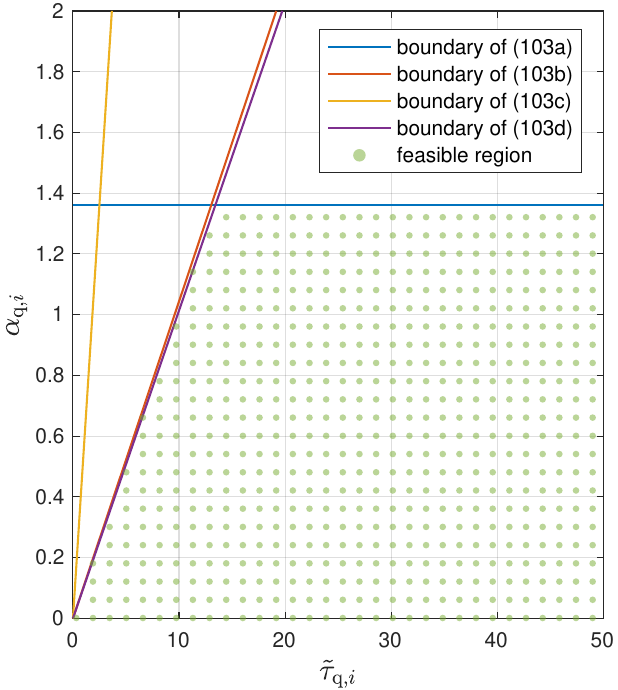}}
\vspace{-5mm}
    \caption{2D plot with $\varepsilon = 0.3$.}
    \vspace{-1mm}
    \label{fig:2d_q}
\end{subfigure}
\hspace{0.25cm}
\begin{subfigure}{0.26\textwidth}
    \centering
\scalebox{0.37}{\includegraphics[]{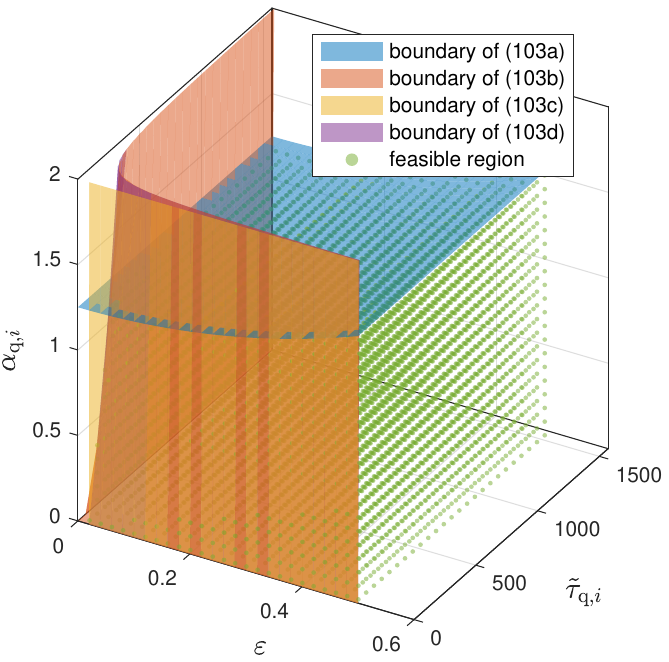}}
\vspace{-1mm}
    \caption{3D plot with variable $\varepsilon$.}
    \vspace{-1mm}
    \label{fig:3d_q}
\end{subfigure}
\caption{Graphical illustration of the stability conditions in \eqref{eq:q_conditions}, where $\alpha_{\mathrm{q},i}=\tfrac{d_{\mathrm{q},i}}{|v|_{0,i}}\sum_{j\ne i}^N B_{ij}$ and $\tilde{\tau}_{\mathrm{q},i}=\tau_{\mathrm{q},i}\omega_0$ (in rad).}
\label{fig:q_conditions_3D}
\vspace{-4mm}
\end{figure}

Given the decoupled structure of $\mathcal{D}'_{i}(s)$, the decentralized stability conditions for each converter $i$ can be categorized into decoupled constraints on the active power-frequency droop control and the reactive power-voltage droop control. For each control scheme, these conditions can be visualized in either a two-dimensional (for fixed  $\varepsilon$) or a three-dimensional (for variable $\varepsilon$) coordinate system, as shown in Figures \ref{fig:p_conditions_3D} and \ref{fig:q_conditions_3D}. The coordinate axes represent scaled versions of the local droop gains and time constants (in 2D) and the global network parameter $\varepsilon$ (in 3D), with typical value ranges. The resulting feasible parameter sets are indicated by green dots. As illustrated in Fig. \ref{fig:p_conditions_3D}, for small values of $\varepsilon$, closed-loop stability is ensured when the droop gain $d_{\mathrm{p},i}$ and/or the self-susceptance $\textstyle\sum_{j\ne i}^N B_{ij}$ are sufficiently small, irrespective of the time constant $\tau_{\mathrm{p},i}$. Notably, this suggests potential instability issues in future transmission systems with a \emph{high} density of devices interconnected by \emph{short lines}, which correspond to large values of $\textstyle\sum_{j\ne i}^N B_{ij}$. Further, interpreting the filter time constant $\tau_{\mathrm{p},i}$ as virtual inertia reveals that increasing virtual inertia (i.e., larger $\tau_{\mathrm{p},i}$) does not necessarily enhance system stability. These findings align with the small-signal stability conditions derived for active power droop control in conjunction with SISO frequency dynamics in \cite{gross2022compensating}. 

Similarly, Fig. \ref{fig:q_conditions_3D} shows that closed-loop stability is guaranteed when the droop gain $d_{\mathrm{q},i}$ and/or the self-susceptance $\textstyle\sum_{j\ne i}^N B_{ij}$ are sufficiently small, provided that $\tau_{\mathrm{q},i}$ is nonzero. In particular, in contrast to $\tau_{\mathrm{p},i}$, a larger $\tau_{\mathrm{q},i}$ can be stabilizing. 

Finally, from the 3D plots in Figures \ref{fig:p_conditions_3D} and \ref{fig:q_conditions_3D}, it becomes apparent how an increasing $\varepsilon$ allows for larger local droop gains of both the active and reactive power droop control.  

In what follows, we provide an algebraic parametrization of the conditions in Figures \ref{fig:p_conditions_3D} and \ref{fig:q_conditions_3D}, which ensure strict passivity of $\mathcal{D}'_{\mathrm{p},i}(s)$ and $\mathcal{D}'_{\mathrm{q},i}(s)$ (a detailed derivation is provided in \cite{haberle2026decentralized}). Namely, for the active power-frequency droop control, we require
\eqref{eq:full_network_polar_matrix_blocks_level0}, the sufficient conditions are
\begin{subequations}
\label{eq:p_conditions}
\begin{align}
\alpha_{\mathrm{p},i}
&<
c_{1,\varepsilon}
\left(
2\varepsilon
+
\widetilde{\tau}_{\mathrm{p},i}
(1+\varepsilon^{2})
\right),
\label{eq:condition1_p}
\\
\alpha_{\mathrm{p},i}
&<
c_{2,\varepsilon},
\label{eq:condition2_p}
\\
\alpha_{\mathrm{p},i}
&<
c_{3,\varepsilon}
\tfrac{
\widetilde{\tau}_{\mathrm{p},i}
\left[
\widetilde{\tau}_{\mathrm{p},i}
(1+\varepsilon^{2})
+
2\varepsilon
\right]
+1
}{
4\widetilde{\tau}_{\mathrm{p},i}
\varepsilon(1+\varepsilon^{2})
+
2\varepsilon^{2}
+
5
},
\label{eq:condition3_p}
\\
\alpha_{\mathrm{p},i}
&<
c_{4,\varepsilon}.
\label{eq:condition4_p}
\end{align}
\end{subequations}
Here, $\alpha_{\mathrm{p},i}
=
\tfrac{d_{\mathrm{p},i}}{\omega_0}
|v|_{\max}^{2}
\textstyle\sum_{j\neq i}^{N}B_{ij}$,
$\widetilde{\tau}_{\mathrm{p},i}
=
\tau_{\mathrm{p},i}\omega_0,$
and the coefficients $c_{\ell,\varepsilon}$,
$\ell\in\{1,2,3,4\}$, depend only on the network ratio
$\varepsilon$.

For the reactive-power--voltage channel, the corresponding sufficient conditions are
\begin{subequations}
\label{eq:q_conditions}
\begin{align}
\alpha_{\mathrm{q},i}
&<
c_{5,\varepsilon},
\label{eq:c5_rho}
\\
\alpha_{\mathrm{q},i}
&<
2c_{6,\varepsilon}
\varepsilon
\widetilde{\tau}_{\mathrm{q},i},
\\
\alpha_{\mathrm{q},i}
&<
c_{7,\varepsilon}
\widetilde{\tau}_{\mathrm{q},i},
\\
0
&<
c_{8,\varepsilon}\alpha_{\mathrm{q},i}^{2}
+
c_{9,\varepsilon}
\alpha_{\mathrm{q},i}
\widetilde{\tau}_{\mathrm{q},i}
+
2\varepsilon
\widetilde{\tau}_{\mathrm{q},i}^{2}.
\end{align}
\end{subequations}
In this case, $\alpha_{\mathrm{q},i}
=
\tfrac{d_{\mathrm{q},i}}{|v|_{0,i}}
\textstyle\sum_{j\neq i}^{N}B_{ij}$, $\widetilde{\tau}_{\mathrm{q},i}
=
\tau_{\mathrm{q},i}\omega_0.$
The quantities $c_{\ell,\varepsilon}$,
$\ell\in\{5,\ldots,14\}$, are functions of $\varepsilon$ whose explicit expressions are provided in
\cite{haberle2026decentralized}, but we neglect them here for interpretability.

The conditions simplify substantially under the quasi-stationary approximation
\eqref{eq:full_network_polar_matrix_blocks_level1}. In this case, it is sufficient that
\begin{align}
\label{eq:level1_conditions}
\alpha_{\mathrm{q},i}
<
\tfrac{5(1+\varepsilon^{2})}{4}
=
c_{5,\varepsilon},
\qquad
\tau_{\mathrm{p},i}>0,
\end{align}
for every converter $i\in\{1,\ldots,N\}$. Thus, the detailed active-power conditions in
\eqref{eq:p_conditions} reduce to positivity of the active-power time constant, while the reactive-power channel retains an explicit upper bound on its normalized droop gain.

Finally, for the zero-power-flow model in
\eqref{eq:full_network_polar_matrix_blocks_level2}, internal stability of the feedback interconnection is guaranteed without imposing additional converter-side conditions, within the assumptions underlying this simplified model.

These results illustrate the tradeoff between model fidelity and the complexity of the resulting decentralized certificate. For the dynamic network model, stability depends jointly on the local droop gains
$d_{\mathrm{p},i}$ and $d_{\mathrm{q},i}$, the filter time constants
$\tau_{\mathrm{p},i}$ and $\tau_{\mathrm{q},i}$, and locally available network quantities such as the aggregate neighboring susceptance
$\sum_{j\neq i}B_{ij}$, the line $R/X$ ratio $\varepsilon$, the local steady-state voltage $|v|_{0,i}$, and the voltage bound
$|v|_{\max}$. Because each condition involves only the parameters of converter $i$ and its neighboring network, the inequalities can be evaluated and enforced locally. They therefore provide explicit decentralized tuning rules and constitute candidate building blocks for future grid-code stability certificates.
\color{black}

\subsubsection{Case Studies}

We evaluate the decentralized stability conditions through detailed electromagnetic transient (EMT) simulations of the three-converter 9-bus system shown in Figure~\ref{fig:9bus_topology}, where the infinite bus is omitted, and line impedances are slightly adjusted to illustrate informative edge cases for the stability conditions in \eqref{eq:p_conditions}–\eqref{eq:level1_conditions}. The network is represented using dynamic transmission-line models, thereby capturing the full nonlinear network dynamics. Each GFM converter employs first-order active- and reactive-power droop controllers as in \eqref{eq:GFM_model}, embedded within a conventional cascaded converter-control architecture (see \cite{haberle2026decentralized} for details). Consequently, the EMT simulation model captures both detailed nonlinear network dynamics and converter dynamics that are not explicitly represented in the derivation of the decentralized stability conditions. We consider a Kron-reduced network with a uniform resistance-to-inductance ratio $\varepsilon=0.05$, a maximum steady-state voltage magnitude $|v|_{\mathrm{max}}=1.1$~pu, and identical self-susceptances $\textstyle\sum_{i\ne j}^3 b_{ij}=5$~pu at all converter nodes $i\in\{1,2,3\}$. GFM~1 and GFM~2 employ fixed controllers that satisfy the conditions in \eqref{eq:p_conditions} and \eqref{eq:q_conditions} throughout, whereas the control parameters of GFM~3 are varied to compare controllers satisfying different levels of the derived stability conditions.

\begin{figure}[!t]
\centering
\includegraphics[width=1\linewidth]{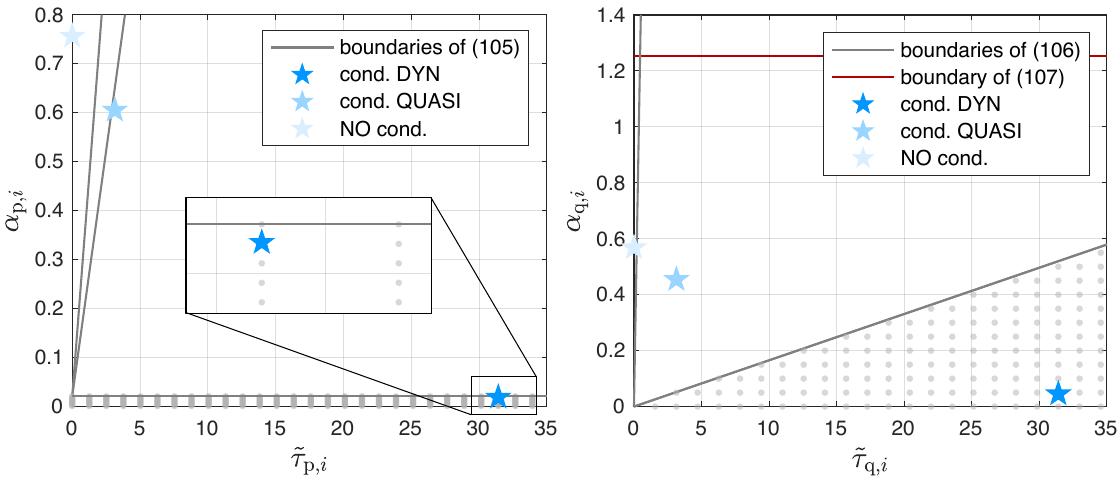}
\caption{2D stability region for the conditions in \eqref{eq:p_conditions} and \eqref{eq:q_conditions} applied to the modified 9-bus system ($\varepsilon = 0.05$). The stars indicate the GFM 3 control parameters in the global per unit system.}
\label{fig:parametric_stability_plots_case_study}
\end{figure}

\begin{figure}[!t]
\centering
\includegraphics[width=1\linewidth]{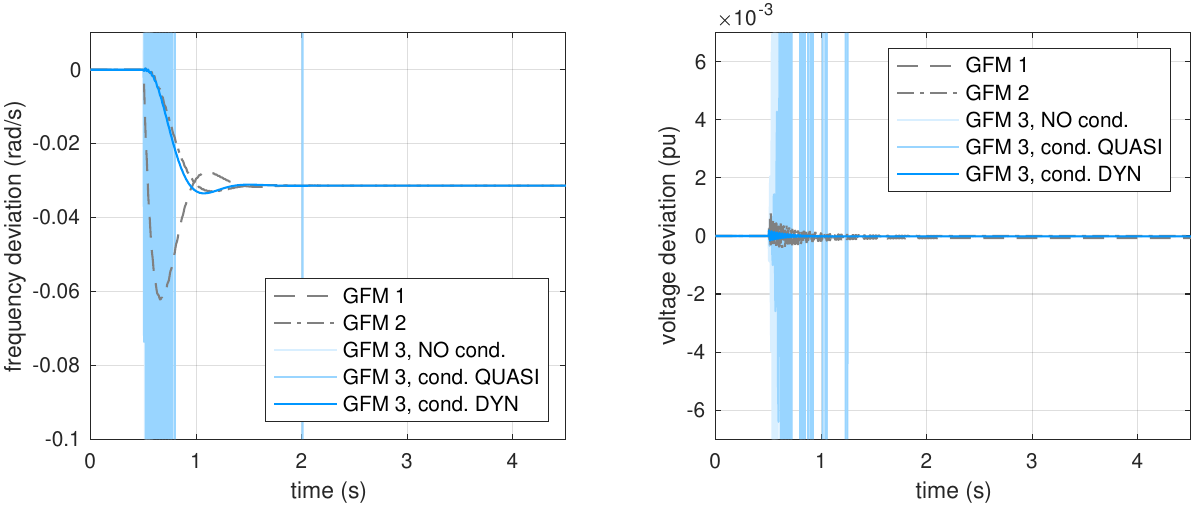}
\caption{System response of the modified 9-bus system with nonlinear network dynamics where all GFM controllers satisfy the conditions in \eqref{eq:p_conditions} and \eqref{eq:q_conditions} (cond. DYN). We also indicate the unstable dynamics of GFM 3 when the controller satisfies no conditions (NO cond.), and the conditions in \eqref{eq:level1_conditions} (cond. QUASI).}
\label{fig:parametric_stability_response_case_study}
\end{figure}

The simulation results for a small load increase in Figure~\ref{fig:parametric_stability_response_case_study} demonstrate the implications of these different conditions. When the GFM~3 controller satisfies the decentralized stability conditions in \eqref{eq:p_conditions} and \eqref{eq:q_conditions} (``cond. DYN'' in Figure~\ref{fig:parametric_stability_plots_case_study}), the system maintains stable operation in the nonlinear EMT model. In contrast, when the GFM~3 controller satisfies only the relaxed conditions in \eqref{eq:level1_conditions} (``cond. QUASI'' in Figure~\ref{fig:parametric_stability_plots_case_study}), or does not satisfy any of the derived stability conditions (``NO cond.'' in Figure~\ref{fig:parametric_stability_plots_case_study}), interactions between the converter and network dynamics can lead to instability. These results highlight two distinct requirements for reliable stability assessment. First, the relevant network dynamics must be retained in the small-signal model used for deriving the stability conditions. Neglecting these dynamics, as in simplified quasi-stationary or static network representations, can lead to overly optimistic stability predictions because potentially destabilizing converter--network interactions are no longer captured. Second, since the proposed conditions are derived from a small-signal model, their validity for the nonlinear EMT system remains local to the operating point around which the system is linearized. Accordingly, the conditions in \eqref{eq:p_conditions} and \eqref{eq:q_conditions} remain effective for the detailed nonlinear circuit model provided that the system operates sufficiently close to the nominal operating point for the small-signal approximation to remain accurate.

\subsection{Decentralized Nonlinear Stability Conditions}

The stability certificates discussed above are derived from linearized models and thus characterize local stability around an operating equilibrium. However, dissipativity theory and the DW shell framework can also be extended to nonlinear stability analysis. In particular, suitable converter control architectures may enable the derivation of nonlinear stability certificates, providing systematic guidelines for stability-oriented controller design and tuning.

In this subsection, we demonstrate how dissipativity theory can be used to derive decentralized nonlinear stability results for GFM converters. Since the nonlinear dissipativity property generally depends on the controller design, and it has not been established for many conventional GFM control designs, we focus on the emerging dispatchable virtual oscillator control (dVOC) \cite{colombino2019global,he2024quantitative}. The dVOC control is reviewed in ``\nameref{sb-dvoc}". We first establish its passivity property and then derive the corresponding decentralized stability results.

\begin{figure}[!t]
\centering
\includegraphics{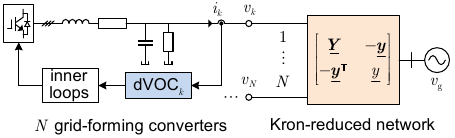}
\caption{Multi-converter grid-connected systems with dVOC GFM
control.}
\label{fig:power-plant}
\end{figure}

\begin{sidebar}{}
\setcounter{sequation}{0}
\renewcommand{\thesequation}{S\arabic{sequation}}
\setcounter{stable}{0}
\renewcommand{\thestable}{S\arabic{stable}}
\setcounter{sfigure}{0}
\renewcommand{\thesfigure}{S\arabic{sfigure}}
\section{Principles of dVOC and Connection to Droop Control}
\label{sb-dvoc}

Dispatchable virtual oscillator control (dVOC) can be interpreted as a linear oscillator augmented with power-setpoint feedback and nonlinear voltage-magnitude regulation. Starting from the nominal-frequency oscillator
\begin{equation*}
\dot{v}_k=\omega_0 J v_k,
\end{equation*}
where $ v_k=[v_{\alpha,k}, v_{\beta,k}]^\top \neq \mathbf{0}$ is the controlled voltage vector. The dynamics are augmented with current feedback to achieve active- and reactive-power tracking as well as synchronize with other interconnected oscillators \cite{colombino2019global}:
\begin{equation*}
\dot{ v}_k
=\omega_0 J v_k
+\eta_k R(\varphi)
\left(S_k^\star v_k- i_k\right),
\end{equation*}
where $ i_k$ is the output-current vector, $\eta_k>0$ is the control gain, and $R(\varphi)=\cos\varphi I_2 + \sin\varphi J$ is the rotation matrix to adapt to the network impedance angle. The setpoint matrix is defined as
\begin{equation*}
S_k^\star
=\frac{1}{v_k^{\star 2}}
\begin{bmatrix}
p_k^\star&q_k^\star\\
-q_k^\star&p_k^\star
\end{bmatrix},
\end{equation*}
where $p_k^\star$, $q_k^\star$, and $v_k^\star$ denote the active-power, reactive-power, and voltage-magnitude setpoints, respectively. The feedback error vanishes when the voltage magnitude and power injections reach their setpoints, as the instantaneous active power $p_k$ and reactive power $q_k$ satisfy $\begin{bsmallmatrix} p_k&q_k\\-q_k&p_k\end{bsmallmatrix} v_k = (v_k^\top v_k) i_k$.
To regulate the voltage magnitude, a quadratic correction term is further introduced, yielding the complete dVOC dynamics \cite{he2024quantitative}:
\begin{equation*}
\dot{ v}_k= \omega_0 J v_k +\eta_k R(\varphi) \left(S_k^\star v_k- i_k \right) +\eta_k\alpha_k\frac{v_k^{\star 2}-\norm{v_k}^2}{v_k^{\star 2}} v_k,
\end{equation*}
where $\alpha_k\geq0$ determines the voltage-regulation strength. The last term directly adjusts the voltage magnitude, increasing it when $\norm{v_k}<v_k^\star$ and decreasing it when $\norm{v_k}>v_k^\star$.

Notice that in a network of interconnected oscillators, the output currents are coupled through the network Laplacian $\mathcal{L}$, such that ${i}=\mathcal{L}{v}$, where ${i}$ and ${v}$ collect the output currents and voltages of all oscillators, respectively. This coupling enables oscillator synchronization through current feedback.

To reveal the connection between dVOC and classical droop control, we consider $\varphi = \pi/2$ for purely inductive networks and express the dVOC dynamics in polar coordinates, $\norm{v_k} = \sqrt{v_{\alpha,k}^2 + v_{\beta,k}^2},\ \theta_k = \arctan(v_{\beta,k}/v_{\alpha,k})$, which gives
\begin{equation*}
\begin{aligned}
\frac{1}{\norm{v_k}}\frac{\mathrm{d} \norm{v_k}} {\mathrm{d} t }&=\eta_k 
\left(\frac{q_k^\star}{v_k^{\star 2}} - \frac{q_k}{\norm{v_k}^2} \right) + \eta_k\alpha_k\frac{v_k^{\star 2}-\norm{v_k}^2}{v_k^{\star 2}},\\
\frac{\mathrm{d} \theta_k} {\mathrm{d} t}& = \omega_0 + \eta_k \left(\frac{p_k^\star}{v_k^{\star 2}} - \frac{p_k}{\norm{v_k}^2} \right),
\end{aligned}
\end{equation*}
which resembles the classical droop control. Namely, the active-power deviation regulates frequency, while the reactive-power deviation and voltage-magnitude feedback regulate the voltage magnitude.
To make this connection more explicit, define the complex angle as $\underline{\vartheta}_k \coloneqq  \ln{\norm{v_k}} + j\theta_k$, the complex frequency as $\underline\varpi_k \coloneqq \frac{1}{\norm{v_k}}\frac{\mathrm{d} \norm{v_k}} {\mathrm{d} t }+j\frac{\mathrm{d} \theta_k} {\mathrm{d} t}$, and the complex power feedback as $\underline{\varsigma}_k \coloneqq \frac{p_k - jq_k} {\norm{v_k}^2}$, dVOC can be interpreted as a nonlinear complex-frequency droop controller with complex-power feedback \cite{he2024quantitative}:
\begin{equation*}
    \dot{\underline{\vartheta}_k} = j \omega_0 + \eta_k e^{j\pi/2}(\underline{\varsigma}_k^\star - \underline{\varsigma}_k) + \eta_k\alpha_k\frac{v_k^{\star 2}-(\Re{\underline{\vartheta}_k})^2}{v_k^{\star 2}},
\end{equation*}
This formulation highlights the connection between dVOC and droop control. Unlike conventional linear droop control, dVOC incorporates nonlinear feedback, which underlies its nonlinear passivity and stability property.

\end{sidebar}

\subsubsection{Nonlinear Feedback Interconnection Modeling}

We consider $N$ converters that are connected to a grid bus, and denote by ${v}_k \coloneqq [v_{d,k},\ v_{q,k}]^\top $ and ${i}_k \coloneqq [i_{d,k},\ i_{ q,k}]^\top $ the terminal voltage and output current of converter $k$. The dVOC's voltage dynamics can be expressed in the grid's \textit{global dq reference frame} as \cite{he2024quantitative}
\begin{equation}
\label{eq:dvoc}
    \dot{{v}}_k = \omega_{\Delta} J {v}_k + \eta_k R(\varphi) \bigl( S_k^{\star} {v}_k -  i_k \bigr) + \eta_k \alpha_k \tfrac{v_k^{\star 2} - \norm{v_k}^2}{v_k^{\star 2}} {v}_k,
\end{equation}
where $\omega_{\Delta} \coloneqq \omega_0 - \omega_{\rm g}$ denotes the nominal frequency setpoint $\omega_0$ relative to the grid frequency $\omega_g$, $\eta_k {>0}$ and $\alpha_k {>0}$ are the controller gains, and
\begin{equation*}
\small
    J \coloneqq
    \begin{bmatrix}
        0 & -1 \\
        1 & 0
    \end{bmatrix},\
    R(\varphi) \coloneqq
    \begin{bmatrix}
        \cos{\varphi} & -\sin{\varphi} \\
        \sin{\varphi} & \cos{\varphi}
    \end{bmatrix},\
    S_k^{\star} \coloneqq \frac{1}{v_k^{\star 2}}
    \begin{bmatrix}
        p_k^{\star} & q_k^{\star} \\
        -q_k^{\star} & p_k^{\star}
    \end{bmatrix},
\end{equation*}
and ${\varphi} \in [0,\pi/2]$ is a rotation angle parameter used to transform the feedback current, typically selected according to the network impedance angle. The quantities $p_k^{\star}$, $q_k^{\star}$, and $v_k^{\star}$ specify the active-power, reactive-power, and voltage-magnitude setpoints, respectively. The current-feedback term in \eqref{eq:dvoc} adjusts the relative phase to regulate the power setpoints, while the final term drives the voltage magnitude toward its prescribed value.

For $N$ converters, let ${v} \coloneqq [ v_1,\cdots, v_N]^\top$ and ${i} \coloneqq [ i_1,\cdots, i_N]^\top$. We assume that the electromagnetic dynamics of the network and the inner-loop dynamics of the converters evolve on time scales sufficiently faster than the dVOC dynamics. The following phasor relation can then represent the network:
\begin{equation}
\label{eq:network}
    \begin{bmatrix}
        i_{d,1} + j i_{ q,1} \\
        \vdots\\
        i_{d,N} + j i_{ q,N} \\
    \end{bmatrix}
    = \underline{Y}
    \begin{bmatrix}
        v_{d,1} + j v_{ q,1} \\
        \vdots\\
        v_{d,N} + j v_{ q,N} \\
    \end{bmatrix} - \underline{y} v_{\rm g},
\end{equation}
where the complex-valued admittance matrix $\underline{Y}$ (symmetric) and vector $\underline{y}$ are obtained by Kron reduction that eliminates the intermediate nodes of the collector network, as illustrated in Figure~\ref{fig:power-plant}; $v_{\rm g}$ denotes the grid voltage magnitude. We note that this large-signal, static network model is similar to, but differs from, the small-signal, dynamic model introduced earlier in \eqref{eq:Y_grid}.

Combining \eqref{eq:dvoc} with \eqref{eq:network} gives the closed-loop interconnection. For the passivity analysis, it is convenient to work with an equivalent representation in which the network admittance is rotated by the same angle $\varphi$ used in dVOC. We define the rotated currents by $ i_{\varphi,k} \coloneqq R(\varphi) i_k$, which, in phasor coordinates, is equivalent to $ i_{\varphi d,k} + j i_{\varphi q,k} = e^{j \varphi} (i_{ d,k} + j i_{ q,k})$. Accordingly, we denote the network's rotated admittance matrix by $e^{j \varphi} \underline{Y}$ and $e^{j \varphi} \underline{y}$.
The node and network equations in \eqref{eq:dvoc} and \eqref{eq:network} then take the equivalent form:
\begin{align}
    \dot{{v}}_k &= \eta_k \bigl[\tfrac{\omega_{\Delta}}{\eta_k} J {v}_k + R(\varphi)S_k^{\star} {v}_k -  i_{\varphi,k} + \alpha_k \tfrac{v_k^{\star 2} - \norm{v_k} ^2}{v_k^{\star 2}} I_2 {v}_k \bigr],
\label{eq:rotated-dvoc} \\
\begin{split}
    {i}_{\varphi} &= \left[\Re\{e^{j \varphi} \underline{Y}\} \otimes I_2 +  \Im\{e^{j \varphi} \underline{Y}\} \otimes J \right] {v}  \\
    &\quad - \left[\Re\{e^{j \varphi} \underline{y}\} \otimes I_2 +  \Im\{e^{j \varphi} \underline{y}\} \otimes J\right] [v_{\rm g}\ \ 0]^\top.
\end{split}
\label{eq:rotated-network}
\end{align}

% =================================================================================
\subsubsection{Nonlinear Passivity Results}
\label{sec:passivity analysis}

We use the subscript $\rm s$ to denote steady-state quantities, with ${v}_{\rm s} = [{v}_{{\rm s},1}, \cdots, {v}_{{\rm s},N}]^\top$ and ${i}_{\varphi{\rm s}} = [{i}_{\varphi{\rm s},1}, \cdots, {i}_{\varphi{\rm s},N}]^\top$. The incremental variables relative to this steady-state equilibrium point are $\Tilde{{v}} \coloneqq {v} - {v}_{\rm s}$ and $\Tilde{{i}}_{\varphi} \coloneqq {i}_{\varphi} - {i}_{\varphi{\rm s}}$. The passivity property of the dVOC's nonlinear dynamics with respect to the steady-state equilibrium is described as follows.

\begin{highlightbox}
\textbf{Passivity Index of a Node's Nonlinear Dynamics}

A node's terminal dynamics controlled by dVOC in \eqref{eq:rotated-dvoc}, with input $- i_{\varphi,k}$ and output ${v}_k$, are output-feedback passive around an equilibrium ${v}_{{\rm s},k}$, with the OFP index:
\begin{equation*}
    \rho_{\mathrm{node},k} = - \left(\tfrac{p_k^\star \cos\varphi + q_k^\star \sin\varphi}{v_k^{\star 2}} + \alpha_k - \alpha_k \tfrac{1}{2 v_k^{\star 2}}  \norm{v_{{\rm s},k}}^2 \right).
\end{equation*}
\end{highlightbox}

To show the output-feedback passivity, we consider the following positive-definite storage function:
\begin{equation*}
    V_k \coloneqq \tfrac{1}{2 \eta_k} \norm{v_k -  v_{{\rm s},k}}^2,
\end{equation*}
By the same argument as in \cite{he2024passivity}, we obtain
\begin{equation}
\label{eq:bound}
\begin{aligned}
     {\dot V}_k & \leq (\tfrac{p_k^\star \cos\varphi + q_k^\star \sin\varphi}{v_k^{\star 2}} + \alpha_k - \alpha_k \tfrac{1}{2 v_k^{\star 2}} \norm {v_{{\rm s},k}}^2) \norm{v_k -  v_{{\rm s},k} }^2 \\
     &\ \quad + ( v_k - v_{{\rm s},k})^\top ( i_{\varphi {\rm s},k} -  i_{\varphi,k}),
\end{aligned}
\end{equation}
which indicates that the node dynamics with incremental input $ u_k \coloneqq -(  i_{\varphi,k} -  i_{\varphi {\rm s},k} )$ and output $ y_k \coloneqq ( v_k -  v_{{\rm s},k})$ is output-feedback passive, as defined in the standard form $ u ^\top  y \geq \dot{V} + \rho_{\mathrm{node},k}  y ^\top  y$ in \eqref{eq:passivity_index_supply}, where the node output-feedback passivity index is given by $\rho_{\mathrm{node},k} = - \left(\tfrac{p_k^\star \cos\varphi + q_k^\star \sin\varphi}{v_k^{\star 2}} + \alpha_k - \alpha_k \tfrac{1}{2 v_k^{\star 2}}  \norm{v_{{\rm s},k}}^2 \right)$.

Inequality \eqref{eq:bound} admits a direct physical interpretation: the power entering the converter accounts for both the change in the controller storage function and the passivity excess or deficit measured by the passivity index. The index further connects this excess or deficit to the controller setpoints, gains, and steady-state voltage magnitude. The passivity excess is characterized by $\rho_{\mathrm{node},k} > 0$, whereas $\rho_{\mathrm{node},k} < 0$ indicates a passivity shortage. Typically, $\rho_{\mathrm{node},k} < 0$ occurs since $\alpha_k > 0$. A larger power setpoint tends to reduce passivity, whereas a higher voltage level has the opposite effect, consistent with engineering experience. Moreover, the last term in $\rho_{\mathrm{node},k}$ can be omitted for a conservative evaluation, making the result independent of global equilibrium information $\norm {v_{{\rm s},k}}$ without compromising rigor because this term is always positive. Alternatively, by approximating $v_k^{\star} \approx  \norm{v_{{\rm s},k}}$ at steady state, we obtain $\rho_{\mathrm{node},k} \approx - \left(\tfrac{p_k^\star \cos\varphi + q_k^\star \sin\varphi}{v_k^{\star 2}} + \frac{1}{2}\alpha_k \right)$, which provides a rough estimate of the passivity index.

Next, we establish the passivity of the network coupling around the steady-state equilibrium.

\begin{highlightbox}
\textbf{Passivity Index of the Network Coupling}

The static network equation, as a memoryless function in \eqref{eq:rotated-network}, with ${v}$ as input and $ i_{\varphi}$ as output, is input-feedforward passive with respect to an equilibrium ${v}_{\rm s}$ and ${i}_{\varphi{\rm s}}$, with the IFP index:
\begin{equation*}
    \nu_{\rm net} = \lambda_{\min} \bigl(\Re \{e^{j\varphi} \underline{Y} \} \bigr).
\end{equation*}
\end{highlightbox}

To show the input-feedforward passivity of the network, we use incremental variables in \eqref{eq:rotated-network}, which gives $\Tilde{{i}}_{\varphi} = \left[\Re\{e^{j \varphi} \underline{Y}\} \otimes I_2 +  \Im\{e^{j \varphi} \underline{Y}\} \otimes J \right] \Tilde{{v}}$. By the same argument as in \cite{he2024passivity}, it follows that
\begin{equation}
\label{eq:network-passivity}
\begin{aligned}
    \Tilde{{v}}^\top \Tilde{{i}}_{\varphi} &= \Tilde{{v}}^\top \left[\Re\{e^{j \varphi} \underline{Y}\} \otimes I_2 +  \Im\{e^{j \varphi} \underline{Y}\} \otimes J \right] \Tilde{{v}} \\
    &= \Tilde{{v}}^\top \left[\Re\{e^{j \varphi} \underline{Y}\} \otimes I_2 \right] \Tilde{{v}} \\
    & \ge \lambda_{\min} \bigl(\Re \{e^{j\varphi} \underline{Y} \}\bigr) \Tilde{{v}}^\top \Tilde{{v}},
\end{aligned}
\end{equation}
where $\underline{Y}$ is a symmetric matrix, $J^\top = -J$, and $\lambda_{\min}$ denotes the minimum eigenvalue. The input-feedforward passivity index of the network is, therefore, given as $\nu_{\rm net} \coloneqq \lambda_{\min} \bigl(\Re \{e^{j\varphi} \underline{Y} \} \bigr) > 0$.
For the network, $\Re \{e^{j\varphi} \underline{Y} \}$ is real symmetric positive-definite because $\underline{Y}$ is nonsingular after removing the row and column associated with the grid node.

Since the admittance matrix $\underline{Y}$ is predominantly inductive, the network's passivity index, $\lambda_{\min} \bigl(\Re \{e^{j\varphi} \underline{Y} \} \bigr)$, provides an important engineering insight: the rotation angle $\varphi$ in the GFM control should be aligned with the impedance angle of the network line to enhance network passivity. Conversely, a mismatch between the two angles reduces network passivity and may therefore fail to compensate for the passivity deficit at the nodes.

\subsubsection{Nonlinear Stability Results}

The following result combines the node and network passivity properties to obtain nonlinear stability conditions for the closed-loop system.

\begin{highlightbox}
\textbf{Decentralized Stability Conditions}

Assume that an equilibrium point exists. The multi-converter grid-connected system described by \eqref{eq:rotated-dvoc} and \eqref{eq:rotated-network} is asymptotically stable about this equilibrium if the following decentralized conditions hold:
\begin{equation}
\label{condi:stability}
    \rho_{\mathrm{node},k} + \nu_{\rm net} > 0,\ \forall k \in \{1,\cdots,N\}.
\end{equation}
If the equilibrium is unique, this stability property is global. 
\end{highlightbox}

\begin{figure}[!t]
\centering
\includegraphics{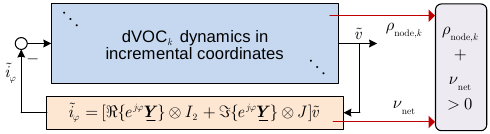}
\caption{The output-feedback passivity of the dVOC-controlled node dynamics and the input-feedforward passivity of the (rotated) network establish decentralized stability conditions: $\rho_{\mathrm{node},k} + \nu_{\mathrm{net}} > 0,\ \forall k$.}
\label{fig:power-plant-passivity}
\end{figure}

Figure~\ref{fig:power-plant-passivity} illustrates the decentralized stability results, in which the passivity of each converter is assessed independently, allowing the stability of the overall system to be certified in a decentralized manner.

To prove the stability conditions, we define the composite Lyapunov function $\nu \coloneqq \sum\nolimits_k V_k$. It is positive definite and radially unbounded with respect to the equilibrium. From \eqref{eq:bound} and \eqref{eq:network-passivity}, it follows that
\begin{equation*}
    \begin{aligned}
        \dot \nu &= \sum\nolimits_k \dot V_k \\
        &\leq \sum\nolimits_k (v_k - v_{{\rm s},k} )^\top (  i_{\varphi {\rm s},k} -  i_{\varphi,k} ) - \rho_{\mathrm{node},k} \norm {v_k -  v_{{\rm s},k}}^2 \\
        &\leq -\Tilde{{v}}^\top \Tilde{{i}}_{\varphi} - \min_k\rho_{\mathrm{node},k}\, \Tilde{{v}} ^\top \Tilde{{v}} \\
        & \leq -\bigl( \lambda_{\min} \bigl(\Re \{e^{j\varphi} \underline{Y} \} \bigr) + \min_k \rho_{\mathrm{node},k} \bigr) \Tilde{{v}} ^\top \Tilde{{v}} \\
        &= -\bigl( \nu_{\rm net} + \min_k \rho_{\mathrm{node},k} \bigr) \Tilde{{v}} ^\top \Tilde{{v}}.
    \end{aligned}
\end{equation*}
Condition \eqref{condi:stability} makes the composite $\dot \nu$ negative definite with respect to the equilibrium. As a result, the equilibrium is globally asymptotically stable (more specifically, globally exponentially stable). The condition does not impose strict passivity on every subsystem. A positive network passivity index can instead offset a negative node passivity index, thereby reducing the conservatism of the decentralized condition.

Alternatively, the dVOC dynamics can be linearized around a steady-state operating point, and frequency-domain passivity analysis can then be used to derive decentralized stability conditions as before. These conditions are less restrictive but apply only to small-signal stability.

% =================================================================================
\subsubsection{Case Studies}

To further validate the decentralized stability condition in \eqref{condi:stability}, we conduct case studies and time-domain simulations using the three-converter 9-bus system shown in Figure~\ref{fig:9bus_topology}. All converters are controlled using dVOC, and the network has a uniform $X/R$ ratio of 10. The voltage setpoints are 1.0~pu, the control gains are $\eta_k = 0.022 \times 100\pi$~rad/s, and $\alpha_k = 1$~pu. The power setpoints of the three converters are $(p^\star_1,, q^\star_1) = (0.8,, 0.1)$~pu, $(p^\star_2,, q^\star_2) = (1.0,, 0.0)$~pu, and $(p^\star_3,, q^\star_3) = (0.5,, 0.2)$~pu.

\begin{figure}[!t]
\centering
\includegraphics[width=0.9\linewidth]{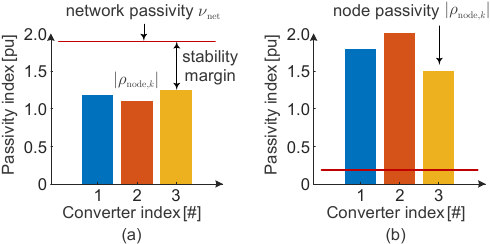}
\caption{Node and network passivity indices in the case study with two different rotation transformation angles: (a) $\varphi = \arctan(X/R)$ and (b) $\varphi = 0$. Note that, since $\rho_{\mathrm{node},k} < 0$, the decentralized stability condition $\rho_{\mathrm{node},k} + \nu_{\rm net} > 0$ is rewritten as $\nu_{\rm net} > \vert \rho_{\mathrm{node},k} \vert$.}
\label{fig:case-study-passivity-index}
\end{figure}

We investigate the impact of the rotation angle $\varphi$, i.e., the control parameter in dVOC used to transform the current feedback, by considering two cases: $\varphi=\arctan(X/R)$ and $\varphi=0$. The calculated passivity indices are shown in Figure~\ref{fig:case-study-passivity-index}. For $\varphi=\arctan(X/R)$ in subfigure (a), which aligns the rotation angle with the network impedance angle, the absolute values of the negative node passivity indices, $\vert \rho_{\mathrm{node},k} \vert$ with $\rho_{\mathrm{node},k} < 0$, are smaller than the network passivity index for all converters. Hence, the decentralized stability condition $\rho_{\mathrm{node},k} + \nu_{\rm net} > 0$, i.e., $\nu_{\rm net} > \vert \rho_{\mathrm{node},k} \vert$, is satisfied for all $k$. The difference between the network passivity index $\nu_{\rm net}$ and the magnitude of the corresponding node passivity deficit $\vert \rho_{\mathrm{node},k} \vert$ provides a measure of the stability margin. In contrast, for $\varphi=0$ in Figure~\ref{fig:case-study-passivity-index}(b), the mismatch between the rotation angle in dVOC and the network impedance angle increases the magnitude of the negative node passivity indices while significantly reducing the network passivity index. The network passivity is therefore insufficient to compensate for the node passivity deficits, and the decentralized stability condition is violated.

\begin{figure}[!t]
\centering
\includegraphics[width=0.9\linewidth]{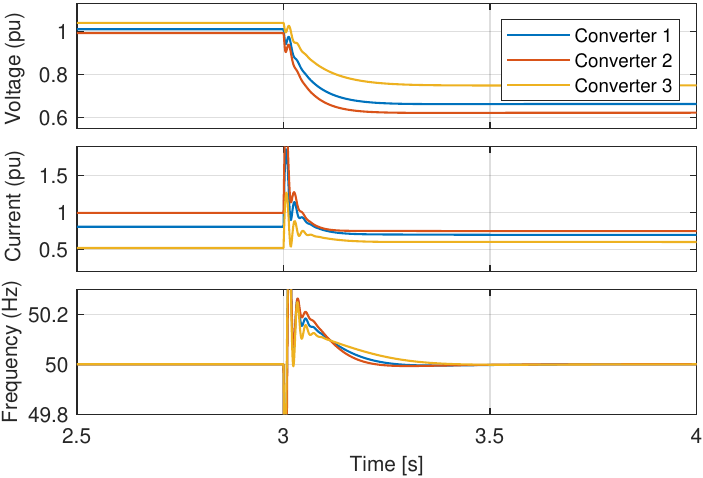}
\caption{The three-GFM-converter system remains stable under a grid voltage dip to 0.5 pu, with the rotation angle $\varphi = \arctan(X/R)$.}
\label{fig:case-study-stable}
\end{figure}

\begin{figure}[!t]
\centering
\includegraphics[width=0.9\linewidth]{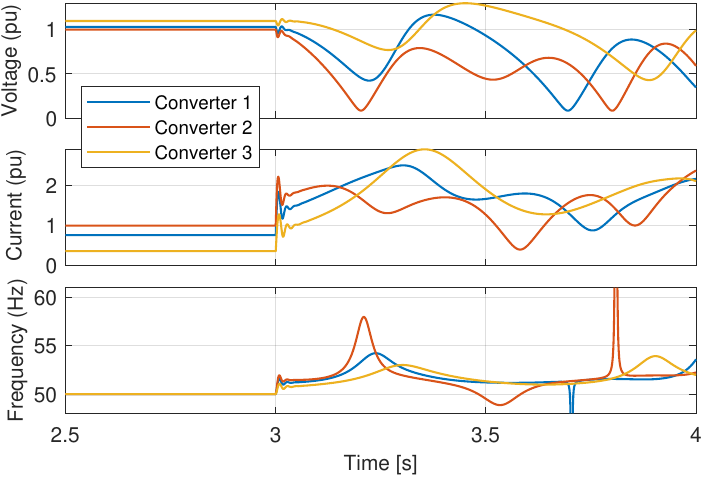}
\caption{The three-GFM-converter system becomes unstable under a grid voltage dip to 0.5 pu, with the rotation angle $\varphi = 0$.}
\label{fig:case-study-unstable}
\end{figure}

The corresponding time-domain simulation results are shown in Figures~\ref{fig:case-study-stable} and \ref{fig:case-study-unstable}. At $t=3$~s, the grid voltage magnitude is reduced from 1.0~pu to 0.5~pu and then maintained at 0.5~pu. For $\varphi=\arctan(X/R)$, the time-domain response in Figure~\ref{fig:case-study-stable} is consistent with the passivity analysis: although the converters experience a substantial current transient following the voltage dip, their voltages and powers remain bounded and converge to a new operating point. In contrast, for $\varphi=0$, Figure~\ref{fig:case-study-unstable} shows that the system loses stability following the same 0.5~pu voltage dip at $t=3$~s. These results demonstrate the consistency between the decentralized passivity certificate and the nonlinear time-domain behavior, and highlight the importance of selecting the rotation angle in accordance with the network impedance.

\section{Conclusion}
\label{sec:conclusion}

Decentralized stability certificates enable the scalable analysis, specification of grid codes, and interoperability-oriented control design of large-scale PE-dominated power systems, where conventional centralized approaches are often analytically and computationally intractable for handling the device-network interactions. We revisit the classic dissipativity conditions, such as passivity and the small-gain theorem, introduce the small-phase theorem that generalizes the passivity condition, and further derive a variety of stability certificates based on DW shell and its projections. We show that DW shell, tracing back to the seminal works by Davis and Wielandt in the 1950s, provides a unified framework to derive decentralized stability certificates for linearized stability, where existing stability results like passivity, the small-gain theorem, and the small-phase theorem can be thought of as special cases of DW shell separation. 
Leveraging these geometric properties, we presented a versatile suite of stability certificates. By offering varying degrees of conservatism and computational complexity, these diverse conditions can be deployed complementarily across different frequency spectra. Ultimately, these non-parametric tools facilitate the modular analysis of complex grids and provide explicit design guidance to guarantee plug-and-play converter interoperability.
% Based on DW shell, we present a set of geometric decentralized stability certificates where different conditions, with different levels of conservatism and computational effort, can be used complementarily at different frequencies. We demonstrate how these non-parametric certificates, based on the input/output behavior, are suited to analyze PE-dominated power systems in a modular and scalable fashion, and guide the design of converters to ensure interoperability. 
We further derive parametric decentralized stability certificates for GFM converters to provide explicit conditions for the tuning and a priori specifications of control parameters. We show that parametric nonlinear stability conditions can be derived based on passivity for dVOC-controlled GFM converters, while it remains an open question how geometric approaches such as DW shell and its projections enable decentralized nonlinear stability conditions with reduced conservatism. 

We envision these decentralized stability certificates evolving into the foundational rulebook--the future grid code--ensuring the reliable, interoperable operation of PE-dominated power systems. In fact, some grid codes, such as the one published by UNIFI, have already involved the passivity criterion. 
The critical challenge ahead lies in bridging the gap between advanced control theory and industrial practice: translating these rigorous mathematical tools into straightforward, intuitive standards that practicing engineers can seamlessly implement within digital controllers. Future work will also include systematic control design methods to enforce the decentralized stability certificates.
% We envision that decentralized stability certificates should eventually be part of the future grid code to ensure interoperability and stable operation of PE-dominated power systems, while it remains a challenge to make such certificates straightforward for engineers to understand and implement in digital controllers. 

% \section*{Acknowledgment}

% The authors wish to thank ...

% The research leading to these results received funding from: the National Natural Science Foundation of China (U24B6008).

% \section*{Author Information}

% \begin{IEEEbiography}{{L}inbin Huang}{\,}(linhuang@ethz.ch)
% is a postdoctoral researcher with the Automatic Control Laboratory at ETH Zürich. He received the B.Eng. and Ph.D. degrees from Zhejiang University, Hangzhou, China, in 2015 and 2020, respectively. His research interests include power system stability, optimal control of power electronics, and data-driven control.
% \end{IEEEbiography}

\bibliographystyle{IEEEtran}
\bibliography{one-bib} 
\longpage

%\processdelayedfloats

\end{document}